\documentstyle [epsf,epsfig,axodraw,xspace,12pt]{article}  
\newcommand{\ra} {$\rightarrow$}

\newcommand{\be}{\begin{equation}}
\newcommand{\ee}{\end{equation}}

\newcommand{\bea}{\begin{eqnarray}}
\newcommand{\eea}{\end{eqnarray}}
\newcommand{\beanon}{\begin{eqnarray*}}
\newcommand{\eeanon}{\end{eqnarray*}}
\newcommand{\ba}{\begin{array}}
\newcommand{\ea}{\end{array}}
\newcommand{\bd}{\begin{description}}
\newcommand{\ed}{\end{description}}
\newcommand{\bi}{\begin{itemize}}
\newcommand{\ei}{\end{itemize}}
\newcommand{\ben}{\begin{enumerate}}
\newcommand{\een}{\end{enumerate}}
\newcommand{\bc}{\begin{center}}
\newcommand{\ec}{\end{center}}

\newcommand{\ol}{\overline}

\newcommand{\toptop}{\mbox{${\mathrm t}{\mathrm \bar t}$}\xspace}
\newcommand{\WWL}{\mbox{${\mathrm W_L}{\mathrm W_L}$}\xspace}
\newcommand{\VVL}{\mbox{${\mathrm V_L}{\mathrm V_L}$}\xspace}

\newcommand{\WWT}{\mbox{${\mathrm W_T}{\mathrm W_T}$}\xspace}
\newcommand{\WW}{\mbox{${\mathrm W}{\mathrm W}$}\xspace}
\newcommand{\ZZ}{\mbox{${\mathrm Z}{\mathrm Z}$}\xspace}
\newcommand{\VV}{\mbox{${\mathrm V}{\mathrm V}$}\xspace}
\newcommand{\ZW}{\mbox{${\mathrm Z}{\mathrm W}$}\xspace}
\newcommand{\VW}{\mbox{${\mathrm V}{\mathrm W}$}\xspace}
\newcommand{\W}{\mbox{${\mathrm W}$}\xspace}
\newcommand{\V}{\mbox{${\mathrm V}$}\xspace}
\newcommand{\VL}{\mbox{${\mathrm V_L}$}\xspace}
\newcommand{\Z}{\mbox{${\mathrm Z}$}\xspace}
\newcommand{\pt}{\mbox{${\mathrm p_T}$}\xspace}
\newcommand{\GeV}{\mbox{${\mathrm GeV}$}\xspace}
\newcommand{\TeV}{\mbox{${\mathrm TeV}$}\xspace}

\newcommand{\eqn}[1]{Eq.(\ref{#1})}

\newcommand{\tbn}[1]{Tab.~\ref{#1}}

\newcommand{\fig}[1]{Fig.~\ref{#1}}

\newcommand{\sect}[1]{Sect.~\ref{#1}}
\newcommand{\subsect}[1]{Sub-Sect.~\ref{#1}}

\renewcommand{\O}{{\mathcal O}}

\newcommand{\Phase}{{\tt PHASE}\xspace}
\newcommand{\Pythia}{{\tt PYTHIA}\xspace}
\newcommand{\MadEvent}{{\tt MADEVENT}\xspace}

\def\pl #1 #2 #3 {{\it Phys.~Lett.} {\bf#1} (#2) #3}   
\def\np #1 #2 #3 {{\it Nucl.~Phys.} {\bf#1} (#2) #3}
\def\zp #1 #2 #3 {{\it Z.~Phys.} {\bf#1} (#2) #3}
\def\pr #1 #2 #3 {{\it Phys.~Rev.} {\bf#1} (#2) #3}
\def\prep #1 #2 #3 {{\it Phys.~Rep.} {\bf#1} (#2) #3}
\def\prl #1 #2 #3 {{\it Phys.~Rev.~Lett.} {\bf#1} (#2) #3}
\def\intj #1 #2 #3 {{\it Int. J. Mod. Phys.} {\bf#1} (#2) #3}
\def\mpl #1 #2 #3 {{\it Mod.~Phys.~Lett.} {\bf#1} (#2) #3}
\def\rmp #1 #2 #3 {{\it Rev. Mod. Phys.} {\bf#1} (#2) #3}
\def\cpc #1 #2 #3 {{\it Comp. Phys. Commun.} {\bf#1} (#2) #3}
\def\epj #1 #2 #3 {{\it Eur. Phys. J.} {\bf#1} (#2) #3}
\def\jhep #1 #2 #3 {{\it JHEP} {\bf#1} (#2) #3}

\begin{document}

\title{Boson-boson scattering and Higgs production at the LHC
   from a six fermion point of view: four jets + l$\nu$
       processes at $\O(\alpha_{em}^6)$.}

\author{E. Accomando$^{1,2}$,
        A. Ballestrero$^{1,2}$,
        S. Bolognesi$^{1,3}$,
        E. Maina$^{1,2}$,
        C. Mariotti$^{1}$\\
{\it $^1$INFN, Sezione di Torino, Italy}\\
{\it $^2$Dipartimento di Fisica Teorica, Universit\`a di Torino, Italy}\\
{\it $^3$Dipartimento di Fisica Sperimentale, Universit\`a di Torino,
Italy}
}

\maketitle

\begin{abstract}
Boson-boson scattering and Higgs production in
boson-boson fusion hold the key to electroweak symmetry breaking.
In order to analyze these essential features of the Standard Model
we have performed a partonic level study of all processes
$q_1 q_2 \rightarrow q_3 q_4 q_5 q_6 l \nu$ at the LHC using the exact matrix
elements at $\O(\alpha_{em}^6)$ provided by \Phase, a new MC generator.
These processes include also three boson
production and the purely electroweak contribution to \toptop production
as well as all irreducible backgrounds. 
Kinematical cuts have been studied in order to enhance the VV scattering
signal over background. \Phase has been compared with different Monte Carlo's
showing that a complete calculation is necessary for a correct
description of the process.
\end{abstract}



\hfill DFTT 37/2005
\vfill\noindent
{\small E.A. is supported by the Italian Ministero dell'Istruzione, 
dell'Universit\`a e della Ricerca (MIUR) under contract Decreto MIUR 
26-01-2001 N.13 ``Incentivazione alla mobilit\`a di studiosi stranieri ed 
italiani residenti all'estero''.\\
Work supported by MIUR under contract 2004021808\_009.}

\clearpage
 
\section{Introduction}
\label{sec:intro}
The nature of Electro--Weak Symmetry Breaking (EWSB) will be investigated at the
LHC\footnote{Detailed reviews and extensive
bibliographies can be found in 
Refs.\cite{HiggsLHC,djouadi-rev1,ATLAS-TDR,Houches2003}}.
The Standard Model (SM) provides the simplest and most economical
explanation of the phenomenon and all its predictions have been verified with
unprecedented accuracy.
The only missing ingredient is the Higgs boson which is essential to the
renormalizability of the theory. 
The fit of precision EW data to the SM currently gives an upper limit on the
Higgs mass of about 200 \GeV, depending on the preferred value of the top
mass \cite{lepewwg}.
A heavier Higgs mass can however be accomodated by a number of models
which in general would be distinguishable from the SM by improved precision data
or by other observations at future colliders \cite{peskin-wells-01} 

In the SM the Higgs is also crucial to 
ensure that perturbative unitarity bounds are not violated in high energy
reactions. Scattering processes between longitudinally polarized vector bosons
are particularly sensitive in this regard and without a Higgs they
violate unitarity at about one \TeV.

It should be noted that the Goldstone theorem and the Higgs mechanism do not
require
the existence of elementary scalars. It is conceiveable and widely discussed in
the literature that bound states are responsible for EWSB.

Since unitarity is essentially a statement of conservation of total probability
it cannot be violated in Nature. Violation of perturbative unitarity implies
that the SM becomes a strongly interacting theory at high energy.
The nature of the interaction between longitudinally polarized vector bosons
and the Higgs mass, or possibly the absence of
the Higgs particle, are strongly related: if a relatively light Higgs exists than
the \VL's  are weakly coupled, while they are strongly interacting if the Higgs
mass is large or the Higgs is nonexistent \cite{reviews}.
In the latter case, 
by analogy with low energy QCD, which can be expressed by exactly the same
formalism which describes the Higgs sector in the SM,
or adopting one of the many schemes for turning
perturbative scattering amplitudes into amplitudes which satisfy by construction
the unitarity constraints, one is led to expect the presence of resonances in 
\WWL scattering. Unfortunately the mass, spin and even number of these
resonances are not uniquely determined \cite{unitarization}.
If, on the other hand, 
a Higgs particle is discovered it will nonetheless
be necessary to verify that
indeed longitudinally polarized vector bosons are weakly coupled at high energy
by studying boson boson scattering in full detail.
If the Higgs mass is not far from the present experimental lower limits it will
take several years of data taking before enough statistics could be
accumulated for a reliable discovery.
In the meantime high energy vector boson scattering could provide an alternative
method for investigating the existence of a light Higgs scalar, providing a strong
incentive to probe harder the low mass range if no sign of strong scattering is
detected.

At the LHC no beam of on shell EW bosons will be available. Incoming quarks will
emit spacelike virtual bosons which will then scatter among themselves and 
finally decay.
These processes have been scrutinized since
a long time \cite{history1,history2}.
All previous studies of boson boson scattering at high energy hadron colliders
have resorted to some approximation, either the Equivalent Vector Boson
Approximation (EVBA) \cite{EVBA},
or a production times decay approach, supplementing a calculation of 
\begin{equation}
\label{2f2b}
q_1 q_2 \rightarrow q_3 q_4 V_1 V_2  
\end{equation}
processes with the on shell decay of the two vector bosons.
This is not surprising since the number of reactions required by a full six
fermion calculation is very large and often approximate results are
adequate for a first discussion.
There are however issues that cannot be tackled without a full six fermion
calculation like exact spin correlations between the decays of different heavy
particles, the effect of the non resonant background, the relevance of off mass
shellness of boson decays, the question of interferences between different
subamplitudes. Without a complete calculation it will be impossible to determine the
accuracy of approximate results.

Recently a full fledged six fermion Monte Carlo has become available
\cite{ref:Phase}, based on the methods of Refs. \cite{method,phact},
which in its present version describes at 
{$\O(\alpha_{em}^6)$}, using exact matrix elements,
all processes of the form

\begin{equation}
\label{6f}
PP \rightarrow q_1 q_2 \rightarrow q_3 q_4 q_5 q_6 l \nu  
\end{equation}

\noindent
(where $q_i$ stands for a generic (anti)quark) which can take place at the LHC.
This provides a complete description of \VV scattering and its EW irreducible
background. \eqn{6f} includes about one thousand different reactions.
A complete classification of all processes can be found in \cite{ref:Phase}.
Preliminary results have been presented in Refs. \cite{ACAT-QFTHEP}.

Since in addition to \VVL scattering many other subprocesses are in general
present in the full set of diagrams, it is not a trivial task to separate
boson boson scattering from the EW irreducible background.
In practice one has to deal also with other types of
background to which
QCD interactions contribute, but which however do not include any boson boson
scattering term. We will refer to these processes as QCD
background even though in general they will be a mixture of QCD and EW
interactions. \toptop production and decay is particularly relevant and
dangerous.
In this paper we are neglecting QCD backgrounds, whose calculation is
in progress, and concentrate on {$\O(\alpha_{em}^6)$} processes.
It is clear that obtaining a good signal over EW irreducible background ratio
is a prerequisite to any attempt at dealing with the QCD one.

In the absence of firm predictions in the strong scattering
regime, trying to gauge the possibilities of discovering signals of
new physics at the LHC requires the somewhat arbitrary definition of a model of
\VVL scattering beyond the boundaries of the SM. Some of these models predict
the formation of spectacular resonances which will be easily detected.
For some other set of parameters in the models only rather small effects are
expected \cite{unitarization}.

The simplest approach is to consider the SM in the presence of a very 
heavy Higgs.
While this entails the violation of perturbative unitarity, the linear rise of
the cross section with $\hat{s}$, the invariant mass squared in the hard
scattering, will be swamped by the decrease of the parton luminosities at large
momentum fractions and, as a consequence, will be particularly challenging to
detect.
At the LHC, for $M_H>$10 \TeV, all Born diagrams with Higgs propagators become
completely
negligible in the Unitary gauge, and the expectations for all processes in
\eqn{6f} reduce to those in the $M_H \rightarrow \infty$ limit.
In this paper we will compare this minimalistic definition of
physics beyond the Standard Model with the predictions of the SM for Higgs
masses within the reach of the LHC.
\begin{figure}
\begin{center}
\mbox{\epsfig{file=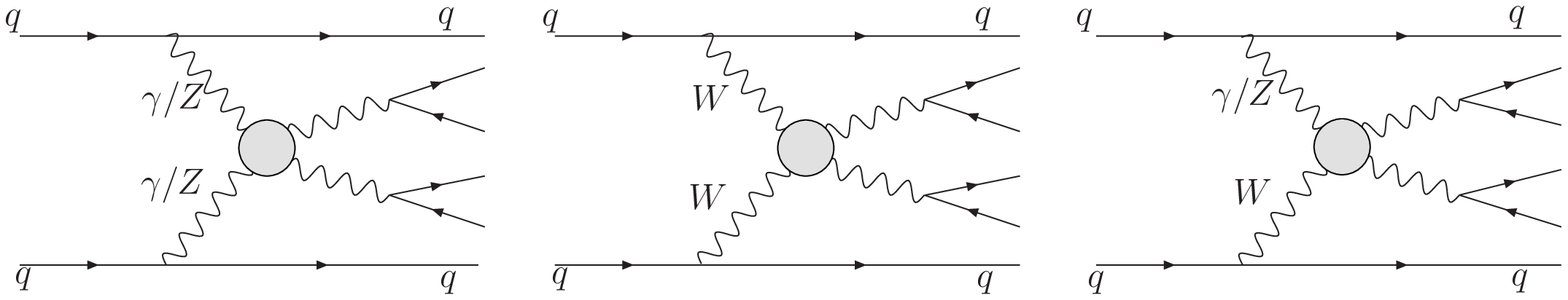,width=12.cm}}
\caption{ Vector boson fusion processes}
\label{VV-diag}
\end{center}
\end{figure}

\begin{figure}
\begin{center}
\mbox{\epsfig{file=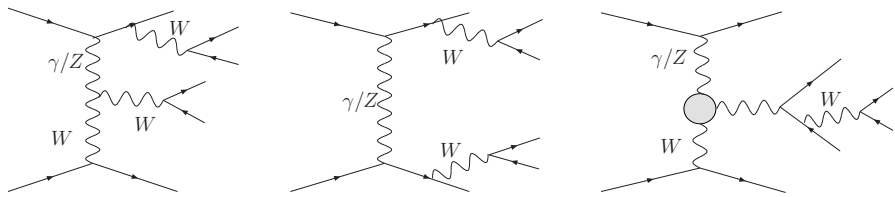,width=12.cm}}
\caption{ Non fusion and non doubly resonant two vector boson production.}
\label{nonreso-diag}
\end{center}
\end{figure}

\begin{figure}
\begin{center}
\mbox{\epsfig{file=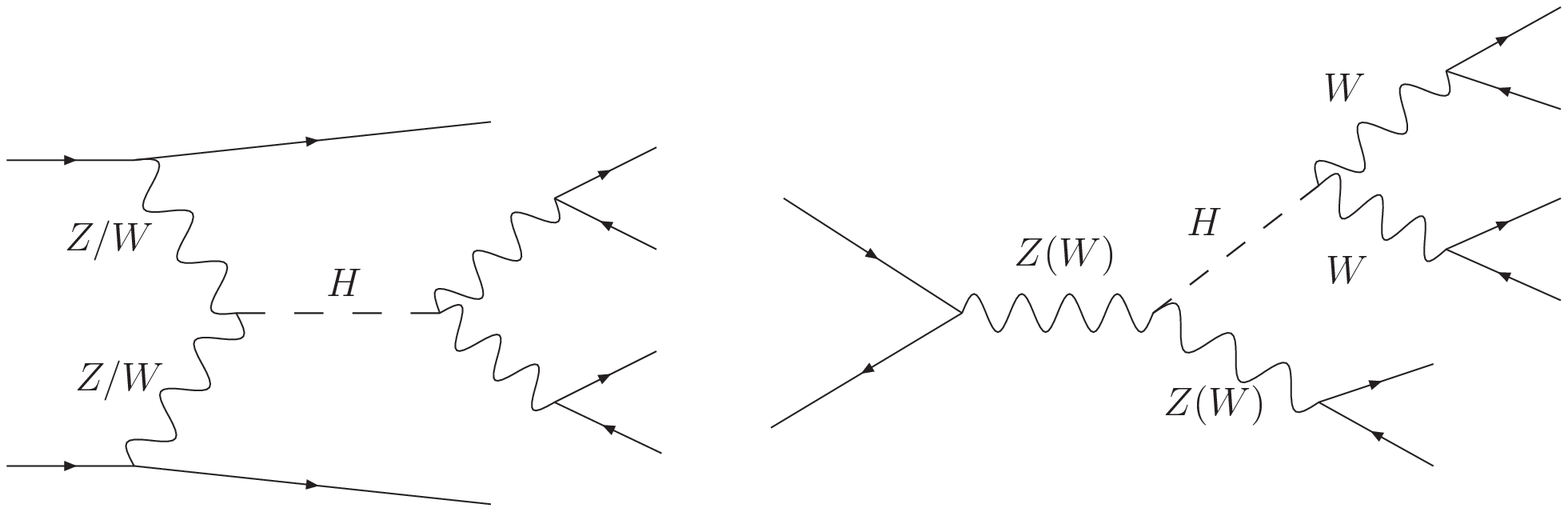,width=12.cm}}
\caption{Higgs boson production via vector boson fusion
  and Higgstrahlung.}
\label{higgs-diag}
\end{center}
\end{figure}

\begin{figure}
\begin{center}
\mbox{\epsfig{file=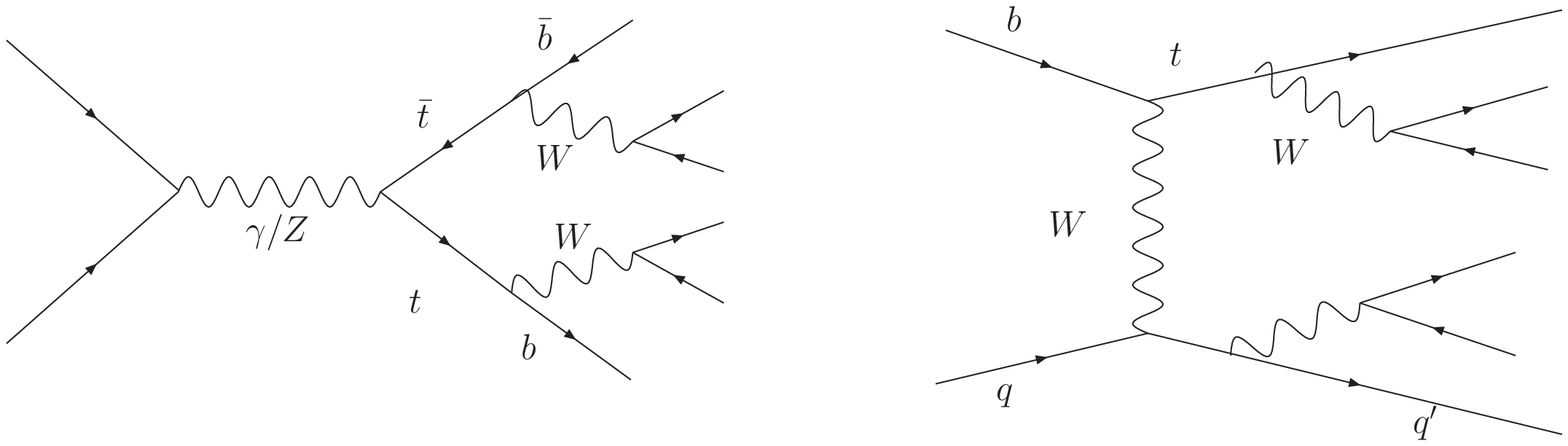,width=12.cm}}
\caption{ Electroweak \toptop and single top production.}
\label{top-diag}
\end{center}
\end{figure}

\begin{figure}
\begin{center}
\mbox{\epsfig{file=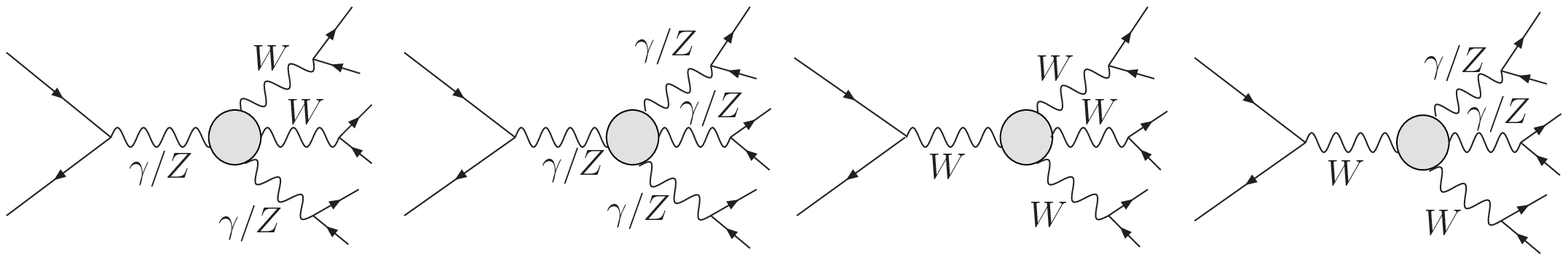,width=14.cm}}
\caption{ Three vector boson production.}
\label{tgc-diag}
\end{center}
\end{figure}

\begin{table}[bth]
\begin{center}
\begin{tabular}{|c|}
\hline
      E(lepton)$>20$ \GeV \\
\hline
      \pt(lepton)$> 10$ \GeV \\
\hline
      $\vert\eta$(lepton)$\vert<3$ \\
\hline
      E(quarks)$>20$ \GeV \\
\hline
      \pt(quarks)$>10$ \GeV \\
\hline
      $\vert\eta$(quark)$\vert<6.5$ \\
\hline
      M(qq)$> 20$ \GeV \\
\hline
\end{tabular}
\caption{Standard acceptance cuts applied in all results. Any pair of colored
fermions must have mass larger than 20 \GeV. Here lepton refers to $l^\pm$ only.} 
\label{standard-cuts}
\end{center}
\end{table}
 
\section{The six fermion final state processes} 
\label{sec:Phase}

Boson-boson scattering and Higgs production in boson-boson fusion
produce intermediate states with two bosons and two quarks as shown
in \fig{VV-diag}. 
In this study we have only considered final states in which one \W boson 
decays leptonically and the other (either \Z or \W) hadronically, which is
regarded as the best channel for probing these processes at the LHC 
\cite{ATLAS-TDR}. 
If both bosons decay hadronically the signal cannot
be distinguished from the QCD non resonant background 
whose cross section is much larger. 
Final states where both vectors decay leptonically 
have a smaller rate and have been left for future studies.

Once vector bosons are decayed we have six fermion final states.
Since the \V's emitted by the initial state particles are spacelike,  
we are forced by gauge invariance to include all diagrams in which final state
\V's are emitted directly by the fermion lines as in the left part of
\fig{nonreso-diag}.
When the finite width of the EW boson is properly
taken into account and the outgoing vector bosons are allowed to be off mass
shell, it becomes necessary, again because of gauge invariance, to
consider the full set of diagrams described by \eqn{6f}.   
As a consequence, in addition to the diagrams which are
related to the process we would like to measure, \VV  fusion, 
shown in \fig{VV-diag},  there will be diagrams 
in which a pair of \V's  are produced  without undergoing \VV scattering, 
as presented in \fig{nonreso-diag}. Furthermore,
diagrams related to Higgs production via Higgstrahlung will also be
present as shown in \fig{higgs-diag}, as well as
diagrams which can be interpreted as \toptop EW
production or as single top production as shown in \fig{top-diag}.
Finally diagrams describing three vector boson production
which include Triple Gauge Coupling and Quartic Gauge Coupling will
contribute as well, as shown in \fig{tgc-diag},
since they produce the same kind of six fermion final states.
 
Depending on the flavour of the quarks in \eqn{6f}
the various subprocesses will contribute and interfere
to a different degree. All processes
will be experimentally indistinguishable, apart from heavy quark tagging,
and will have to be studied simultaneously.
 
A number of samples of events representative of all possible processes in
\eqn{6f} have been
produced with \Phase. In order to comply with the
acceptance of the CMS detector and with the CMS trigger requirements,
the cuts in \tbn{standard-cuts} have
been applied. We have used the CTEQ5L \cite{CTEQ5} PDF set with scale
\be
Q^2 = M_W^2 + \frac{1}{6}\,\sum_{i=1}^6 p_{Ti}^2.
\label{scale}
\ee
where $p_{Ti}$ denotes the transverse momentum of the $i$--th final state
particle.
 
\begin{figure}
\begin{center}
\mbox{
{\epsfig{file=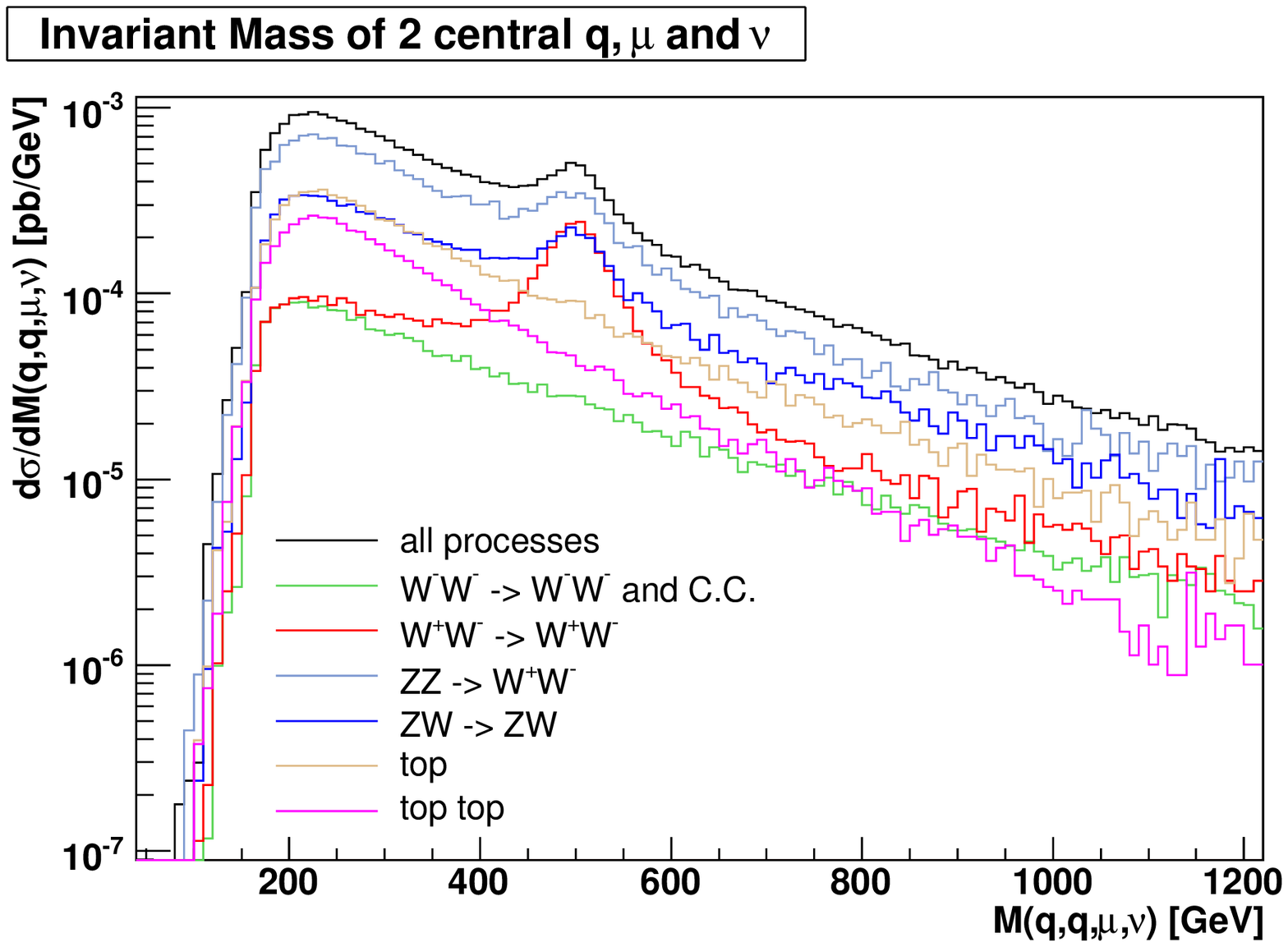,width=13cm}}
} 
\mbox{
{\epsfig{file=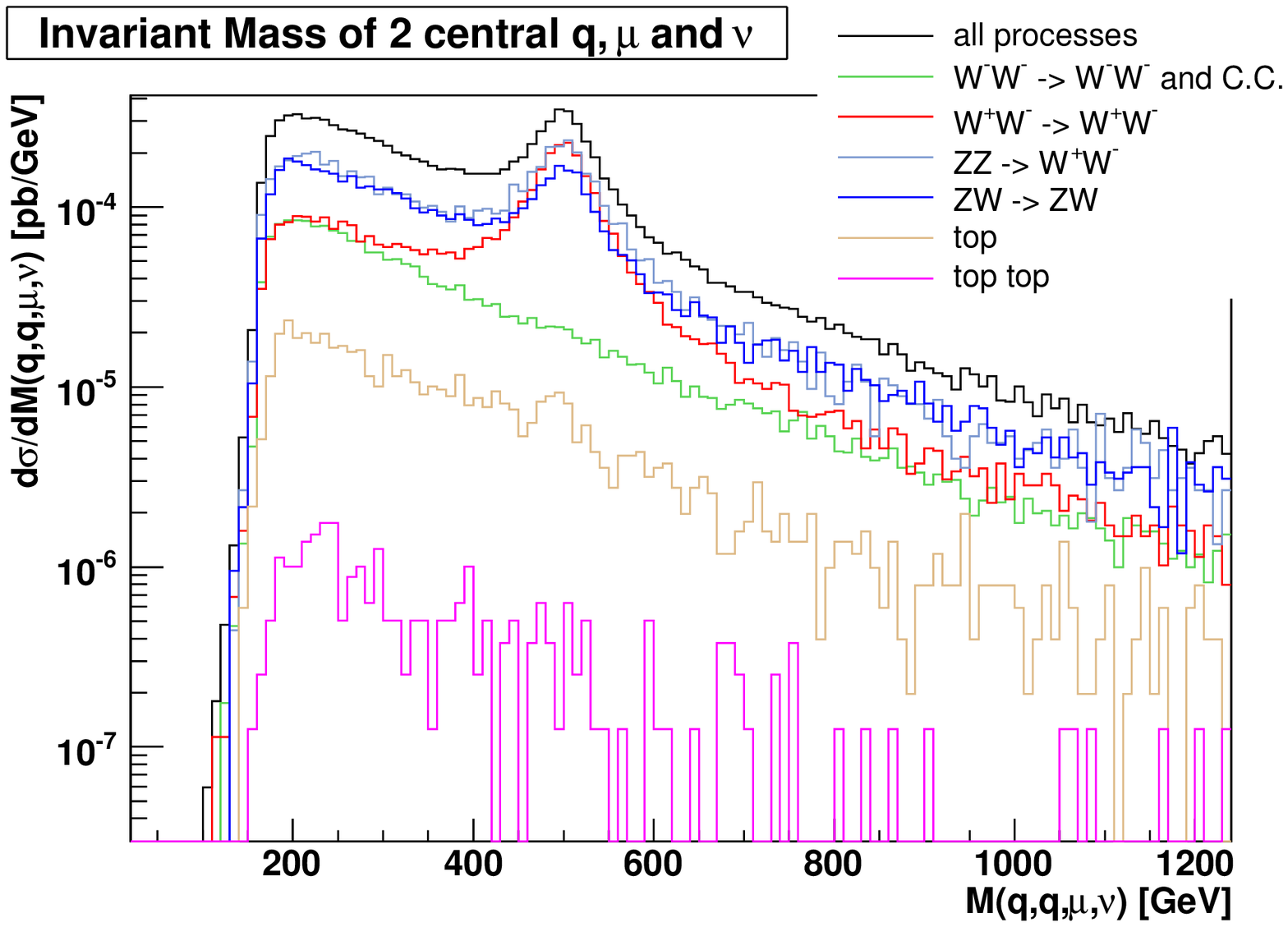,width=13cm}}
} 
\caption{Invariant mass distribution of the 
lepton, neutrino and the two most central quarks, for different sets
of processes. In the upper part with the set of cuts described in the text,
in the lower part after vetoing  top and \toptop production and
requiring the vector bosons to be close to their mass shell.} 
\label{sub-proc}
\end{center}
\end{figure}

\subsection{Physical sub-processes}
\label{PhysSub}

As already mentioned many subprocesses 
(i.e. \WW \ra \WW, \ZW \ra \ZW, \ZZ \ra \WW, \toptop)
will in general contribute to a specific six fermion reaction. 
It is  impossible to separate and compute individually the cross 
section due to a single subprocess, since there are large interference effects
between the different contributions.

We can however select all complete 2\ra6 processes which include a specific set
of subdiagrams.
For instance, \ZW\ra\ZW with on shell bosons is described by 4 Feynman diagrams.
These same diagrams, with all external vector bosons connected to a fermion
line, constitute the  \ZW\ra\ZW set of 2\ra6 diagrams.
Several sets can contribute to a single
process and therefore the same process can appear in different groups.
The upper part of \fig{sub-proc} shows the invariant mass distribution
of the two most central quarks (when ordered in pseudorapidity
$\eta$),the lepton and the neutrino for the reactions which contain the
different subprocesses
as well as the distribution for the complete set of processes.
We assumed M(H)=500 \GeV.
For this figure, in addition to the standard cuts decribed in 
\tbn{standard-cuts}, the following cuts have been applied.
The pseudorapidity of the 
two most central quarks should satisfy $\vert\eta$(central-quark)$\vert<3$,
the pseudorapidity of the two most forward-backward quarks are constrained to
$\vert\eta$(forward-quark)$\vert>1$ and the
transverse mass of the lepton neutrino pair should be smaller than 100
\GeV. These three last requirements have been
released for the studies presented in the following sections.

It should be clear that the total cross section in \fig{sub-proc}
is smaller than the sum of the cross sections for the various groups.
Notice that the Higgs peak is present in the \ZW\ra\ZW curve. This is due to
processes that in addition to the \ZW\ra\ZW set of diagrams include also
diagrams describing Higgs production, e.g. 
$u \overline{u} \rightarrow u \overline{u} u \overline{d} \mu^-\overline{\nu}$.

The groups comprising single top and \toptop diagrams have a large
cross section. An invariant mass analysis reveals that the processes in these
two groups are indeed dominated by top production.
The lower part of \fig{sub-proc} shows
the same distributions after subtraction of single top and \toptop
and with the additional requirement that the two most central quarks have
invariant mass between 60 and 100 \GeV.

Top candidates are identified requiring a b-quark and two other quarks
in the final state of the right flavour combination to
be produced in a W decay, with a total  invariant mass between 160 and 190
\GeV.
Analogously, events in which the muon, the neutrino and a b quark have an
invariant mass between 160 and 190 \GeV are rejected.
These vetoes reduce drastically the EW top background and produce a much sharper
Higgs peak.

\subsection{Identification of the hadronically decaying vector boson}
\label{V-id}

\begin{figure}
\begin{center}
\mbox{
{\epsfig{file=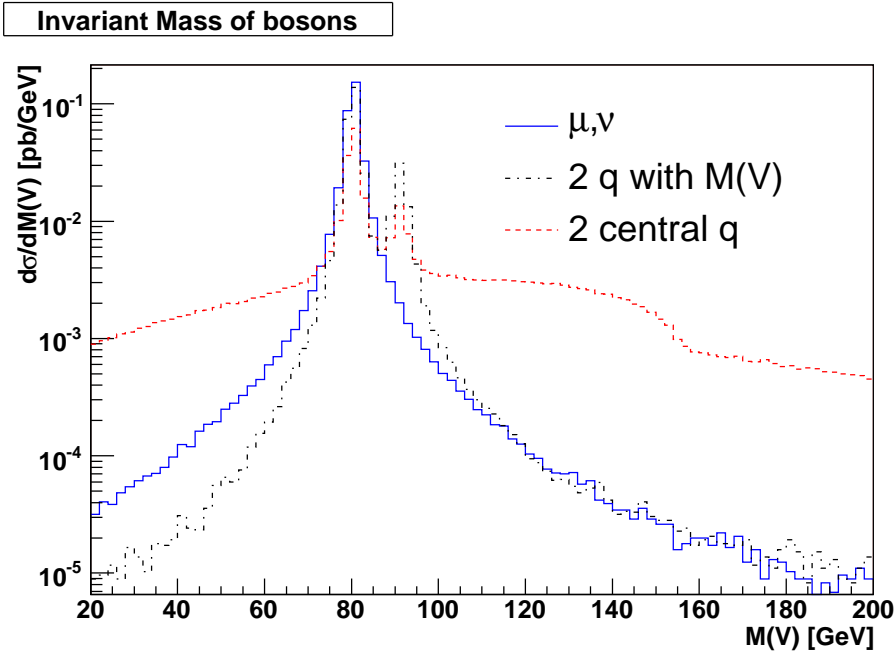,width=13cm}}
} 
\caption{ The invariant mass of the two most central quark in
  pseudorapidity (red/dashed) and for the quarks that gives the best nominal
  vector boson mass (black/dash-dotted) with \Phase. The
  invariant mass of the lepton and neutrino pair is shown in blue/continous 
  for comparison. }
\label{mW-selec}
\end{center}
\end{figure}

In the LHC environment it is not obvious how to identify the jets produced
by the hadronic decay of an EW vector boson among the four which are present
at generator level in the processes we are interested in.
\fig{mW-selec} shows the invariant mass of the candidate vector boson
selected with two different procedures.
In the first one final state quarks are ordered in pseudorapidity. The quarks
with the largest and smallest $\eta$ are identified as tag quarks, while the
remaining two, which will be referred to as ``most
central'' quarks, are identified as the quarks from the \W or \Z
decay\footnote{Notice that these two quarks are actually those
with smaller $\vert \eta \vert$ only in the center of mass of the hard reaction
and not in the laboratory frame. However the ordering in pseudorapidity
is preserved by boosts along the beam line and the selection of the two jets
with intermediate rapidity will be the same in the two reference frames}.
In the second procedure the two quarks among the four that have
the mass closest to a vector boson mass are identified as the quarks from the
decay while the remaining two are identified as tag quarks.
As a comparison the invariant mass of the lepton and neutrino pair is also
shown. It is clear from the figure
that the second choice reproduces much better the lepton--neutrino mass
distribution, giving higher efficiency and purity at generator level,
and will be adopted in the rest of this paper. 

\section{The VV-fusion signal}
\label{sig}

In order to isolate the \VV fusion process from all
other six fermion final states and investigate EWSB, different kinematical 
cuts have been applied to the simulated events.

The main background we want to subtract at parton level is 
single top and \toptop production. These events have been identified as in
\subsect{PhysSub} and rejected. 

The invariant mass of the muon and the neutrino has to reconstruct the mass of
a \W, and  is required to be in the range $M_W \pm 10$ \GeV.
In \VV fusion an additional \W or a \Z decaying hadronically
is expected to be present.
Therefore events are required to contain two quarks with the correct flavours
to be produced in \W or  \Z decay, with an invariant
mass of $\pm$ 10 \GeV around the central value of the appropriate EW bosons.
If more than one combination of two quarks satisfies these requirements, the one
closest to the corresponding central mass value is selected. This combination 
will in the following be assumed to originate from the decay of an EW vector
boson.

In a second step, in order to reject events which can be identified with the
production of three vector bosons, the flavour content and the invariant mass 
of the two remaining quarks is compared with a \W and a \Z. If compatible 
within 10 \GeV with either, the event is rejected.
This happens in about 2\% of the cases. 
The events satisfying all these constraints will constitute the ``signal'' sample.

These requirements are not fully realistic: no flavour information will be
available for light quarks, b's will be tagged only in the central part of
the detector and the invariant mass of the $l\nu$ system will not be directly
available. Our aim however is to define a ``signal'',
in the same spirit as {\tt CC03} was adopted as such at LEP 2,
that is a pseudovariable which could be used to compare the results from the
different collaborations. The signal is not necessarily directly observable
but it should be possible to relate it via Monte Carlo to measurable quantities.
If such a definition is to be useful it must correspond as closely as possible
to the process which needs to be studied and the Monte Carlo corrections
must be small. 
At this stage we want to isolate as much as
possible the \VV fusion signal from all other production channels,
while keeping reasonably close to the experimental practice and taking into
account the full set of diagrams required by gauge invariance.
It becomes then  possible to analyze the differences between the
\VV fusion signal sample and its intrinsic background.
This also provides some preliminary experience at the generator level which 
could guide more realistic and complete studies.

\begin{figure}
\begin{center}
\mbox{\epsfig{file=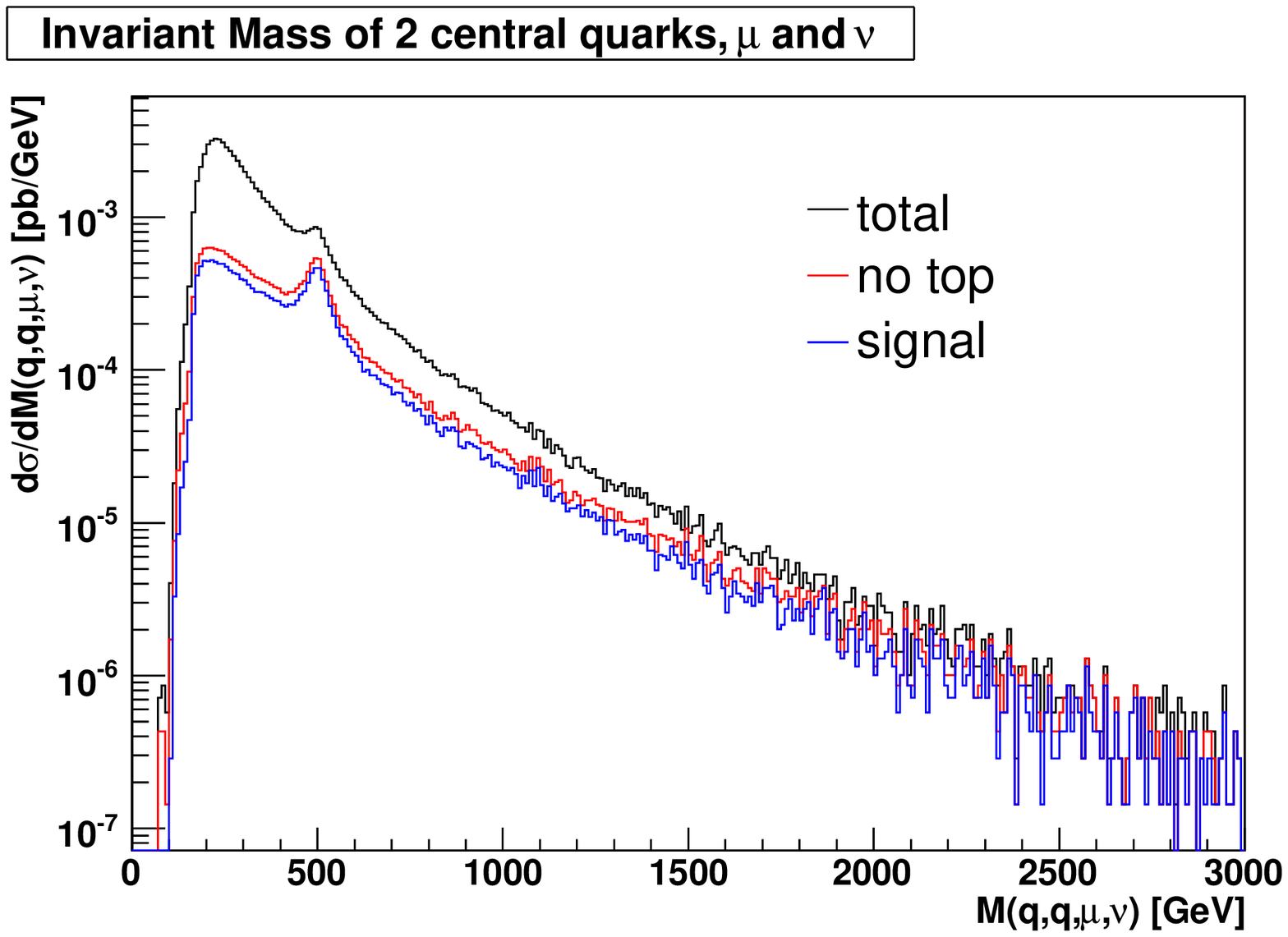,width=8cm}
\epsfig{file=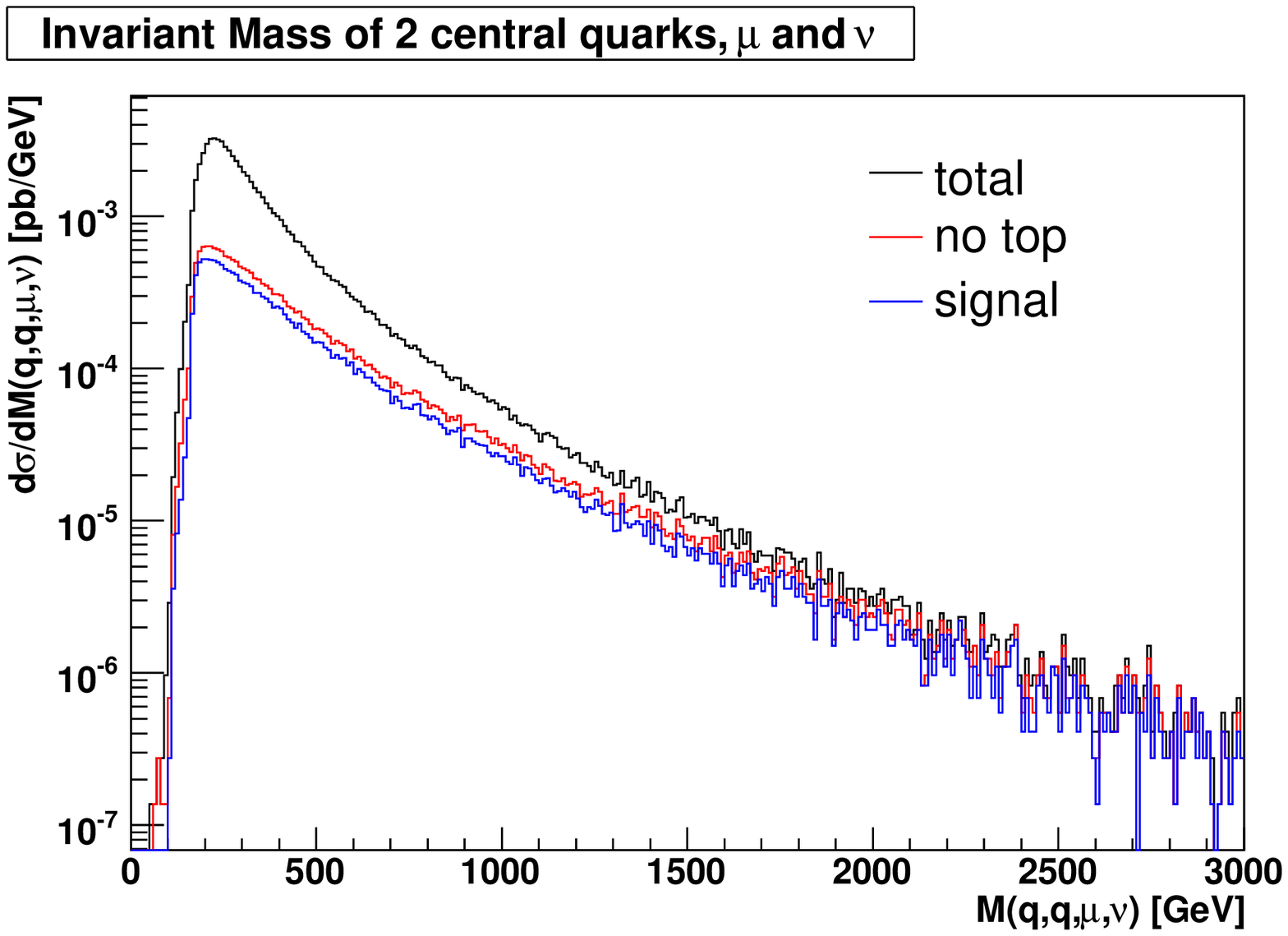,width=8cm}} 
\caption{ 
Distributions of the invariant mass of the 
two central quarks, the muon and the neutrino for M(H)=500 \GeV (left) and for 
the no-Higgs case (right). The black line refers to the full sample.
The red one shows the effects of antitagging on the top.
The blue line corresponds to imposing all the mentioned cuts
and antitagging on the presence of three vector bosons.
}
\label{imvv-cuts}
\end{center}
\end{figure}

\begin{figure}
\begin{center}
\mbox{\epsfig{file=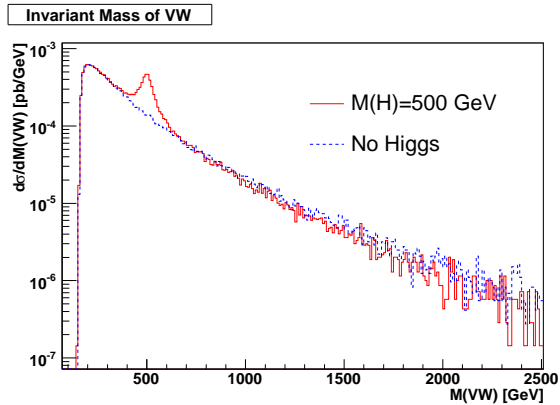,width=8cm}} 
\caption{ Distributions of the invariant mass of the 
two reconstructed vector bosons for M(H)=500 \GeV (red) and for 
the no-Higgs case (blue) after all cuts.In this plot the vector boson
that decays hadronically is reconstructed using the two quarks that give
its best mass.}
\label{imvv-sign}
\end{center}
\end{figure}

In \tbn{cuts} the cross sections obtained for 
each of the described cuts are shown. The corresponding \VV mass distributions
are displayed in \fig{imvv-cuts} and \ref{imvv-sign}.

\fig{imvv-cuts} shows how the invariant
mass distributions for two reconstructed vector bosons is modified
when the cuts described above are sequentially imposed, for
the case where the Higgs boson exists and its mass is 500 \GeV and
for the no-Higgs case. The no-Higgs amplitude is the amplitude in
the limit $M_H \rightarrow \infty$. In the Unitary gauge,
for the processes under investigation, this limit amounts to neglecting
all diagrams with Higgs propagators. 
The black line refers to the full sample, the red one shows the effects of
 antitagging on the top.
The blue line corresponds to imposing all the mentioned cuts
and antitagging on the presence of three vector bosons.
\fig{imvv-sign} compares the reconstructed vector boson invariant mass
distribution after all cuts for the two Higgs cases\footnote{
The black and red line of \fig{imvv-sign} do not reproduce the results of 
\fig{sub-proc} because of a different set of cuts}.

\begin{figure}
\begin{center}
\mbox{
{\epsfig{file=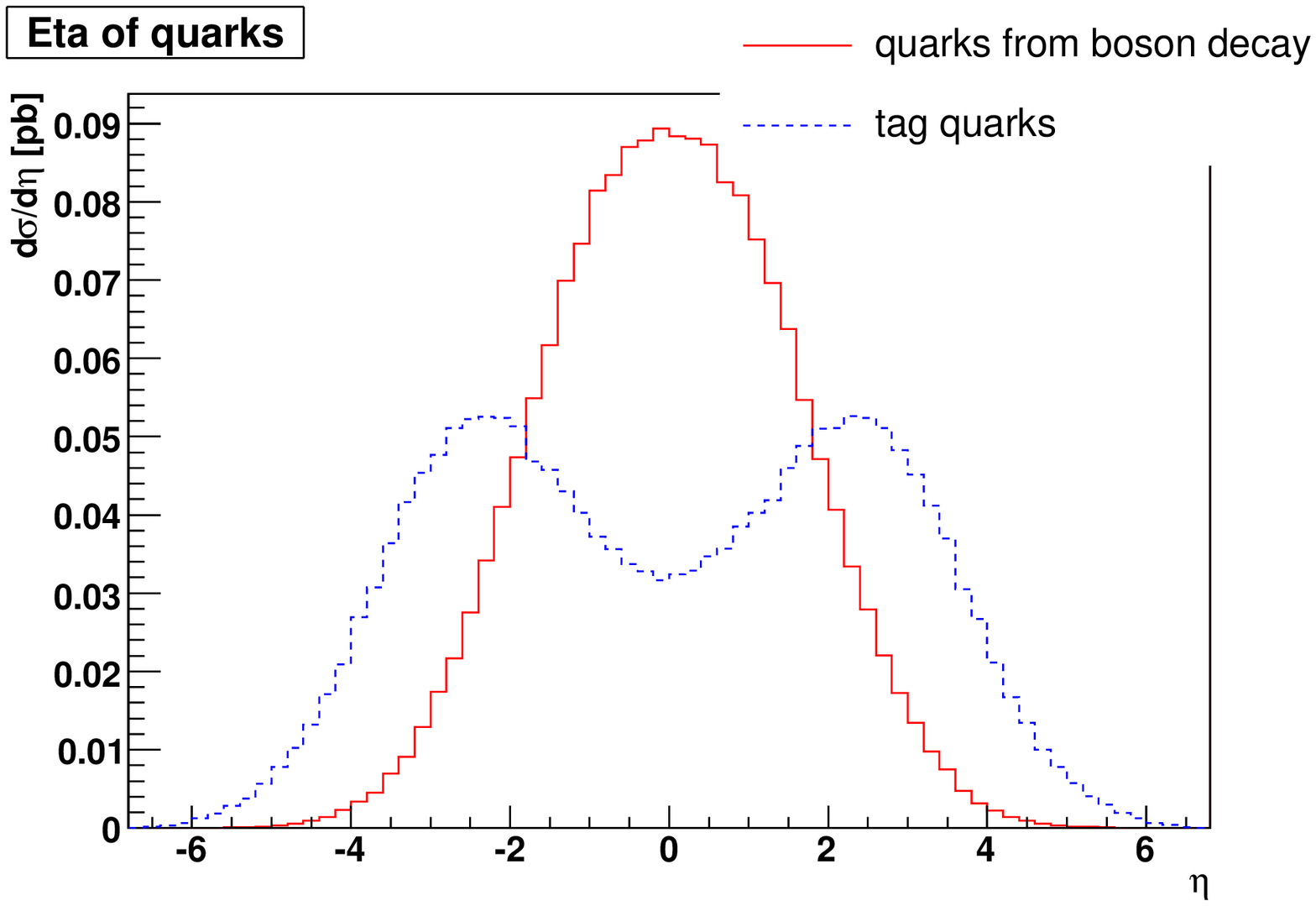,width=8cm}}
{\epsfig{file=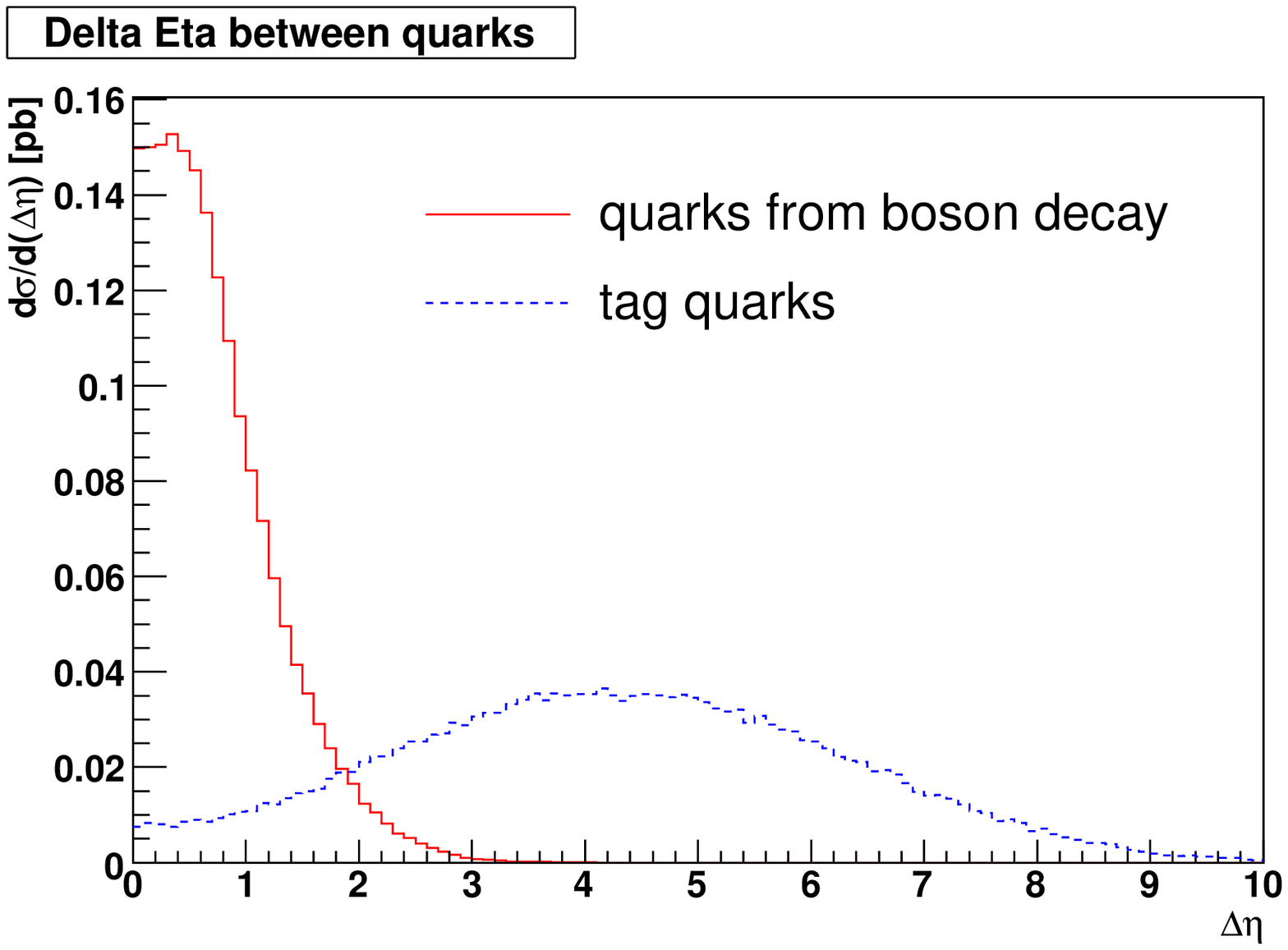,width=8cm}}
}
\mbox{{\epsfig{file=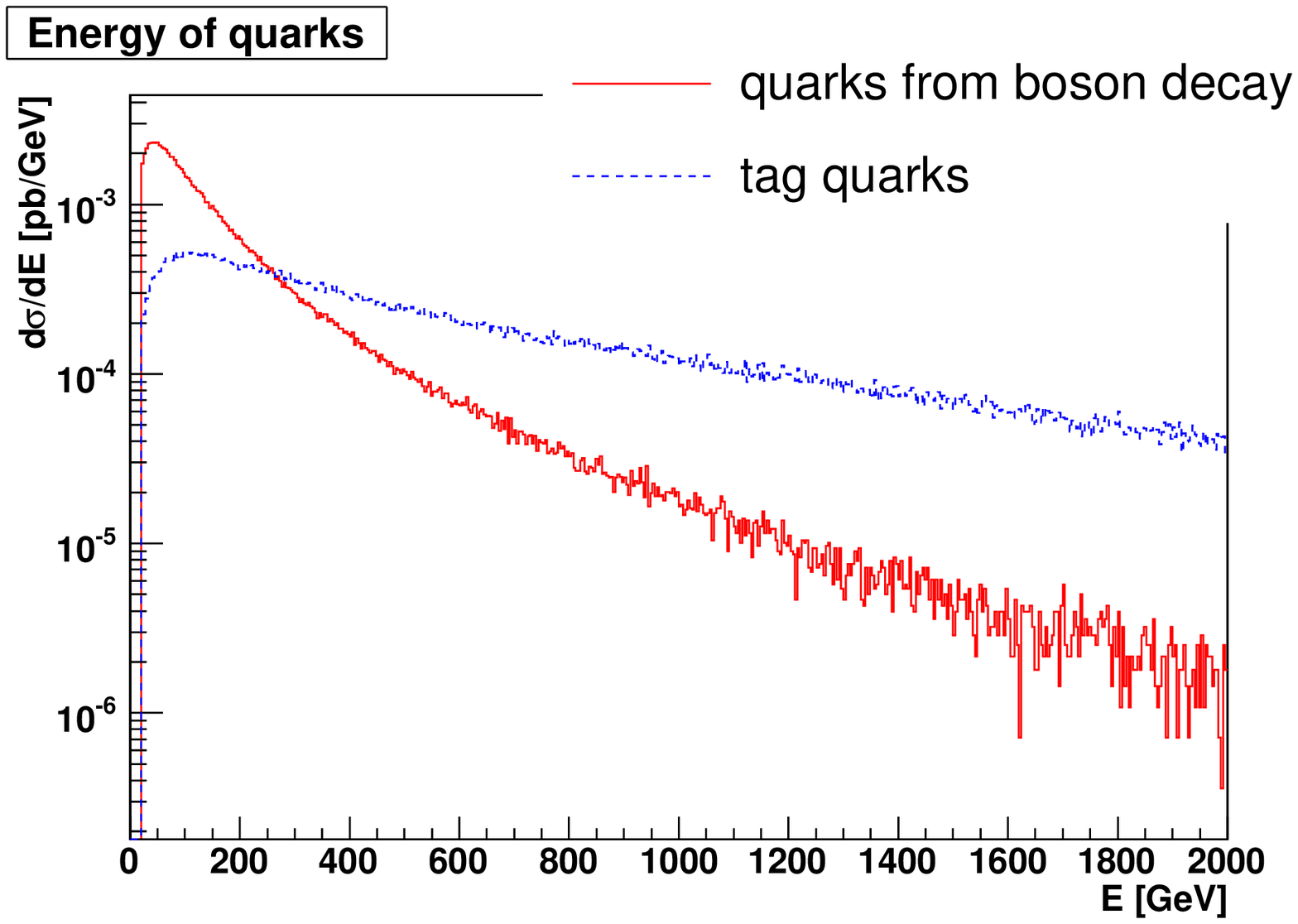,width=8cm}}
{\epsfig{file=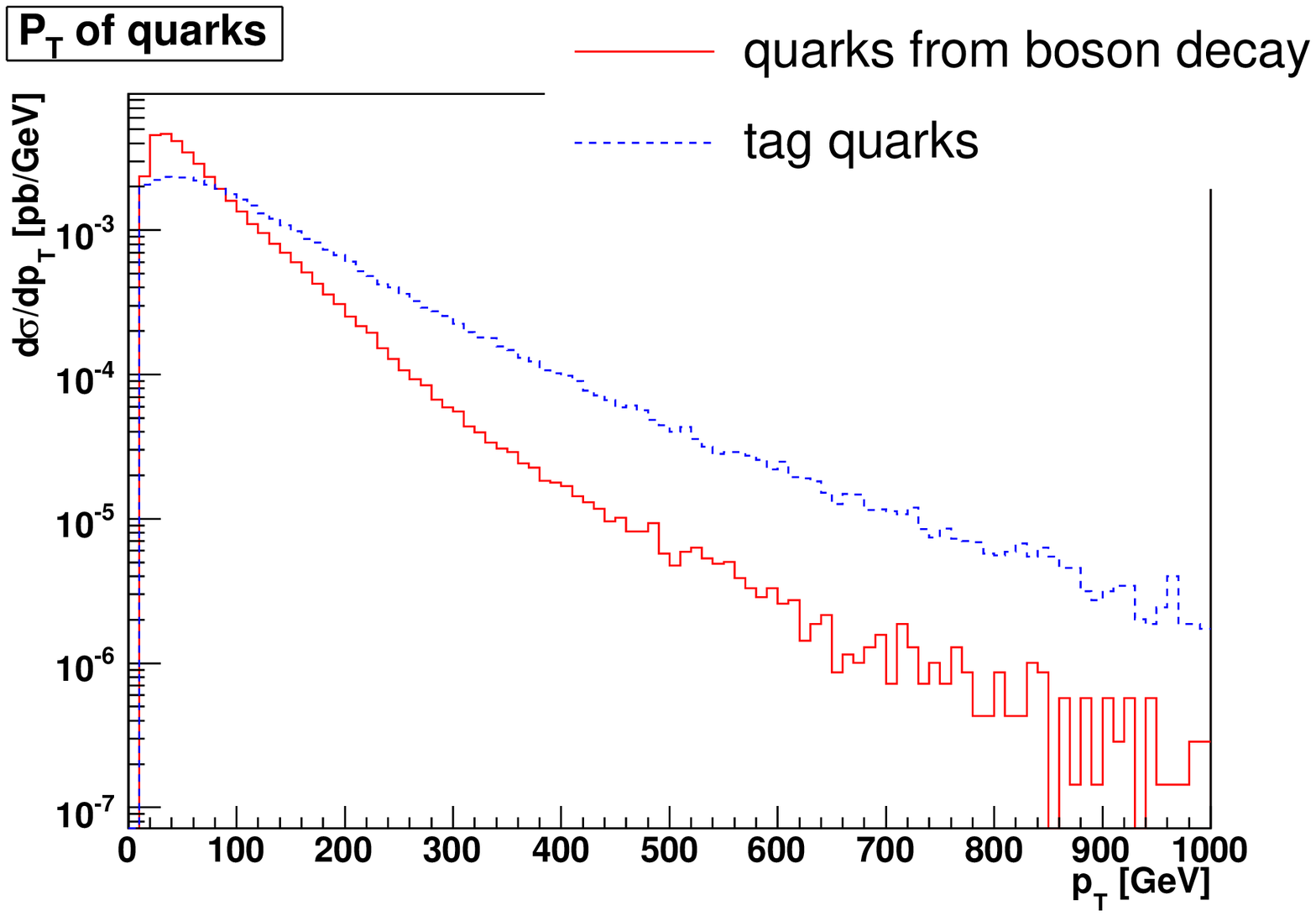,width=8cm}}
}
\caption{Distributions of the pseudorapidity $\eta$,
the difference $\Delta \eta$, the energy and transverse momentum of the quarks.
In red
(full line) for the quarks from the decay of the vector boson and in blue
(dotted line) for the tag quarks.
All plots refer to M(H)=500 \GeV. The distributions for the no-Higgs case lead
to the same conclusions.}
\label{qW_qF}
\end{center}
\end{figure}

Since our definition of signal begins with the identification of the
jets which reconstruct the EW boson rather than with the identification
of the tag jets, we compare in some detail the kinematics of the quarks 
from a vector boson decay with the kinematics of the spectator ones.
\fig{qW_qF} shows that a good separation in phase space between the two pairs of
jet is maintained. 
The distribution of the pseudorapidity $\eta$ of the quarks
coming from the decay of the vector boson and of the tag quarks
shows that the quarks from the \V decay are, as expected, rather 
central (i.e. low $\eta$) while the spectator quarks 
tend to go forward/backward. Consequently the difference in
pseudorapidity $\Delta\eta$
between the quarks from the bosons will be smaller than the $\Delta\eta$
between the tag quarks.
The quarks from the vector boson have less energy and smaller transverse
momentum with respect to the tag quarks.

\begin{figure}
\begin{center}
\mbox{
{\epsfig{file=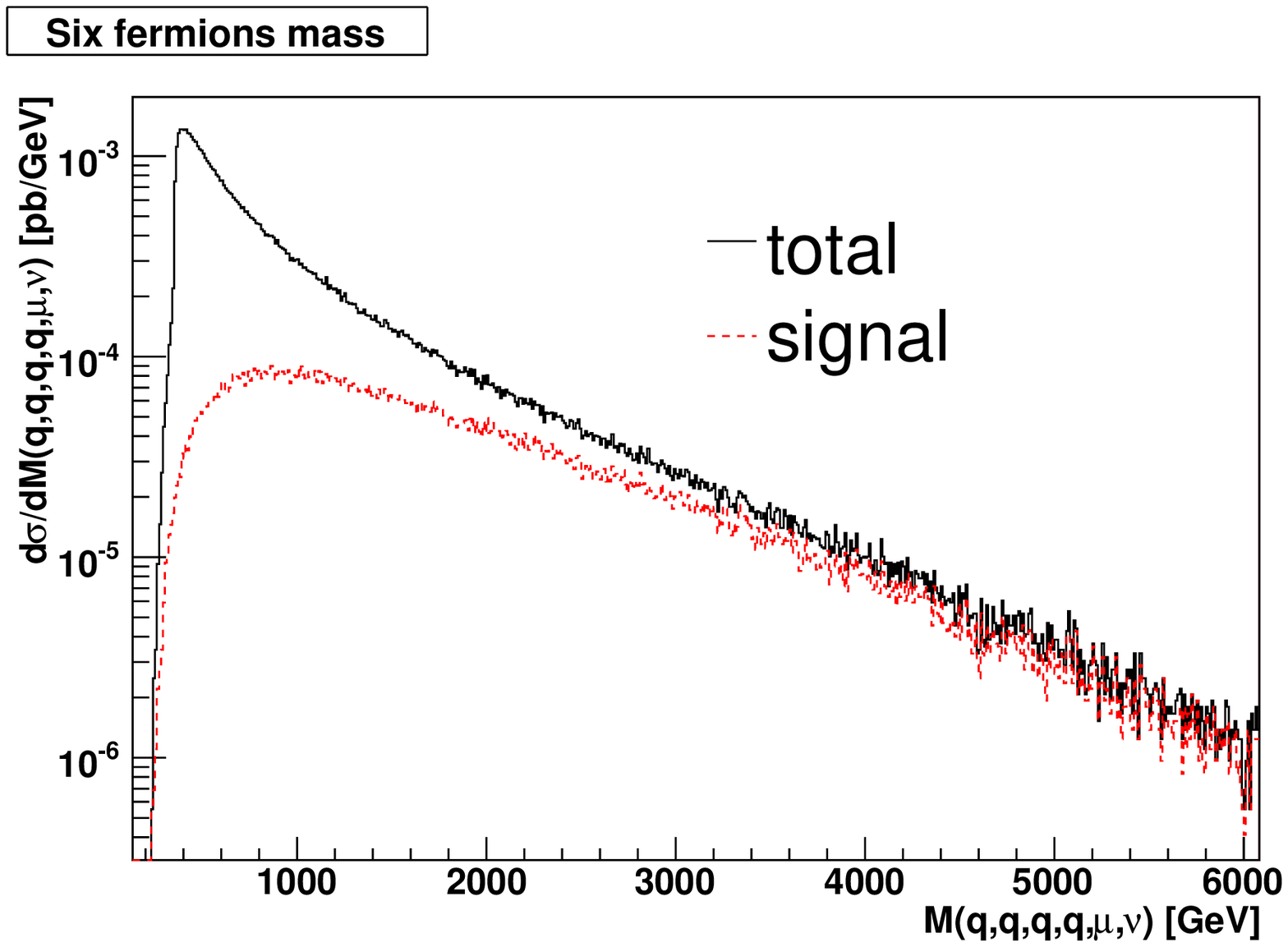,width=8cm}}
{\epsfig{file=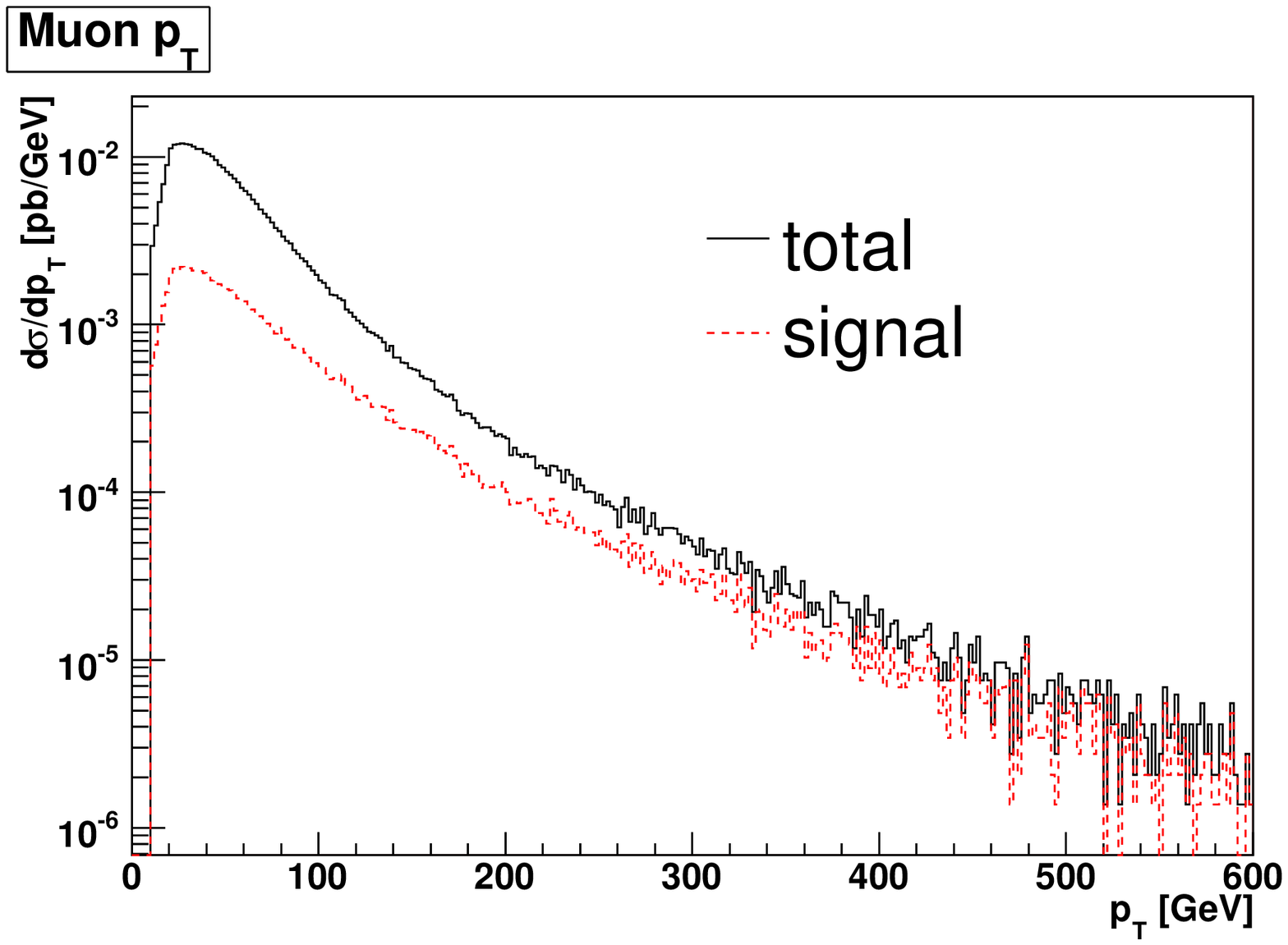,width=8cm}} 
} 
\mbox{
{\epsfig{file=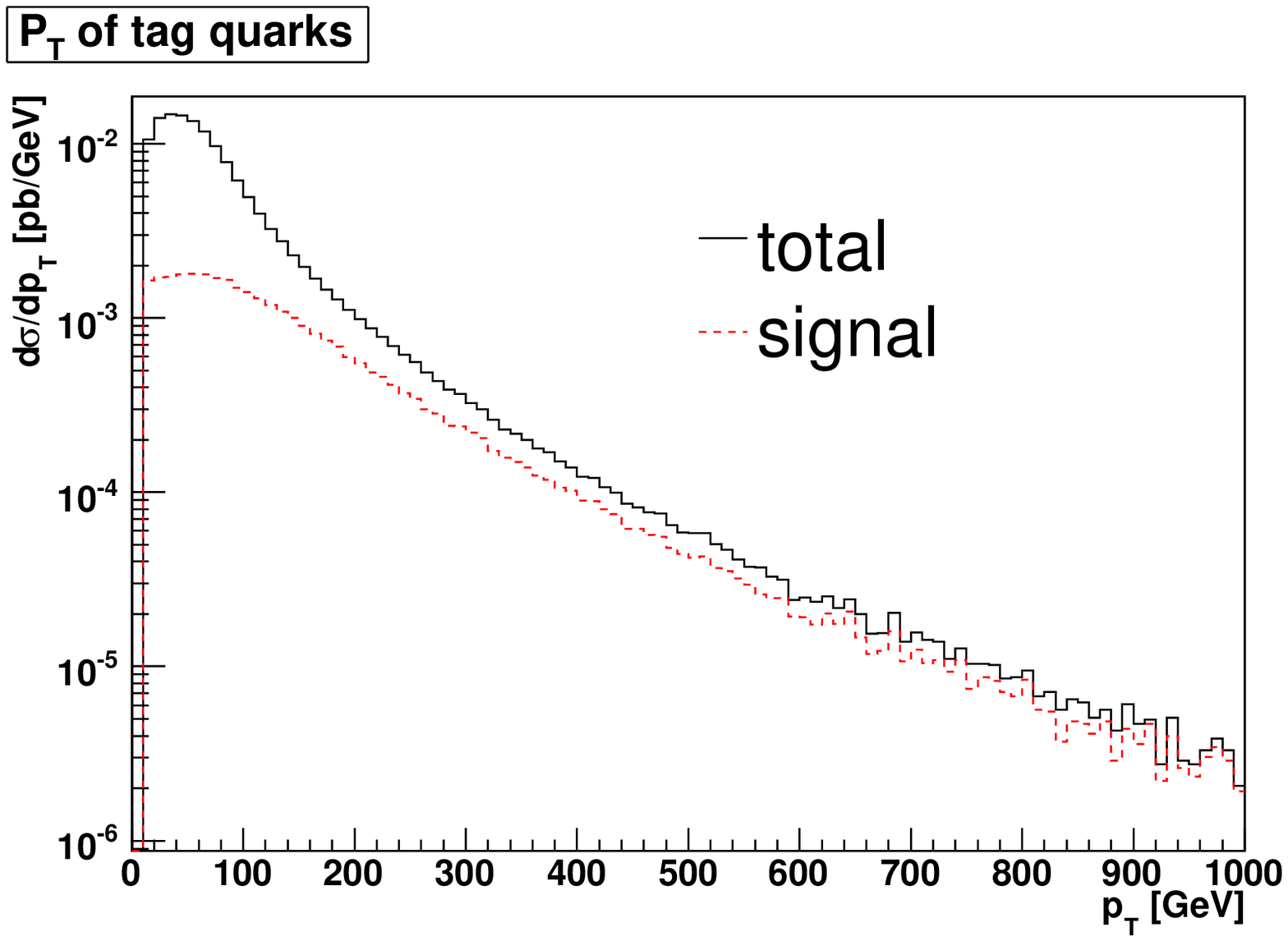,width=8cm}} 
{\epsfig{file=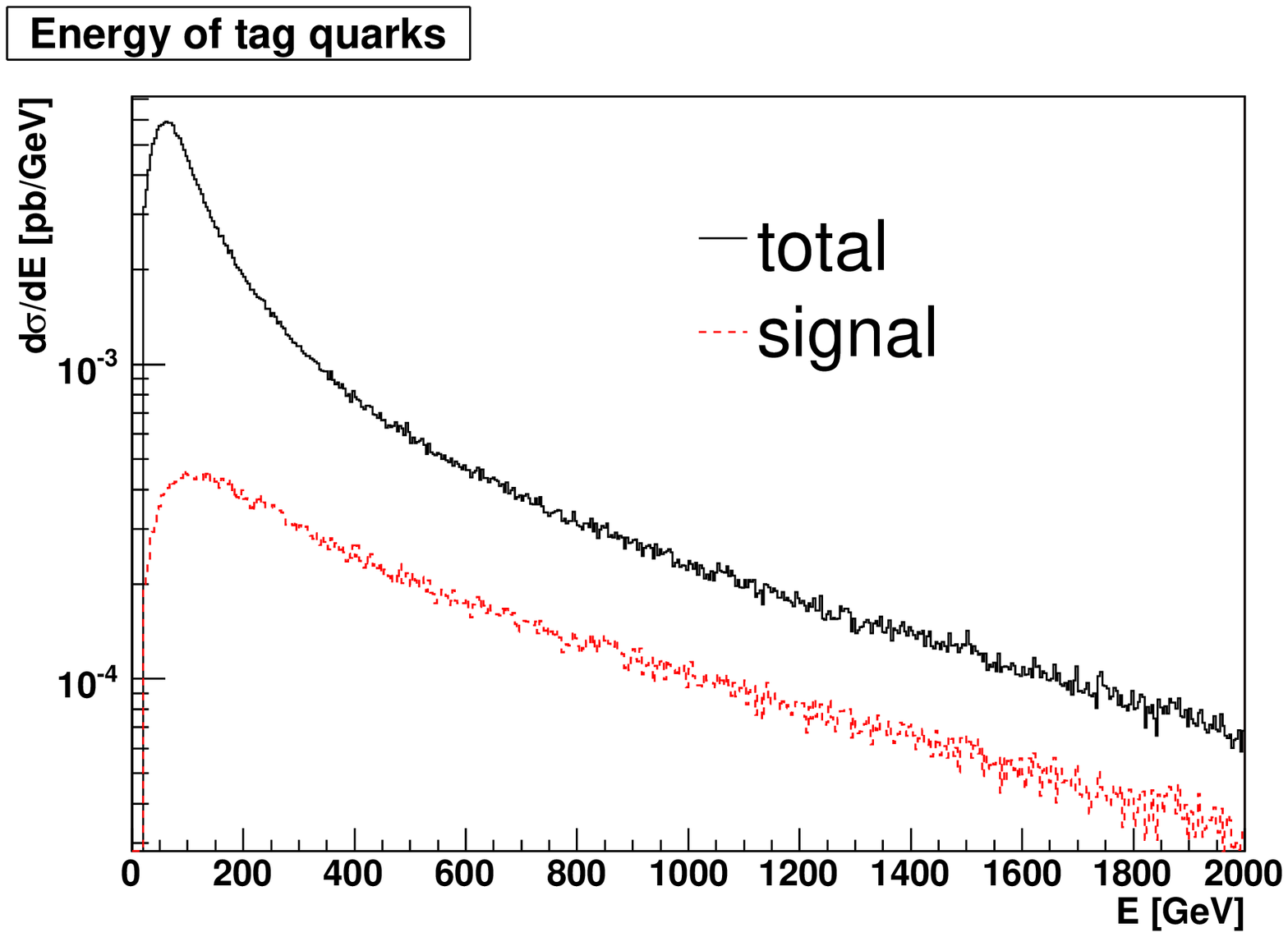,width=8cm}} 
} 
\mbox{
{\epsfig{file=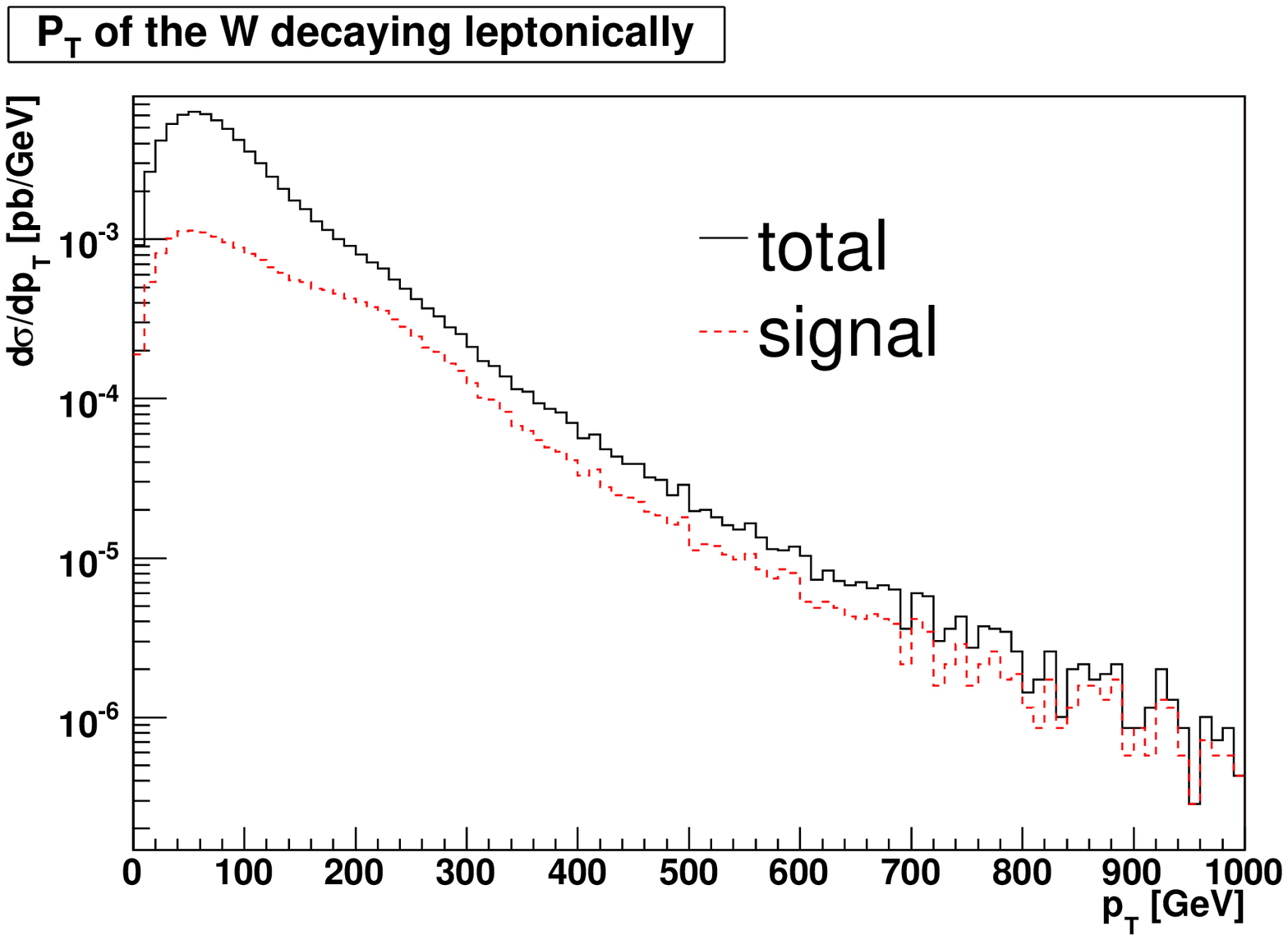,width=8cm}} 
{\epsfig{file=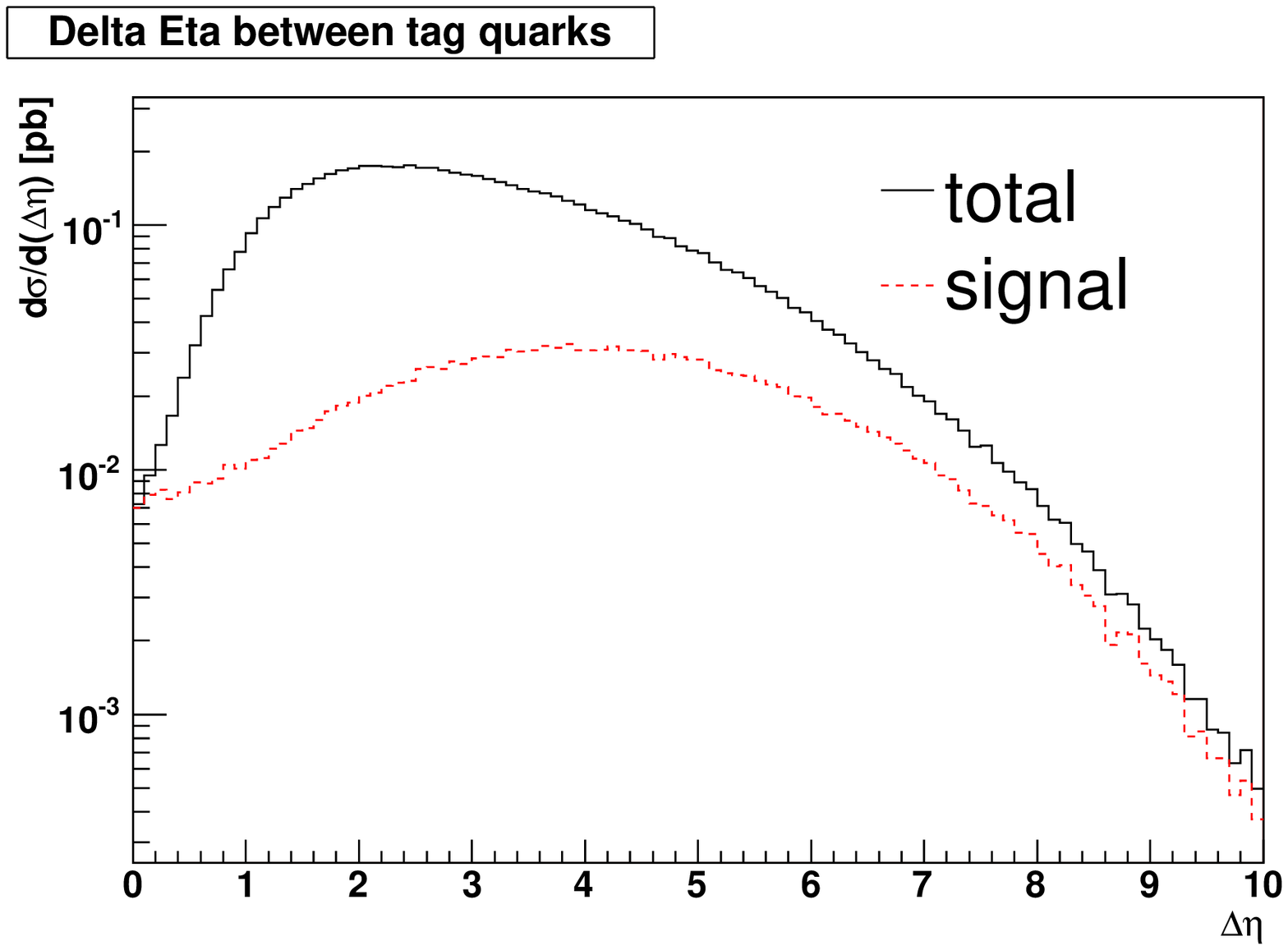,width=8cm}} 
} 
\caption{ Differential cross section as a function of the invariant mass of the 
six fermions in the final state, the transverse momentum of
the muon, the transverse momentum of the two tag quarks,
the energy of the two forward quarks, the transverse momentum of
the W decaying leptonically and the difference in
pseudorapidity between the two tag quarks
for all the events (black) and for the
signal events (red).
All plots refer to the no-Higgs case with the exception of the $p_T$ of the
$W$ which decays leptonically which is for
M(H)=500 \GeV. } 
\label{m6ferm}
\end{center}
\end{figure}

It is interesting to look at possible differences in the kinematics
of the VV-fusion signal with respect to the irreducible background.
In \fig{m6ferm} the ``no-Higgs'' case is chosen as an
example, but there are no significant differences with the case of a 
massive Higgs. Only some variables, which are connected to the mass of the
Higgs boson, show the presence of the resonance. In that case the
figure for the finite mass Higgs is shown.
The total invariant mass of the six fermions in the
final state is presented for the signal and for the full sample: the
signal tends to have a very large final six fermion mass.
The muon from the signal has a larger $p_T$ than the one
from the background, and the same applies to the spectator quarks.
In the background the energy of the tag quarks are peaked
at low values, although there is a long
tail at high energy as for the signal.
The transverse momentum of the W that decays
leptonically is shown for the signal events only and for the total.
We can clearly see the presence of the Higgs resonance around
 $p_T \approx M_H/2$. 
The difference of the $\eta$'s of the
tag quarks is also shown: the signal tends to have a larger
$\Delta \eta$ with respect to the background. 
       
\begin{table}[thb]
\begin{center}
\begin{tabular}{|l|c|c|c|c|c|c|}
\hline
      & \multicolumn{2}{|c|}{no Higgs} & \multicolumn{2}{|c|}{500 GeV} & \multicolumn{2}{|c|}{170 GeV}\\
\cline{2-7}
      & $\sigma$ (pb) & perc. & $\sigma$ (pb) & perc. & $\sigma$ (pb) & perc.\\
\hline
total   & 0.689 & 100\% & 0.718 & 100\% & 1.003 & 100\% \\
\hline
signal  & 0.158  & 23\% & 0.184 & 26\%  & 0.409 & 41\% \\
\hline
top      & 0.495 & 72\%  & 0.494 & 69\%  & 0.496 & 49\% \\
\hline
non \VV resonant & 0.020 & 3\% & 0.023 & 3\% & 0.040 & 4\% \\
\hline
three bosons   & 0.016 & 2\% & 0.017 & 2\% & 0.057 & 6\% \\
\hline
\end{tabular}
\caption{Cross sections and percentages of events for different Higgs masses} 
\label{cuts}
\end{center}
\end{table}

\section{\Phase versus other Monte Carlo's}
\label{sec:mcs}

In this section we compare \Phase with \Pythia \cite{pythia}
and \MadEvent \cite{mad}.
Since \Phase  is the only dedicated code that uses an
exact calculation for the 2 \ra 6 processes, it has been used
as a reference. 
The \Pythia Monte Carlo employs the Equivalent Vector 
Boson Approximation, and includes  only longitudinally polarized
vector bosons (processes 71--77).
The cross section in \Pythia 
depends strongly on the cut applied to the \pt of the outgoing \W's 
in their center of mass. This cut, stored in  CKIN(3), 
is unavoidable since it
eliminates the divergence due to the photon exchange diagram
in the on shell \WW scattering.
Notice that this divergence is an artifact of the on shell
projection of the $2 \rightarrow 2$
amplitude. In the exact six fermion matrix element
the photon divergence is absent.
For the present comparison CKIN(3) was set at 50 \GeV.

\begin{figure}
\begin{center}
\mbox{{\epsfig{file=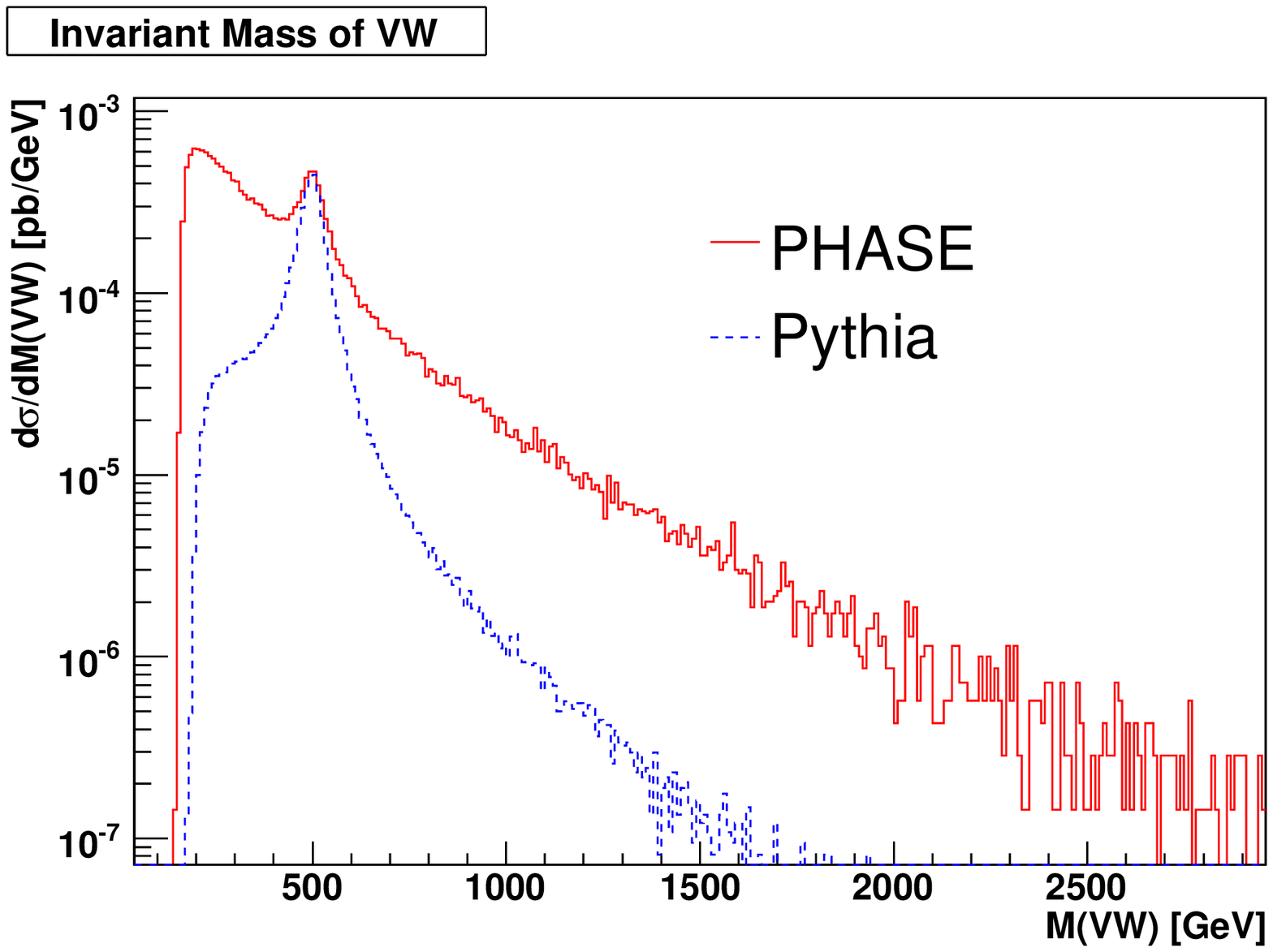,width=13cm}}} 
\mbox{{\epsfig{file=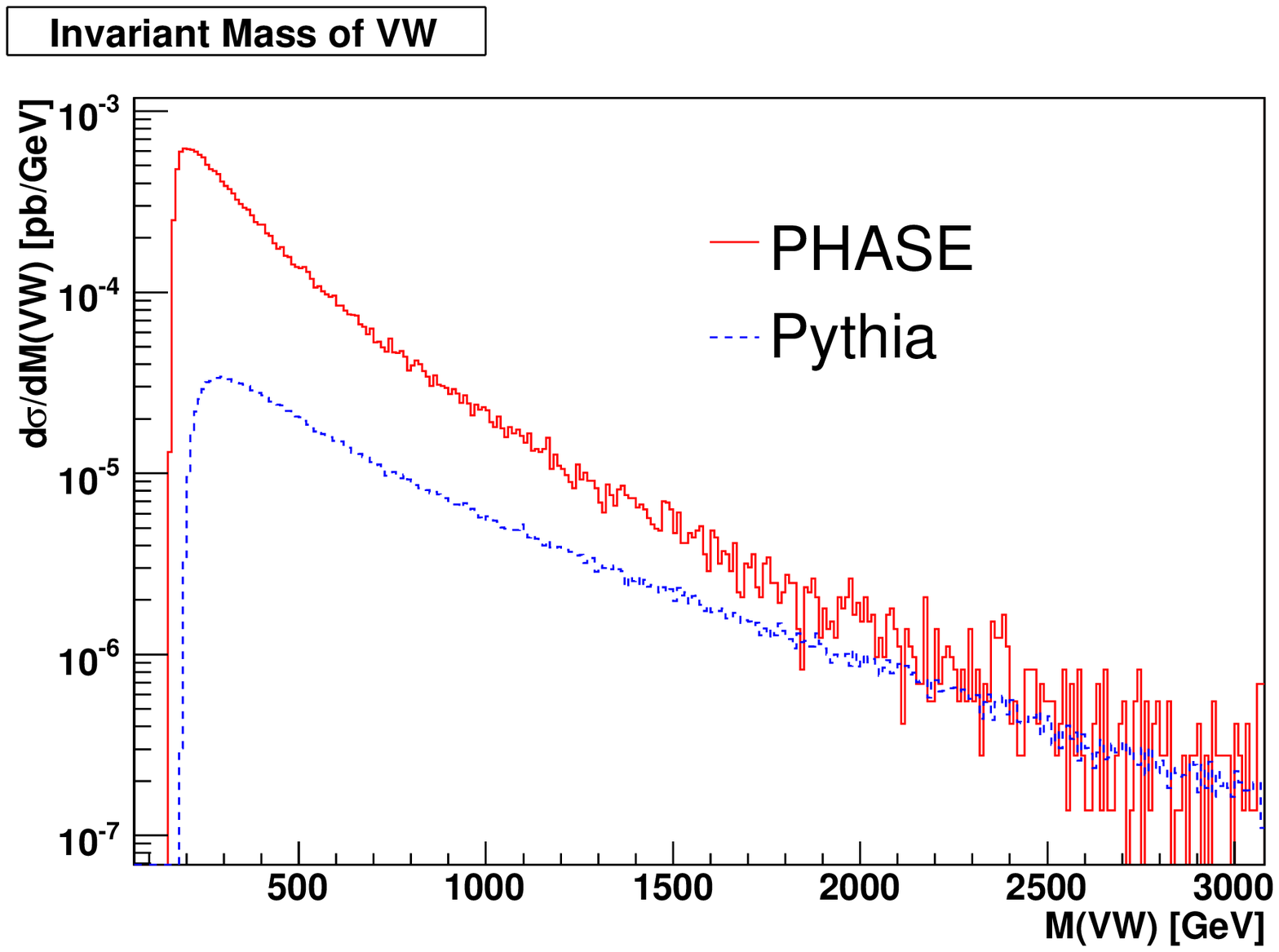,width=13cm}}} 
\caption{ Cross section as a function of the invariant mass of the 
two vector bosons for \Phase (red) and for Pythia (blue).
The upper plot refers to a Higgs mass of 500 \GeV,
the lower one to the  no-Higgs case. }
\label{pythia1}
\end{center}
\end{figure}

\fig{pythia1} shows the comparison between \Phase and \Pythia for the
$M_H =500$ \GeV and for the case
where no Higgs boson is produced (in \Pythia this can be
simulated by asking for a very high Higgs mass).
The same signal selection cuts, including those in \tbn{standard-cuts}
have been applied to both \Phase and \Pythia events.
For a Higgs of 500 \GeV one finds a reasonable 
agreement around the resonance but a large discrepancy in the rest
of the mass spectrum. For very large Higgs masses the two results are in sharp
disagreement over most of the spectrum,  but for the largest \VV
invariant masses.

\begin{figure}[thb]
\begin{center}
\mbox{\epsfig{file=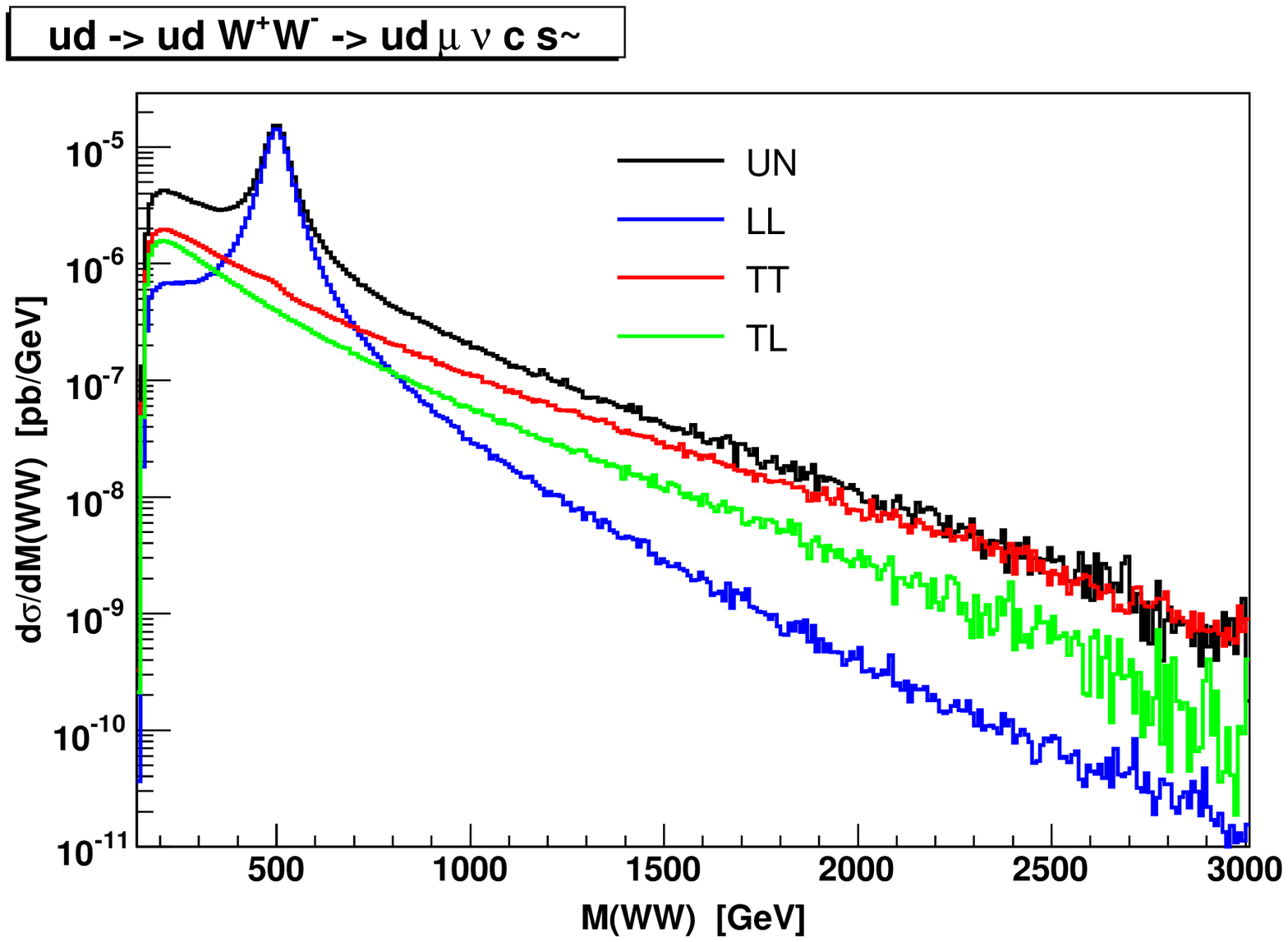,width=13cm}}
\mbox{\epsfig{file=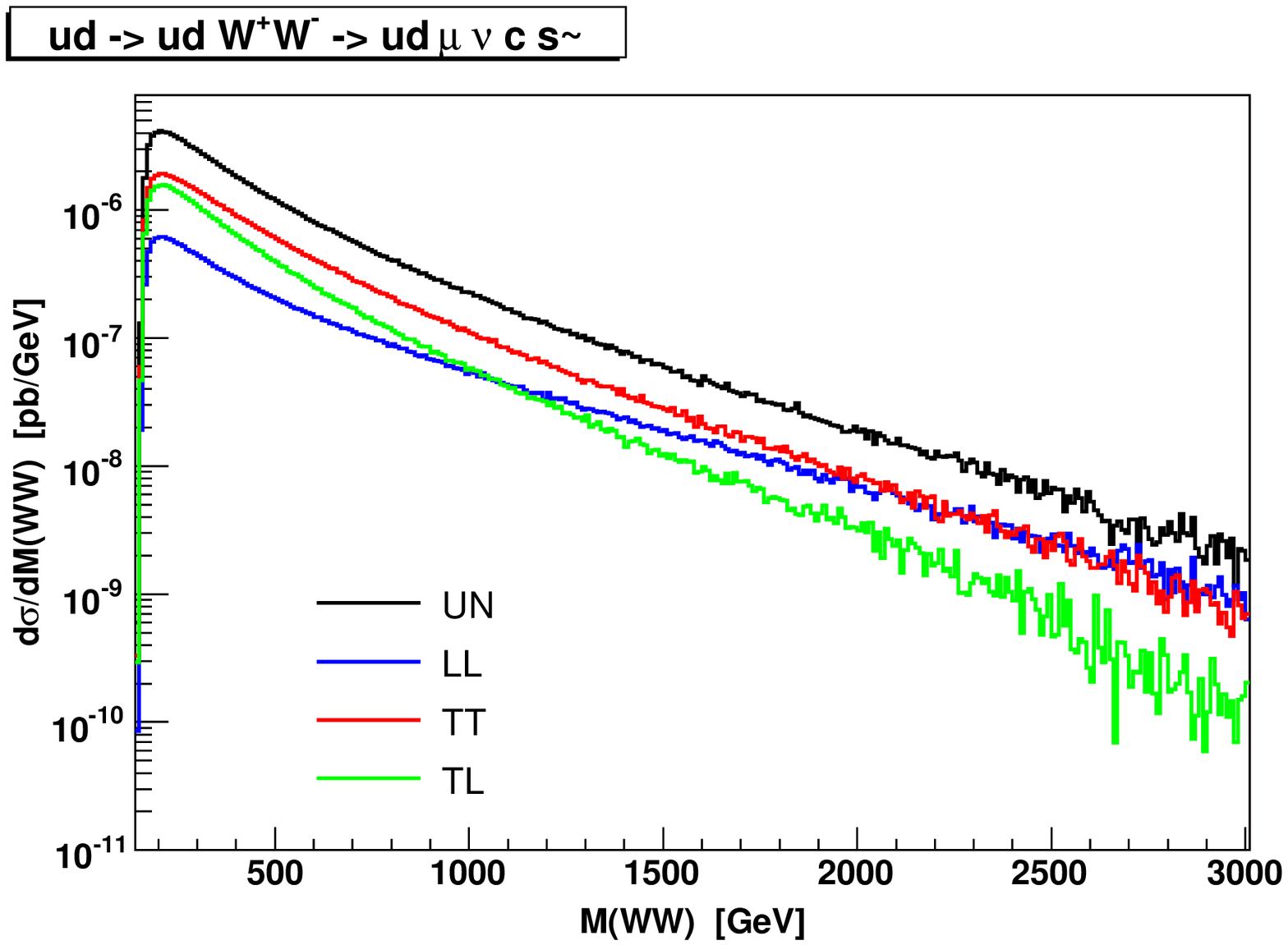,width=13cm}} 
\caption{ Cross section as a function of the invariant mass of the 
two vector bosons for their different polarizations.
The upper plot refers to a Higgs mass of 500 \GeV,
the lower one to the  no-Higgs case.}
\label{polariz1}
\end{center}
\end{figure}

As a check, in \fig{polariz1} the contributions to the cross section due to
the different polarizations in $ud \rightarrow ud c\ol{s}\mu\ol{\nu}$
are shown for the two different scenarios.
Since it is impossible to separate the polarizations of the vector bosons in the
full 2\ra 6 matrix element, we have used a dedicated code that keeps only the
diagrams in which two \W bosons are produced.
This corresponds to convoluting the matrix element for 
$ud \rightarrow ud W^+W^-$ with
the matrix elements for the polarized decay of the \W's and with their
Breit-Wigner distributions. This approach violates gauge invariance, but since
we are restricting the \W's to be close to their mass shell these effects are
expected to be small.
The cuts shown in \tbn{pol:cuts} have been applied in this case, in order to 
enhance $WW \rightarrow WW$ compared to $ZZ \rightarrow WW$.

The TT cross section is essentially
independent of the Higgs mass and it is the largest one,
outside the peak region, for $M_H=500$ \GeV .
In the no-Higgs case LL and TT production contribute equally to the spectrum.
An attempt to experimentally separate LL and TT polarizations
has been done and is described in \sect{sec:highmass}.

For $M_H=500$ \GeV case the disagreement at high invariant mass is due to
the fact that only LL is considered in \Pythia, while the cross
section for TT polarized vector bosons dominates.
For the no-Higgs case, the cross section for LL and TT polarizations
are of the same order. The agreement between \Pythia and \Phase at high
masses is probably accidental, the missing polarizations being compensated by
the growth of the LL cross section in  \Pythia ,
which as the total center of mass energy increases
gets larger contribution from the photon exchange diagram at small scattering
angles. We have not pursued this issue any further since \fig{pythia1} makes it
clear that \Pythia is not an appropriate tool to investigate \VV scattering.

\begin{table}[bh]
\begin{center}
\begin{tabular}{|c|}
\hline
      E(quark,lepton)$>20$ \GeV \\
\hline
      \pt(quark,lepton)$> 10$ \GeV \\
\hline
      $1 < \eta(d) < 5.5$\\
\hline
      $- 1 > \eta(u) > - 5.5$\\
\hline
      70 \GeV $< M(c \ol{s},\mu \ol{\nu}) <$ 90 \GeV \\
\hline
\end{tabular}
\caption{Acceptance cuts for 
$ud \rightarrow ud W^+W^- \rightarrow ud c \ol{s} \mu \ol{\nu}$ in
\fig{polariz1}.
The initial state $u(d)$-quark is in the $z(-z)$ direction.} 
\label{pol:cuts}
\end{center}
\end{table}
 
\MadEvent in principle could simulate all processes using exact matrix
elements. In practice it is far too slow to do so and is used 
in the ``production times decay'' approximation, 
i.e. computing only the processes in which two vector bosons are produced
on their mass shell and then decaying them. Even if this
procedure partially accounts for spin correlations, it neglects all
correlations within decays. This approach overlooks a number of contributions
and cannot be applied in the region below the
\VV threshold which will be actively investigated at the LHC in the quest for
an intermediate mass range Higgs.
Moreover, it leads to double counting in some regions of
phase space. For instance the reaction
\be
u \overline{d} \rightarrow u \overline{d} c \overline{s} \mu^-\overline{\nu}
\ee
would be approximated by the sum of two processes
\be
u \overline{d} \rightarrow u \overline{d} W^+ W^-
\rightarrow u \overline{d} c \overline{s} \mu^-\overline{\nu}
\quad\quad\quad
u \overline{d} \rightarrow W^+ c \overline{s} W^-
\rightarrow u \overline{d} c \overline{s} \mu^-\overline{\nu}.
\ee
While this is a reasonable approximation over most of phase space since the 
$u \overline{d}$ pair from the first reaction will mainly be produced at 
large rapidity separation
and the $u \overline{d}$ pair from the decay of the $W^+$ in the second
reaction will be at small separation, and as a consequence interference
effects will be small, it leads to an
overestimate of the cross section in the region of three vector
boson production where all final state particles are rather central.

\begin{figure}
\begin{center}
\mbox{\epsfig{file=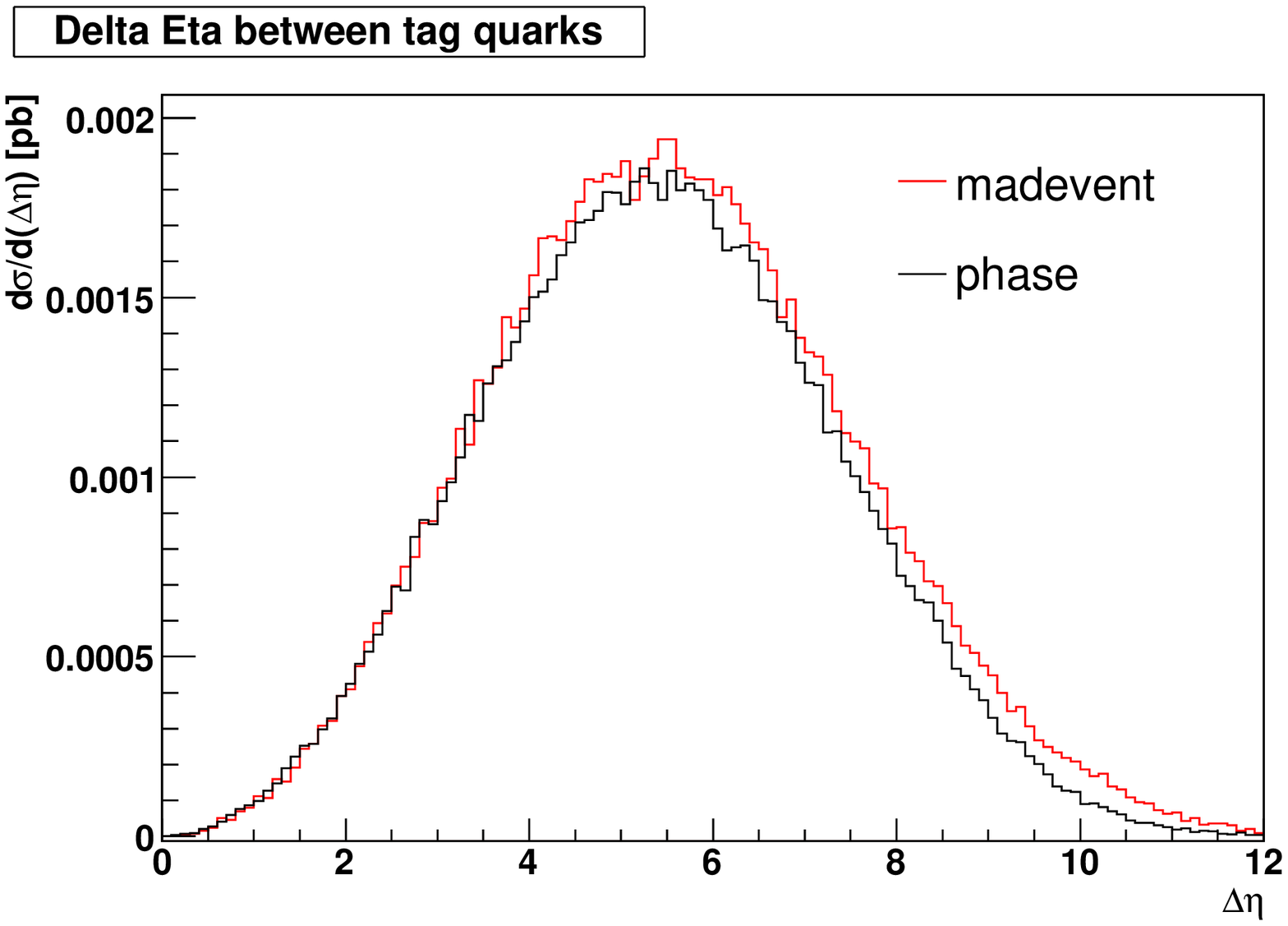,width=8cm}
\epsfig{file=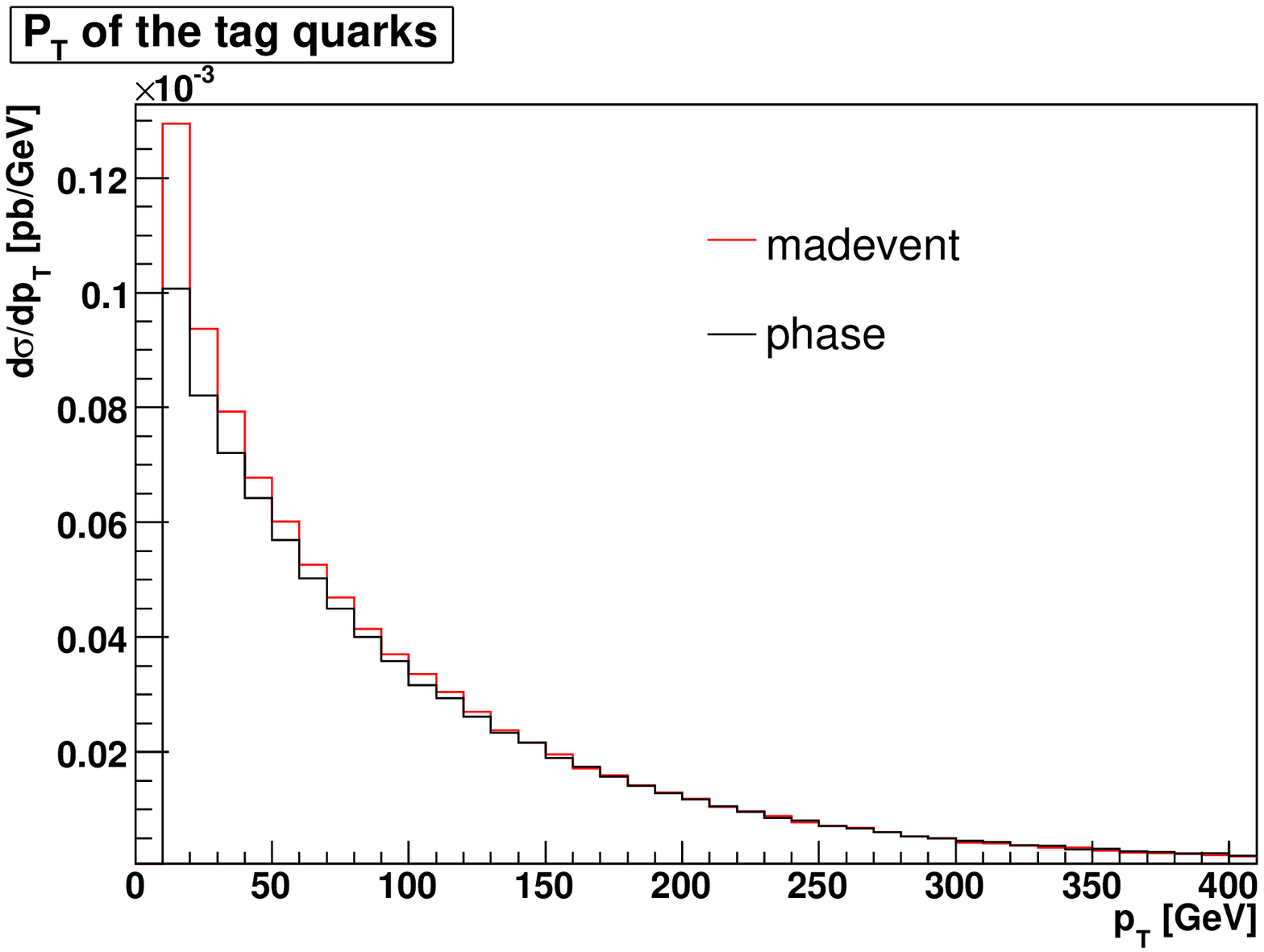,width=8cm}} 
\mbox{\epsfig{file=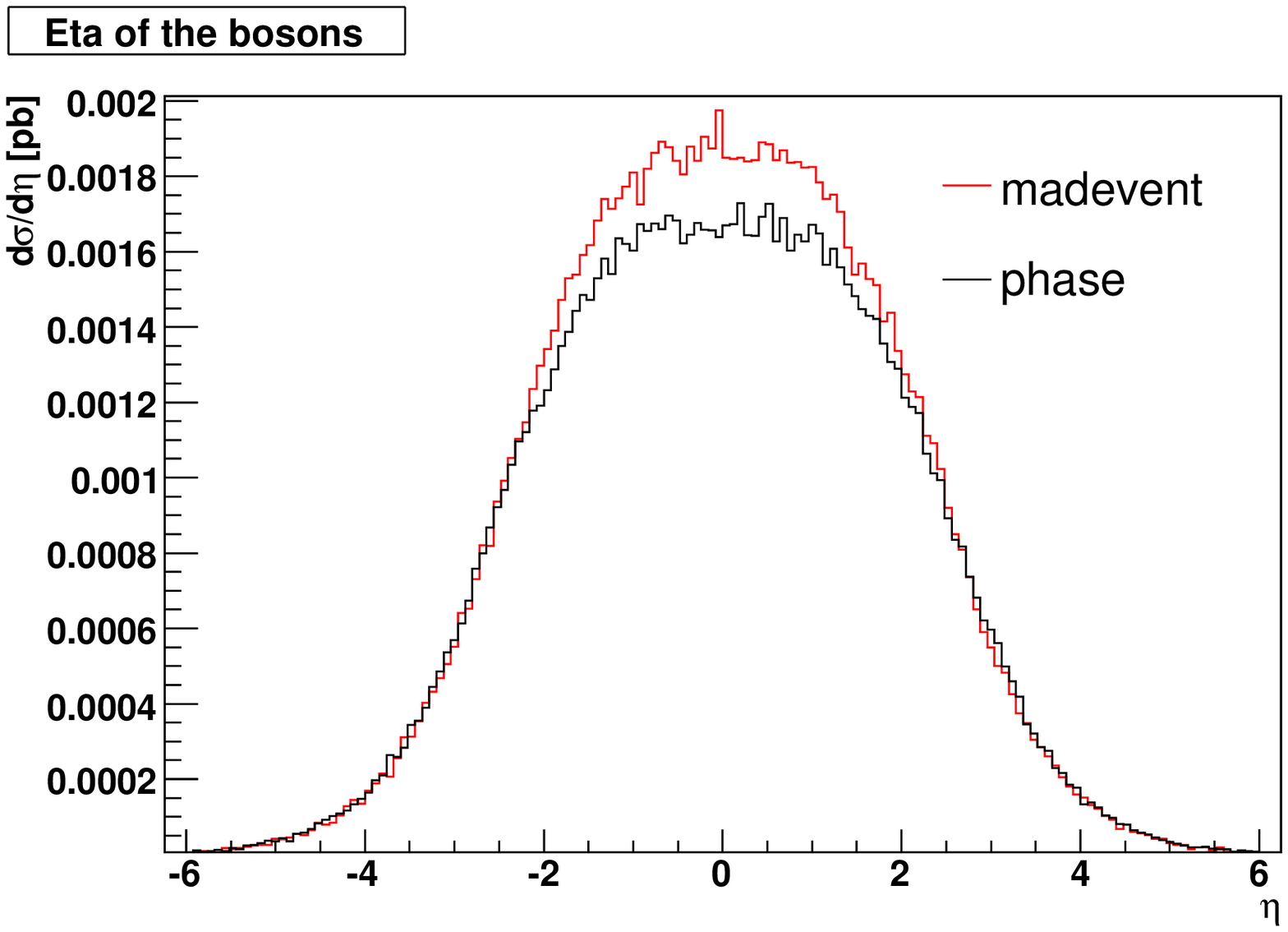,width=8cm}
\epsfig{file=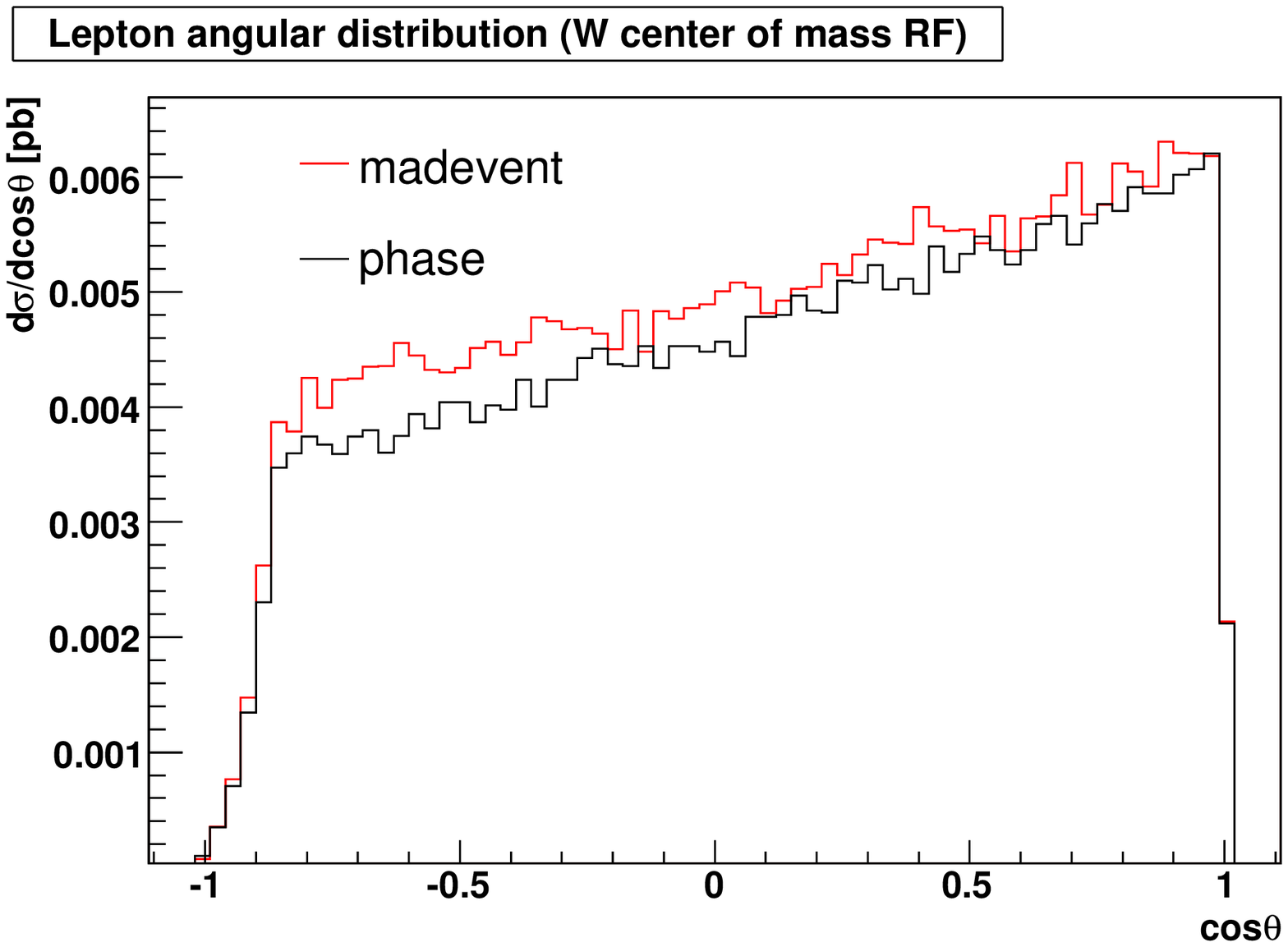,width=8cm}} 
\caption{ Cross section for the processes
$ uu \rightarrow uu \overline{q}q^\prime \mu \nu $ with 
$\overline{q}q^\prime = \overline{d}u + \overline{s}c$
as a function of the separation in
pseudorapidity between the spectator quarks, their transverse momentum,
the pseudorapidity of the reconstructed vector bosons and the angle of
the muon with respect to the \W direction of flight in the \W rest frame
for \Phase (black) and for \MadEvent (red)
for a Higgs mass of 500 \GeV.}
\label{madevent_s}
\end{center}
\end{figure}

\begin{figure}
\begin{center}
\mbox{\epsfig{file=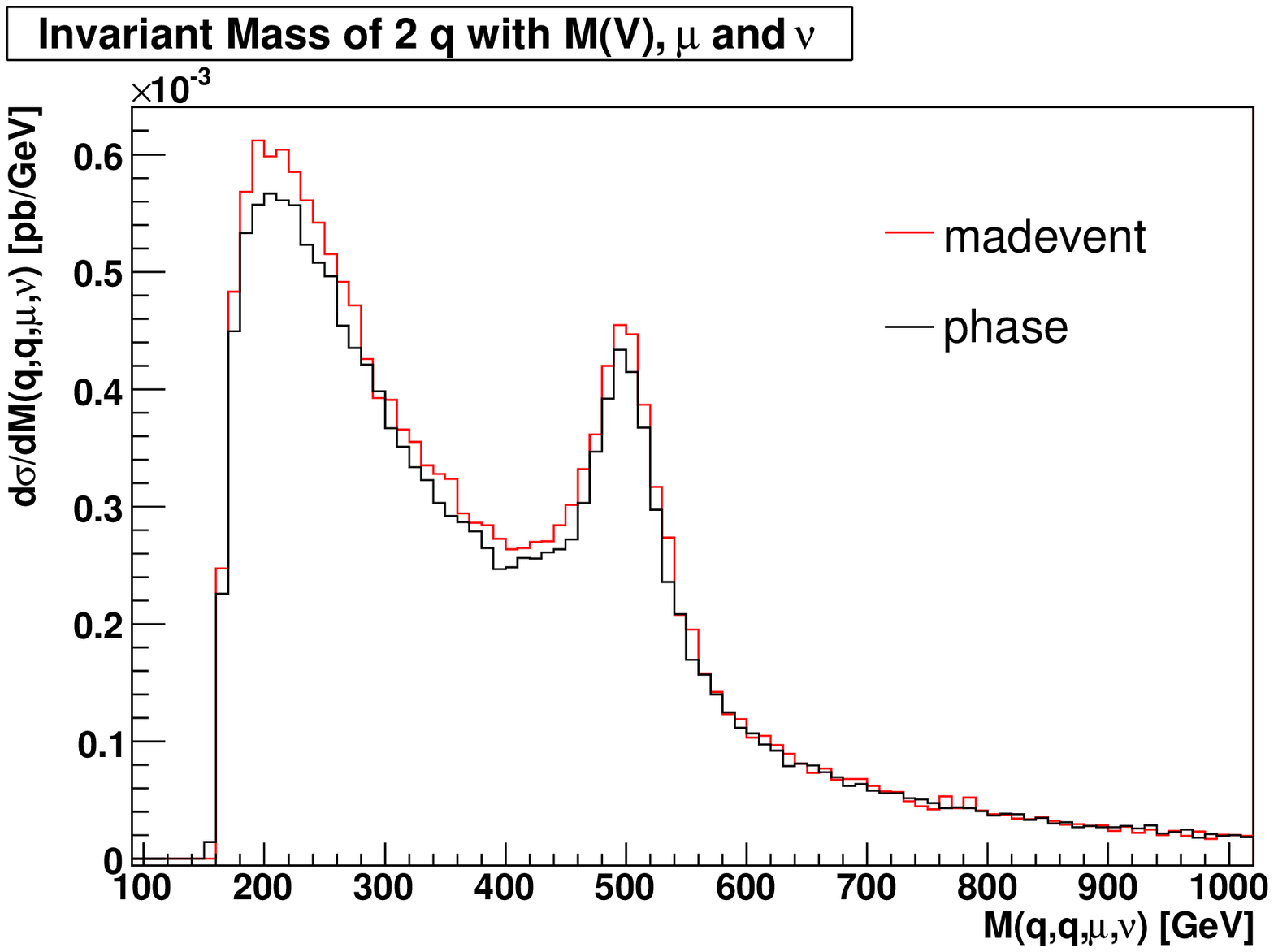,width=8cm}
\epsfig{file=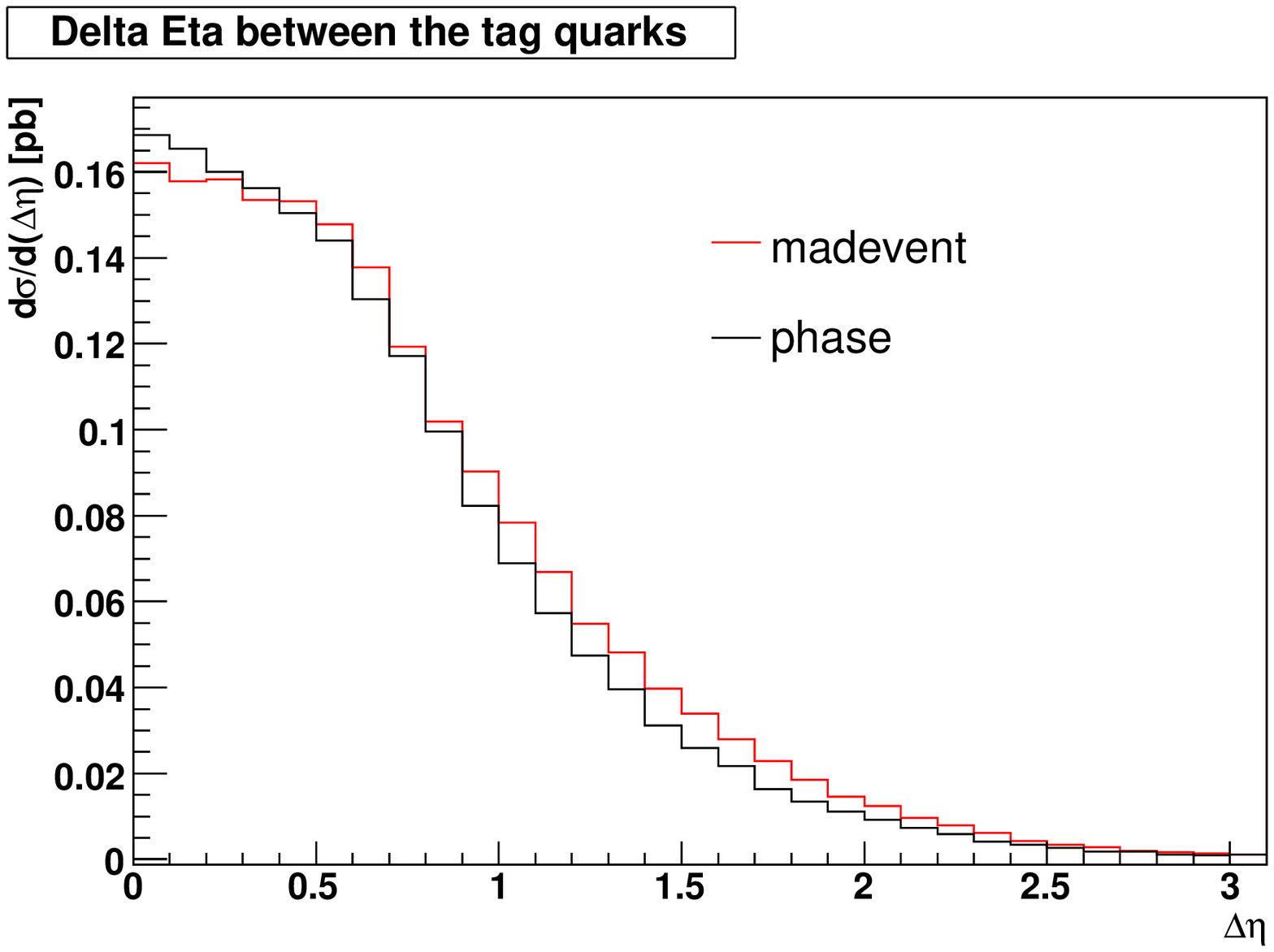,width=8cm}} 
\mbox{\epsfig{file=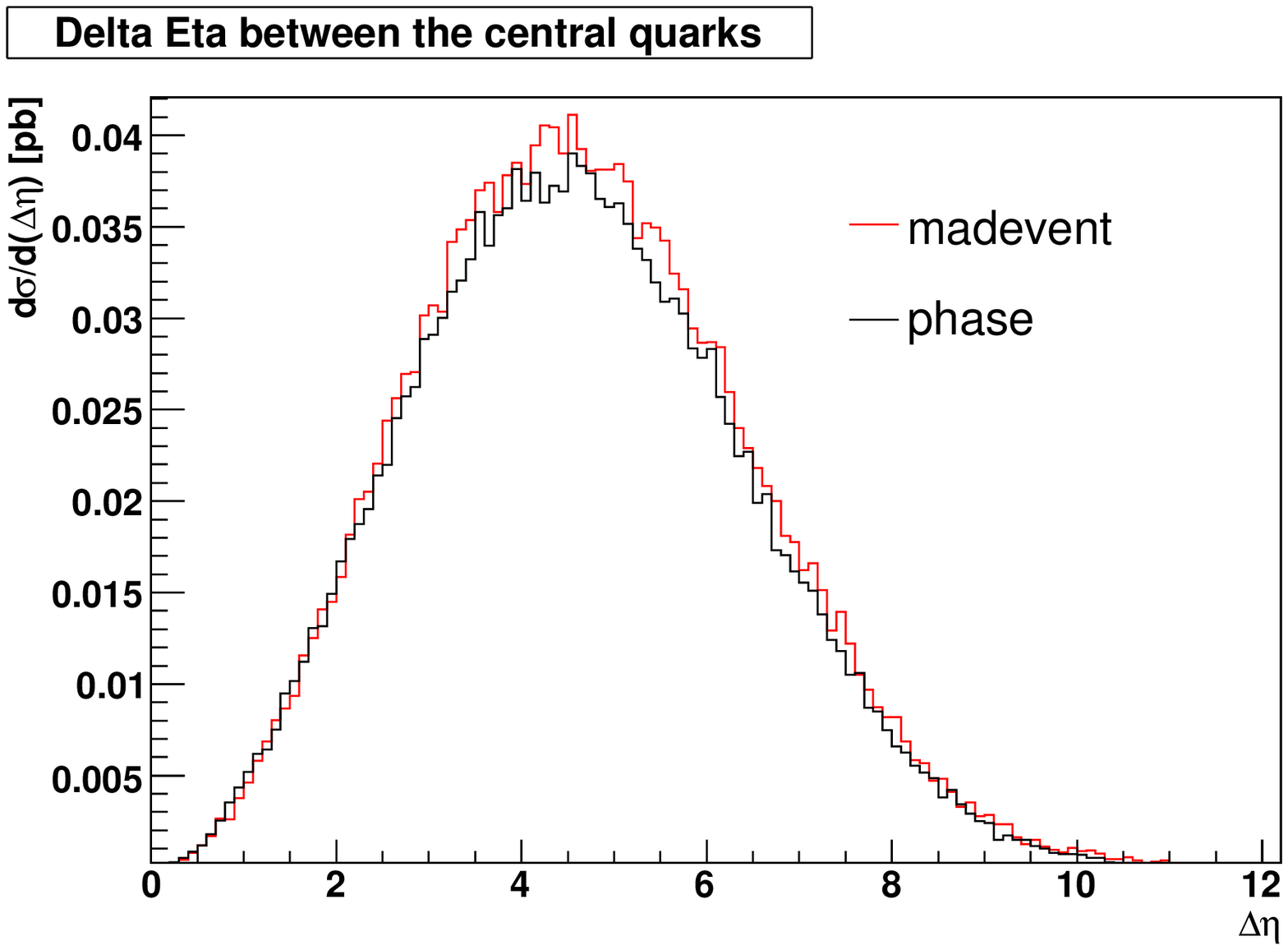,width=8cm}
\epsfig{file=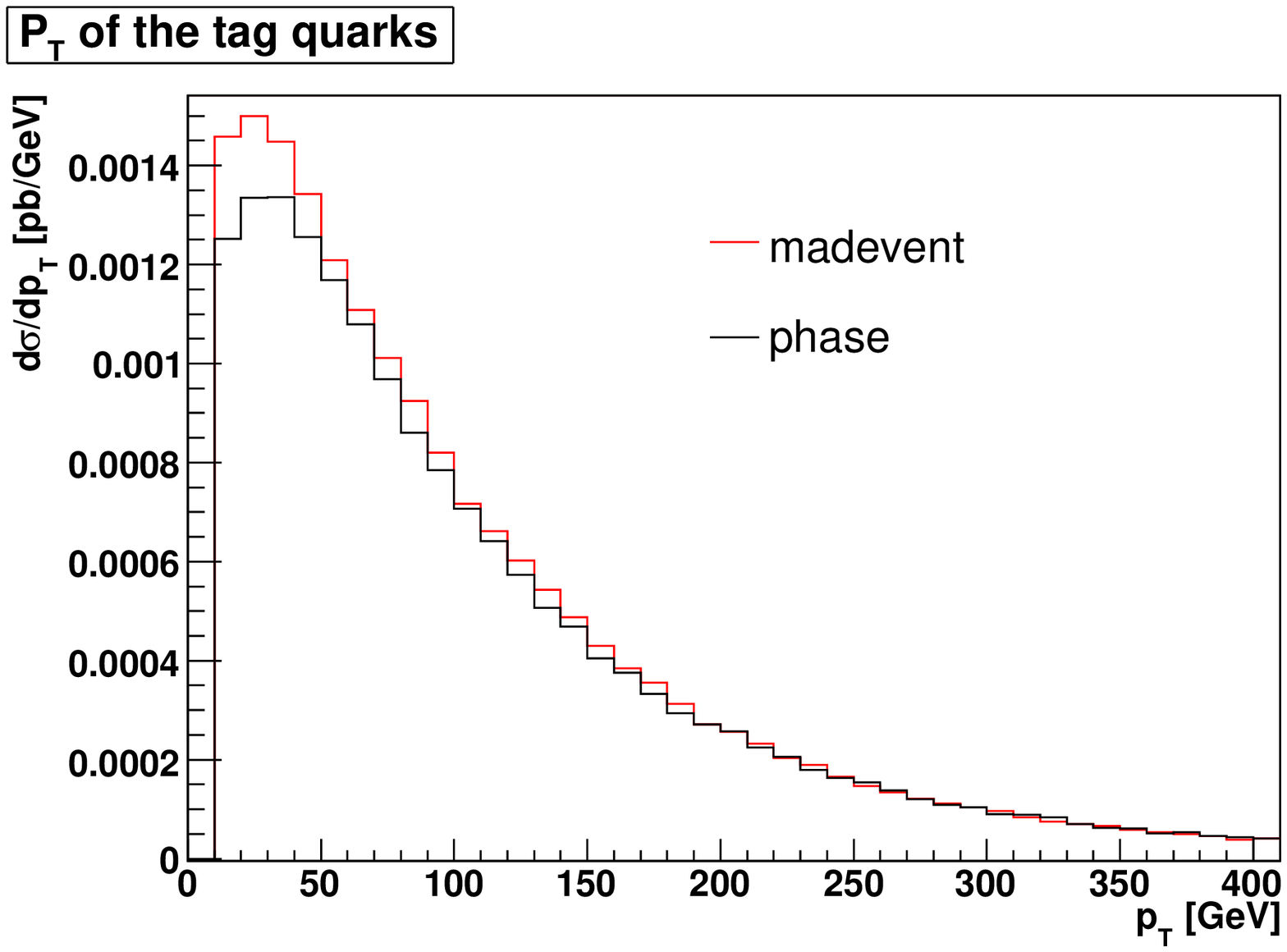,width=8cm}} 
\mbox{\epsfig{file=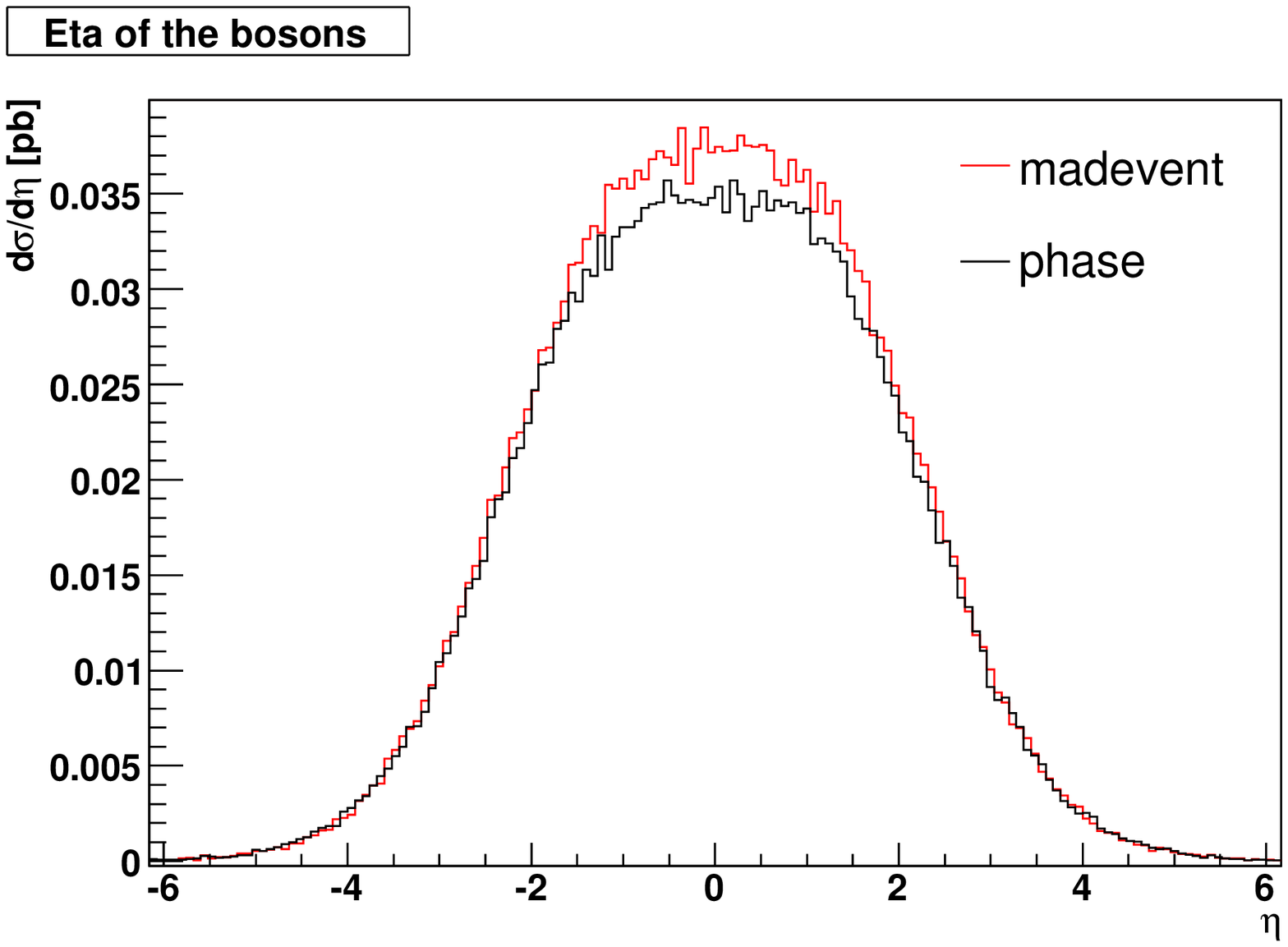,width=8cm}
\epsfig{file=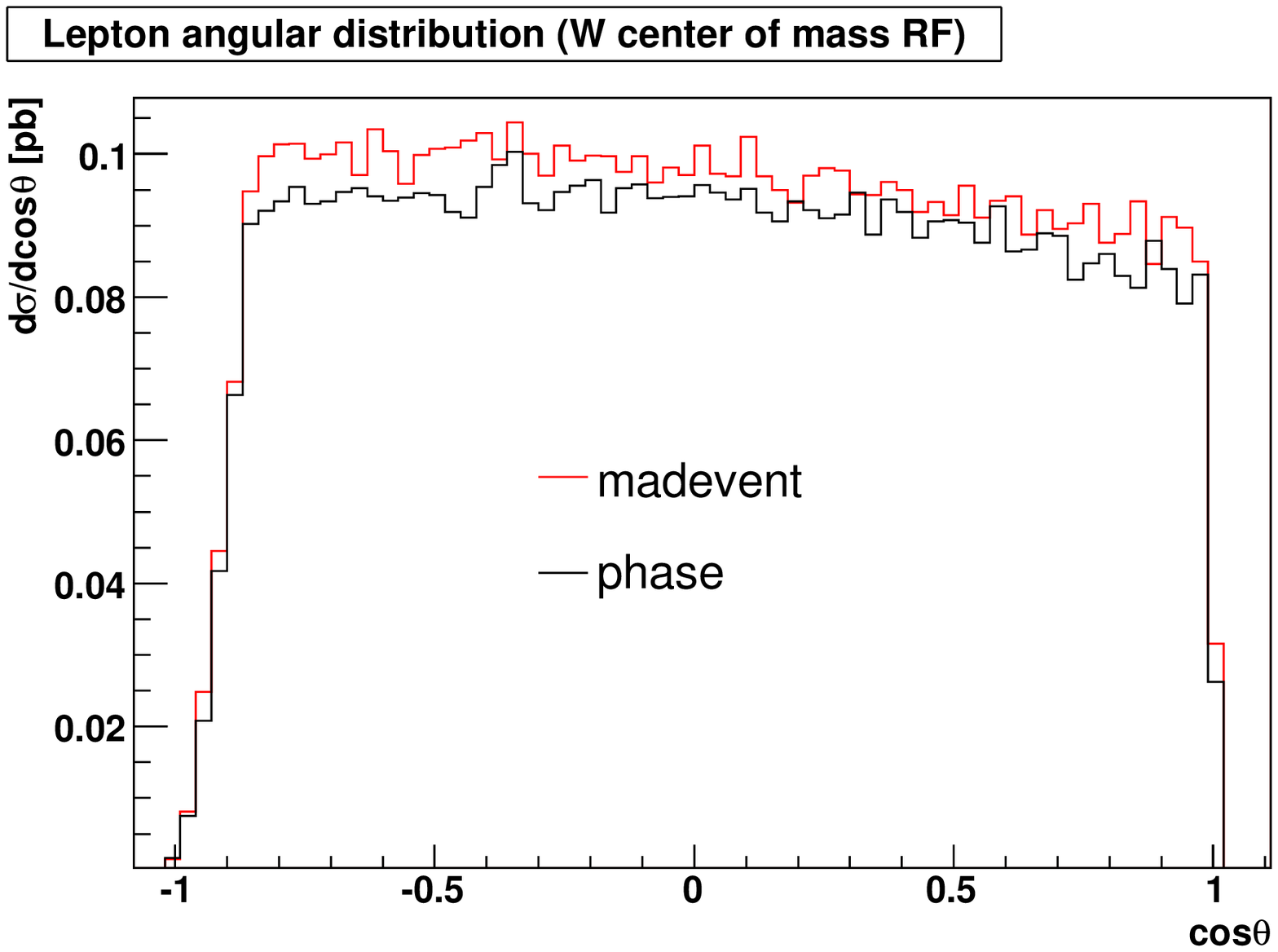,width=8cm}} 
\caption{ 
Cross section for the full sample
as a function of
the invariant mass of the two candidate vector bosons,
the $\Delta\eta$ between the two most central quarks,
the $\Delta\eta$ between spectator quarks, their transverse momentum,
the pseudorapidity of the reconstructed vector bosons and the angle of
the muon with respect to the \W direction of flight in the \W rest frame
for \Phase (black) and for \MadEvent (red)
for a Higgs mass of 500 \GeV.
}
\label{madevent_all}
\end{center}
\end{figure}

We have compared the predictions of \Phase and those of \MadEvent for the
process $ uu \rightarrow uu \overline{q}q^\prime \mu \nu $ with 
$\overline{q}q^\prime = \overline{d}u + \overline{s}c$ and for the sum of all
processes with no $b$ quarks
(b's require a separate generation in \MadEvent).
In this case we have adopted the scale $Q^2 = M_Z^2$ since our standard
choice \eqn{scale} cannot be used in the production times decay approach.
The signal has been selected as described in \sect{sig} therefore eliminating
the region of three vector boson production which is not correctly described by
\MadEvent.
A number of distributions are shown in \fig{madevent_s} and \fig{madevent_all}
respectively.
\fig{madevent_s} shows significant differences between the two results.
The transverse momentum of the spectator quarks is harder in \Phase, as is the
pseudorapidity of the reconstructed vector bosons, while the separation in
pseudorapidity between the spectator quarks tends to be smaller.
In \Phase the muon is more likely to fly in the direction of the \W momentum than
in \MadEvent.
While none of these differences is dramatic their impact should be
carefully assessed. 
 
These differences are also present in \fig{madevent_all}, only slightly diluted in
the more inclusive sample. \fig{madevent_all} presents also the distribution of
the invariant mass of the two candidate vector bosons and of 
the $\Delta\eta$ between the two most central quarks. At large invariant masses the two
predictions coincide while close to the \VV threshold \MadEvent predictions are about 
7\% larger. The difference in pseudorapidity between the two most central quarks
is significantly harder in \MadEvent with differences of about 15\% at large
separation.

It should be noted that the result of the comparison between the two codes
is sensitive to the mass window around the nominal mass of the EW bosons which
is included. In  real simulations this will depend on the necessity of rejecting
backgrounds and on the uncertainty with which invariant masses of two quarks and
two leptons will be reconstructed.
\MadEvent predictions show no dependence on this cut,
since all \W's are produced exactly on mass shell. \Phase, on the contrary,
takes into account the Breit-Wigner shape of the propagator squared
and includes a large number of additional diagrams which become important
when the vector boson are allowed to be off shell. 

\section{Higgs production in \Phase}

Higgs production in \VV fusion followed by Higgs decay to \WW or \ZZ is 
the second most abundant production channel over almost the full range of 
Higgs masses which will be explored at LHC. It is regarded
as the channel with the highest statistical significance for an intermediate
mass Higgs \cite{Atlas_HinWW}. \Phase is capable of simulating Higgs production
in \VV fusion together with all its EW irreducible background
for any Higgs mass and is particularly useful in the intermediate
mass range where only one of the vector bosons can be approximately
treated in a production times decay approach.
For an intermediate mass Higgs the dilepton final state
$H \rightarrow WW^{(\ast)} \rightarrow l\nu l\nu$ is slightly favored with
respect to the $H \rightarrow WW^{(\ast)} \rightarrow l\nu jj$ channel because
of the $W + nj$ background which affects the latter. In both cases the main
background comes from $t \bar t$ production followed in importance
by EW $WWjj$ production
which is estimated to be about 10\%, much larger than QCD $WWjj$ production.
The latter can be reduced with a central jet veto which does not affect the
former..
The cross section after selection cuts is of the order of $1\div 4\ fb$ and with
such a small number of events it is important to have a precise simulation of
the complete final state with full spin correlations which are complicated by
the fact that in this range of Higgs masses one of the $W$ is far off shell.

\begin{figure}[thb]
\begin{center}
\mbox{\epsfig{file=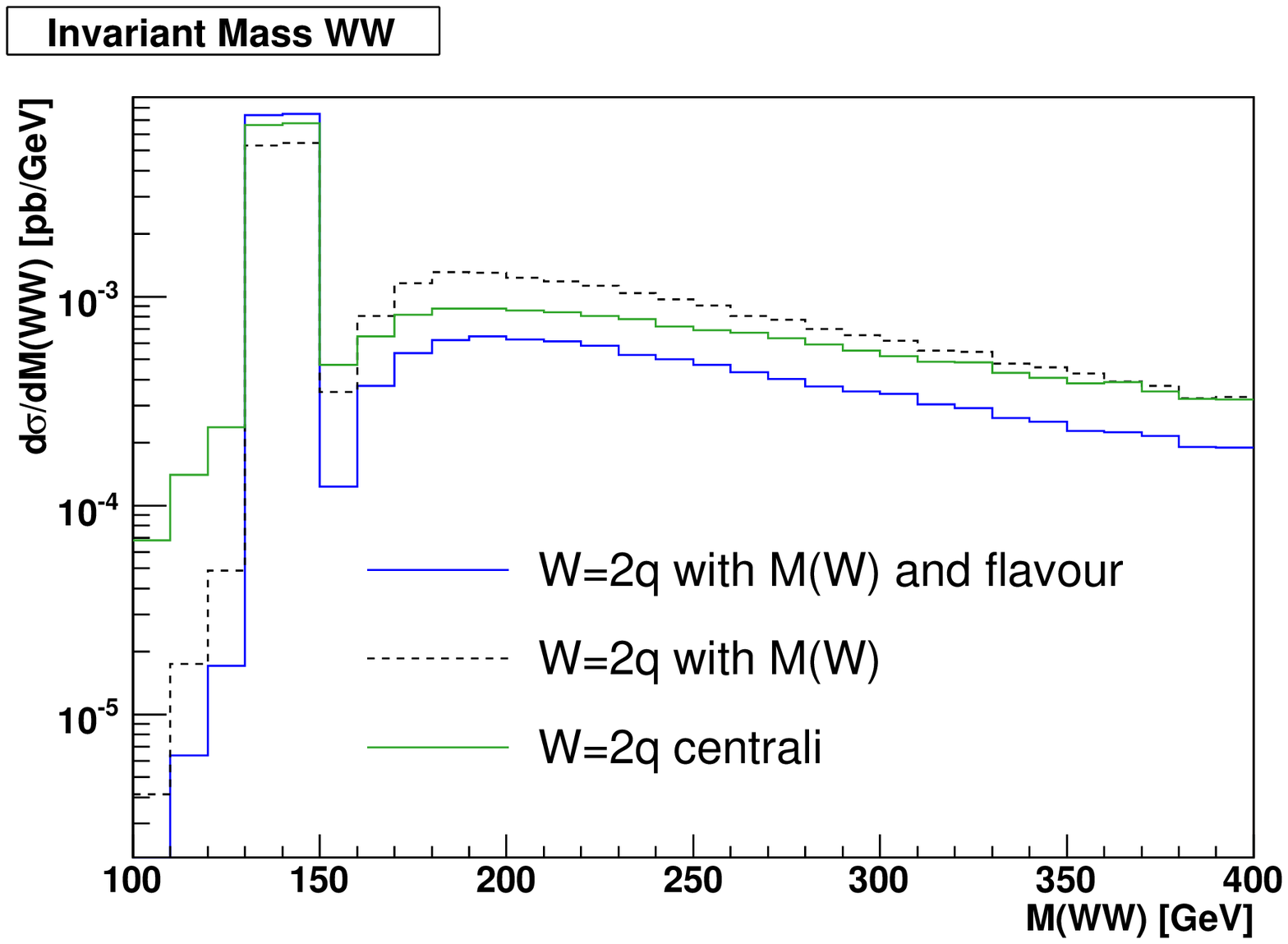,width=8cm}\epsfig{file=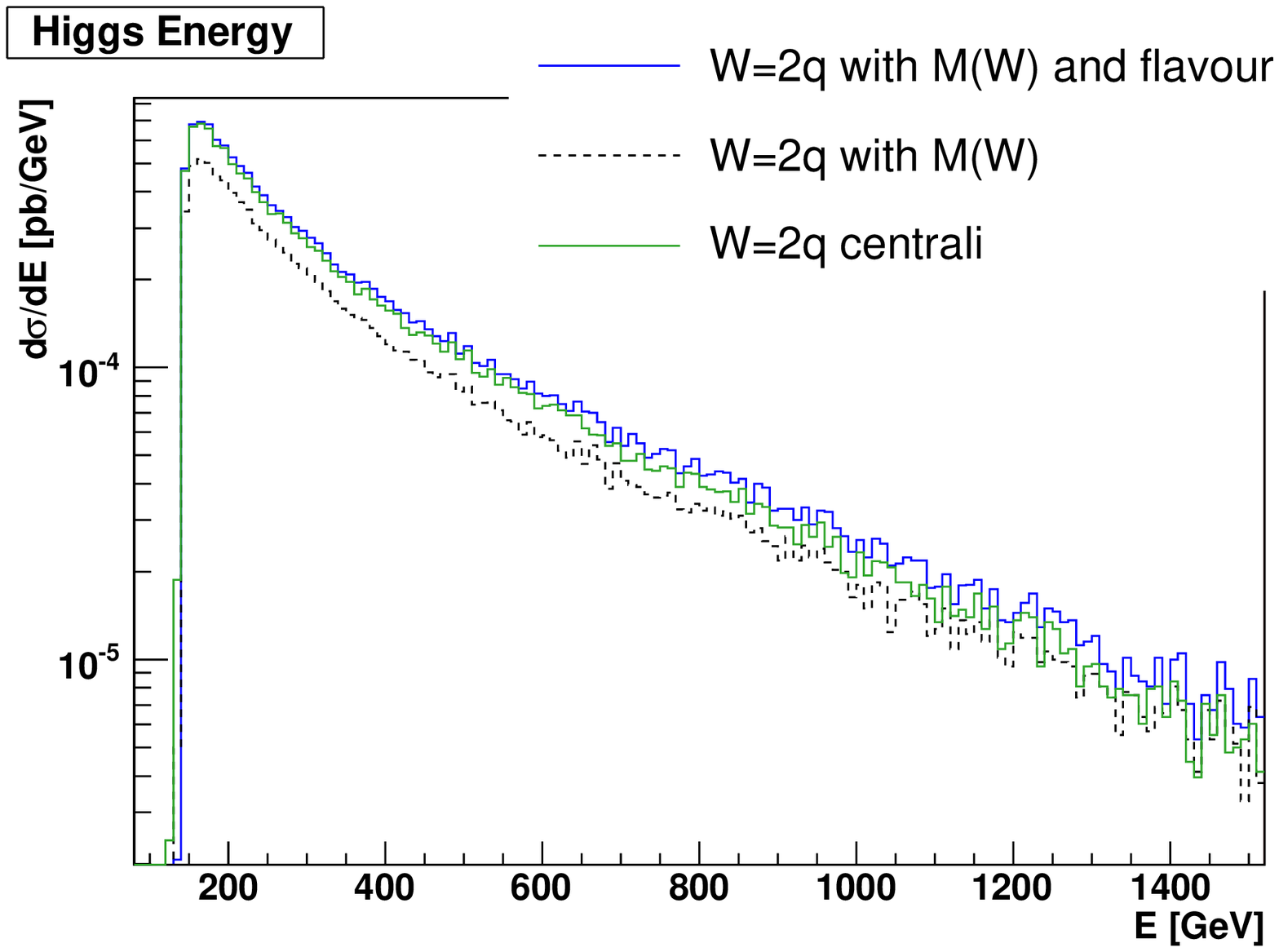 ,width=8cm}}
\mbox{\epsfig{file=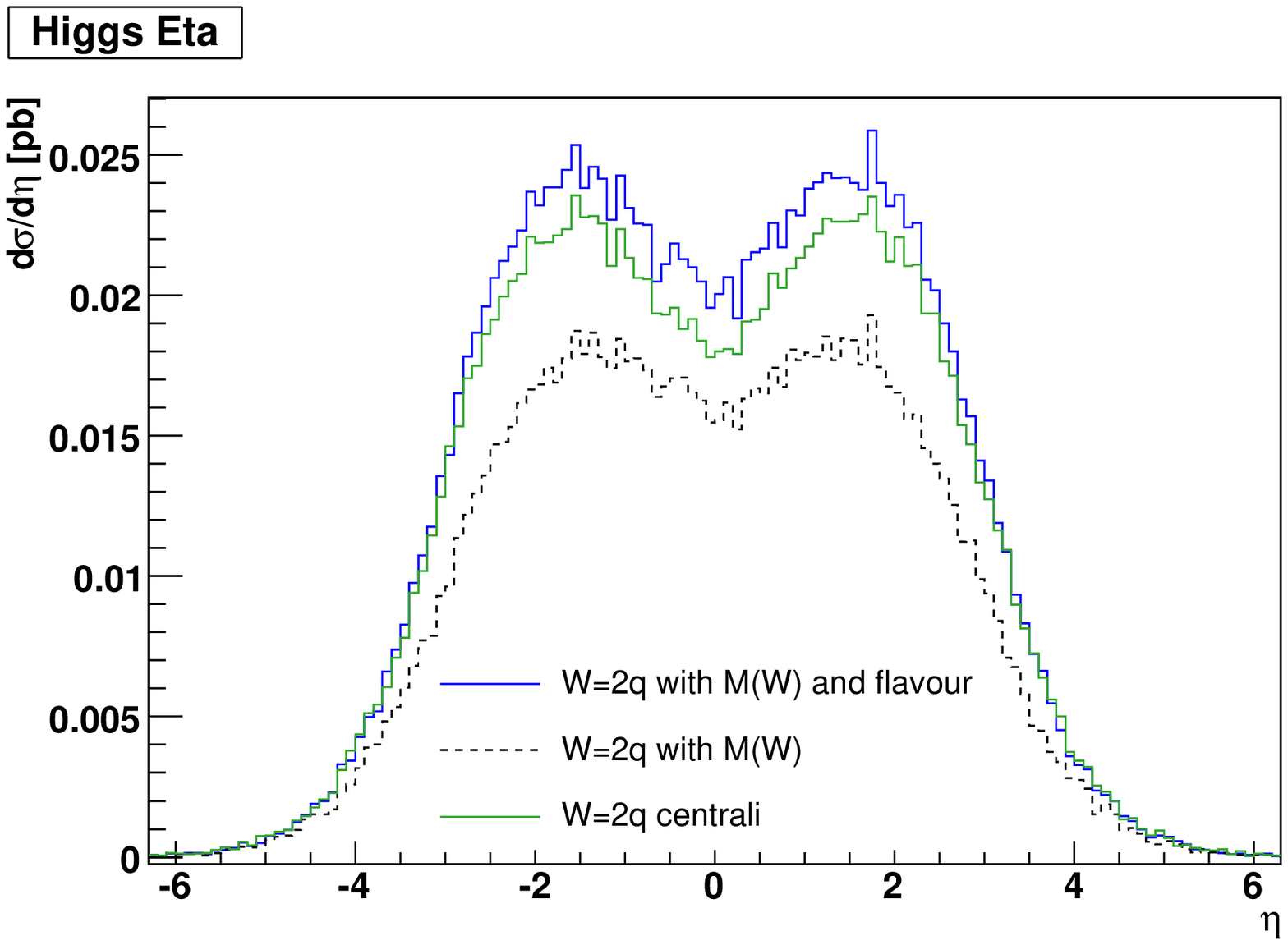,width=8cm}\epsfig{file=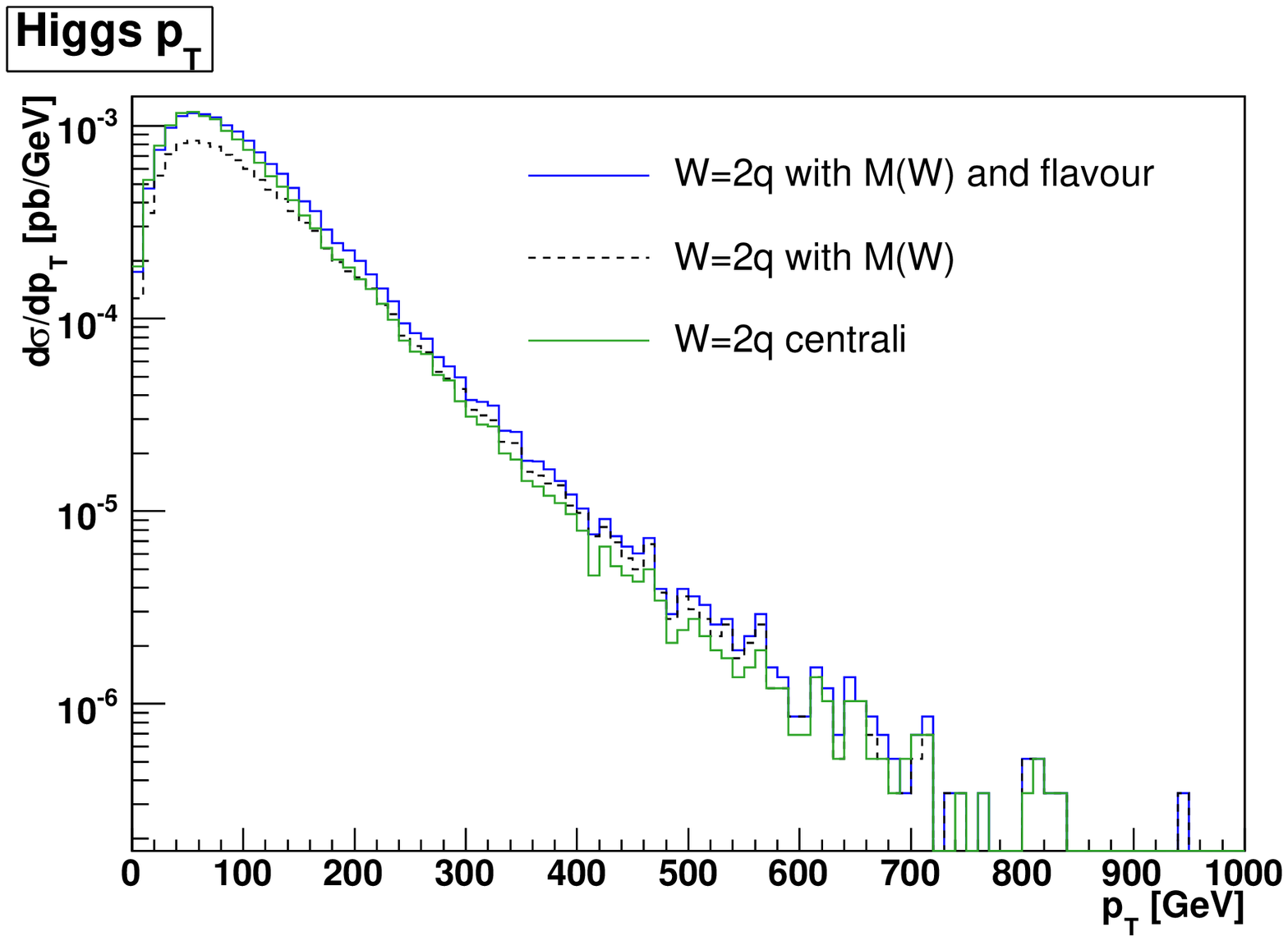,width=8cm}}
\caption{ Distribution of the invariant mass $M_{\VV}$
of the two candidate vector bosons, their energy, pseudorapidity and
transverse momentum for a Higgs mass of 140 \GeV. In the last three plots only
events with $120 < M_{\VV} < 160$ \GeV are included.}
\label{mh140}
\end{center}
\end{figure}
In \fig{mh140} we present a number of kinematical distributions
for M(H)=140 \GeV
with different selection procedures. The green line refers to
identifying the two
most central quarks in pseudorapidity as the quarks from \W 
decay while the black histogram refers to selecting the two
quarks among the four that have
the mass closest to the \W boson mass. The blue line is obtained 
selecting the quark pairs which have the correct flavour content
to be produced in a \W decay and which
have the mass closest to the \W mass.
The first procedure, which is also the most robust from an experimental point of
view, agrees well with the third one, which is more appropriate
from a theoretical perspective but which can only be applied at generator
level, over the full range of all variables.
It should be pointed out that the
actual ratio of signal over background depends cruciallly on the experimental
resolution and on the set of additional selection cuts \cite{Atlas_HinWW}.
For reference, the invariant mass distribution in \fig{mh140}
is plotted with a bin size of 10 \GeV.

\begin{figure}
\vspace{-2cm}
\begin{center}
\mbox{
{\epsfig{file= 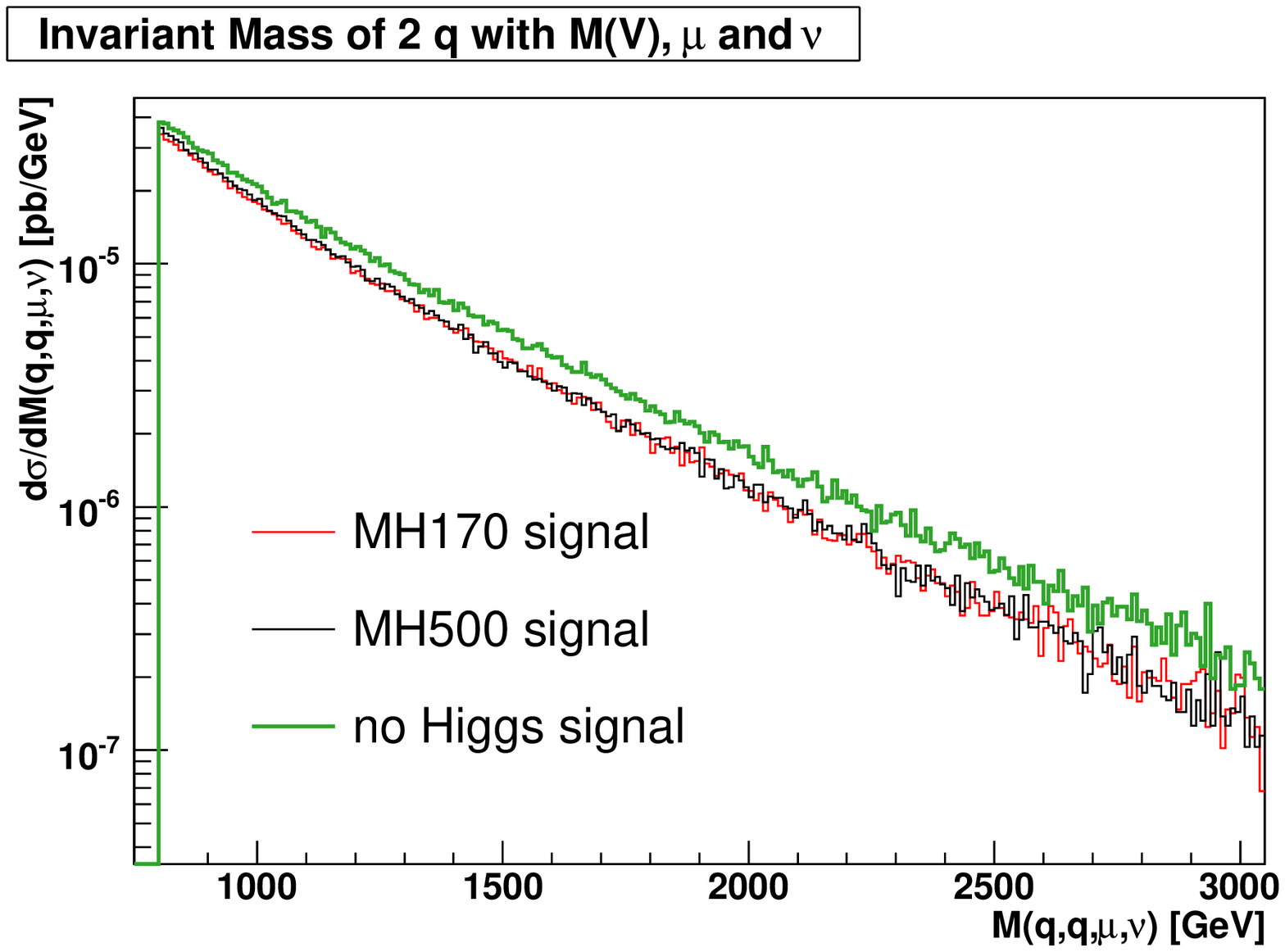,width=8cm}} 
{\epsfig{file=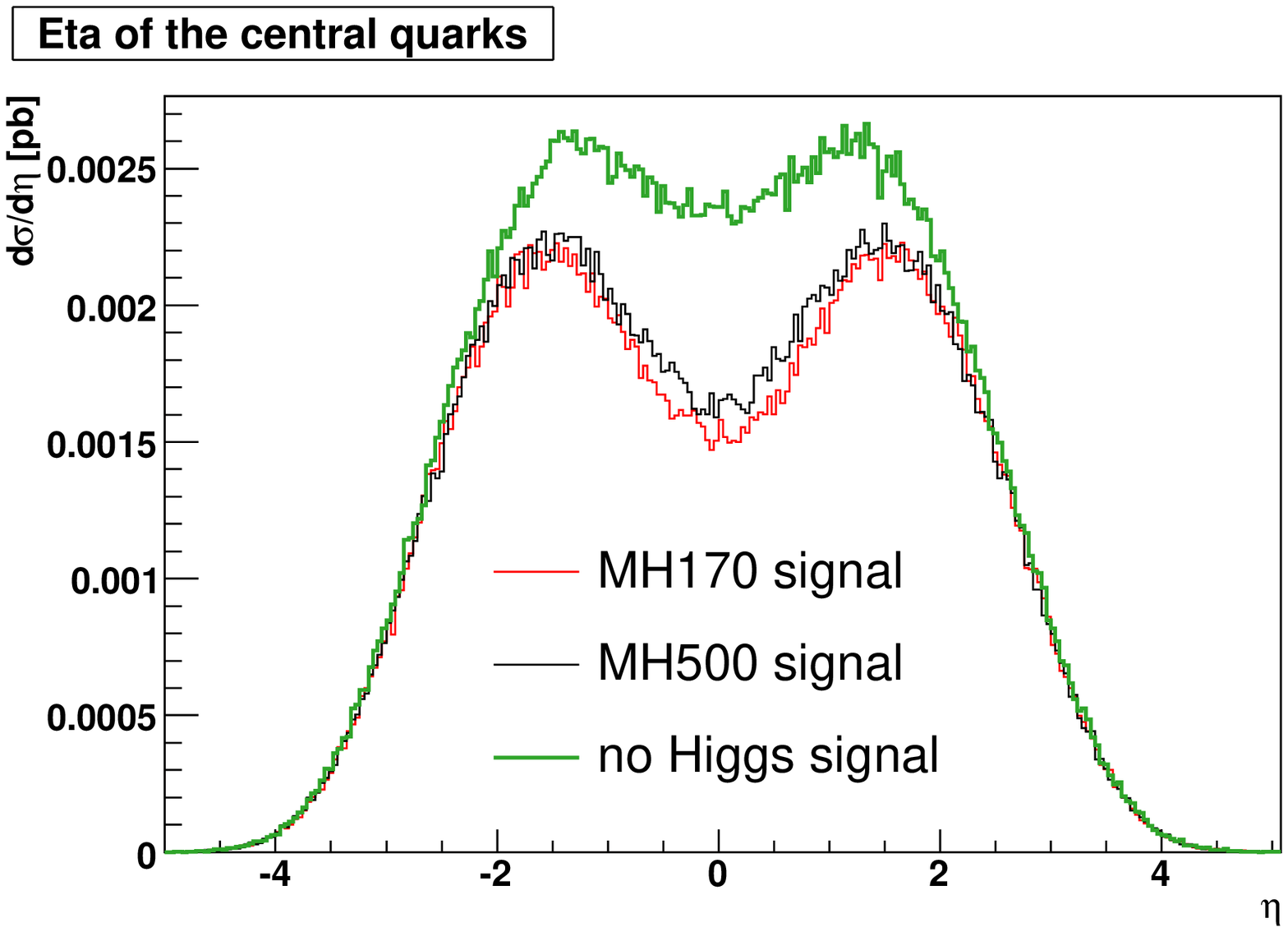,width=8cm}} 
}
\mbox{
{\epsfig{file= 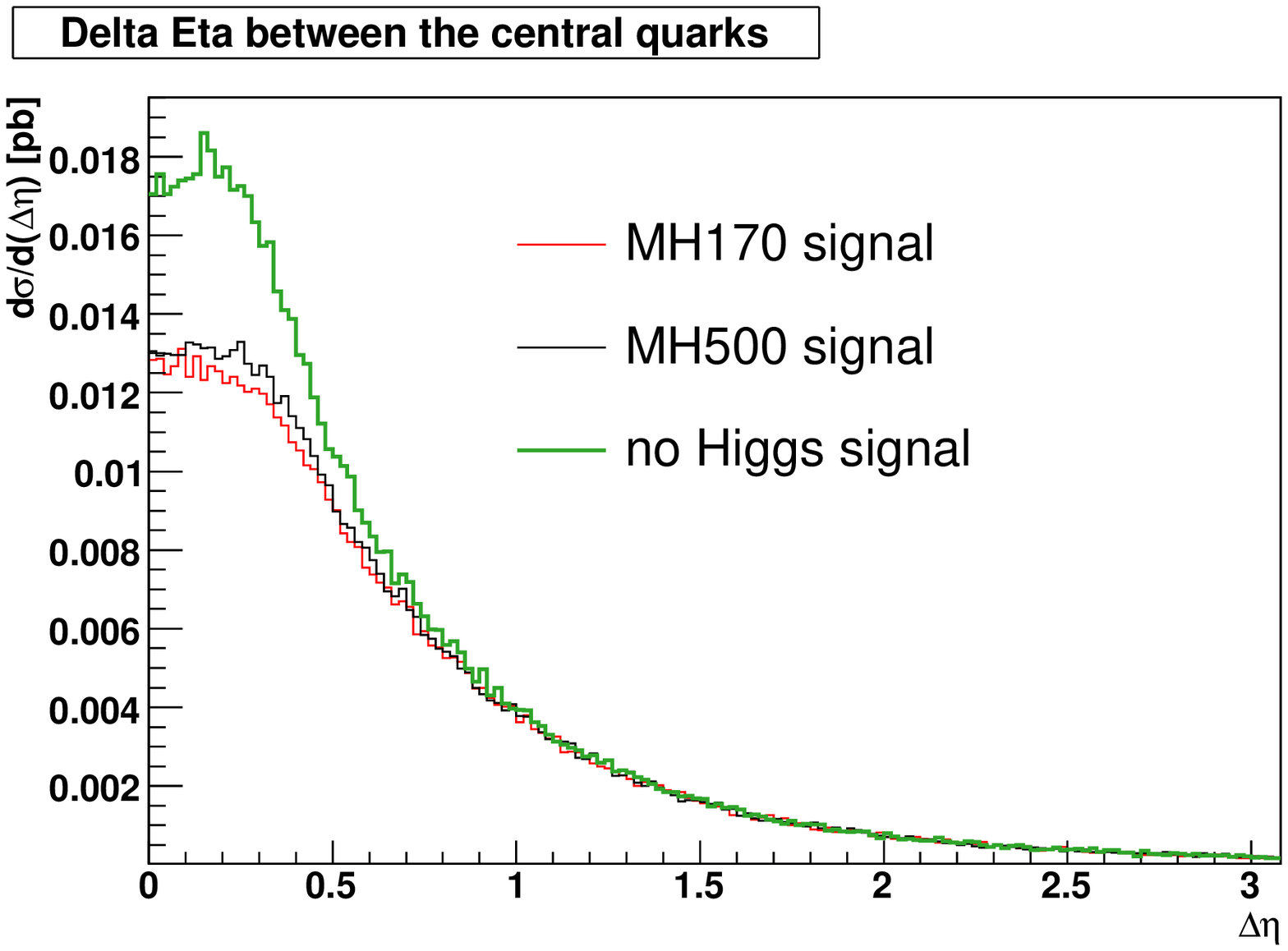,width=8cm}} 
{\epsfig{file=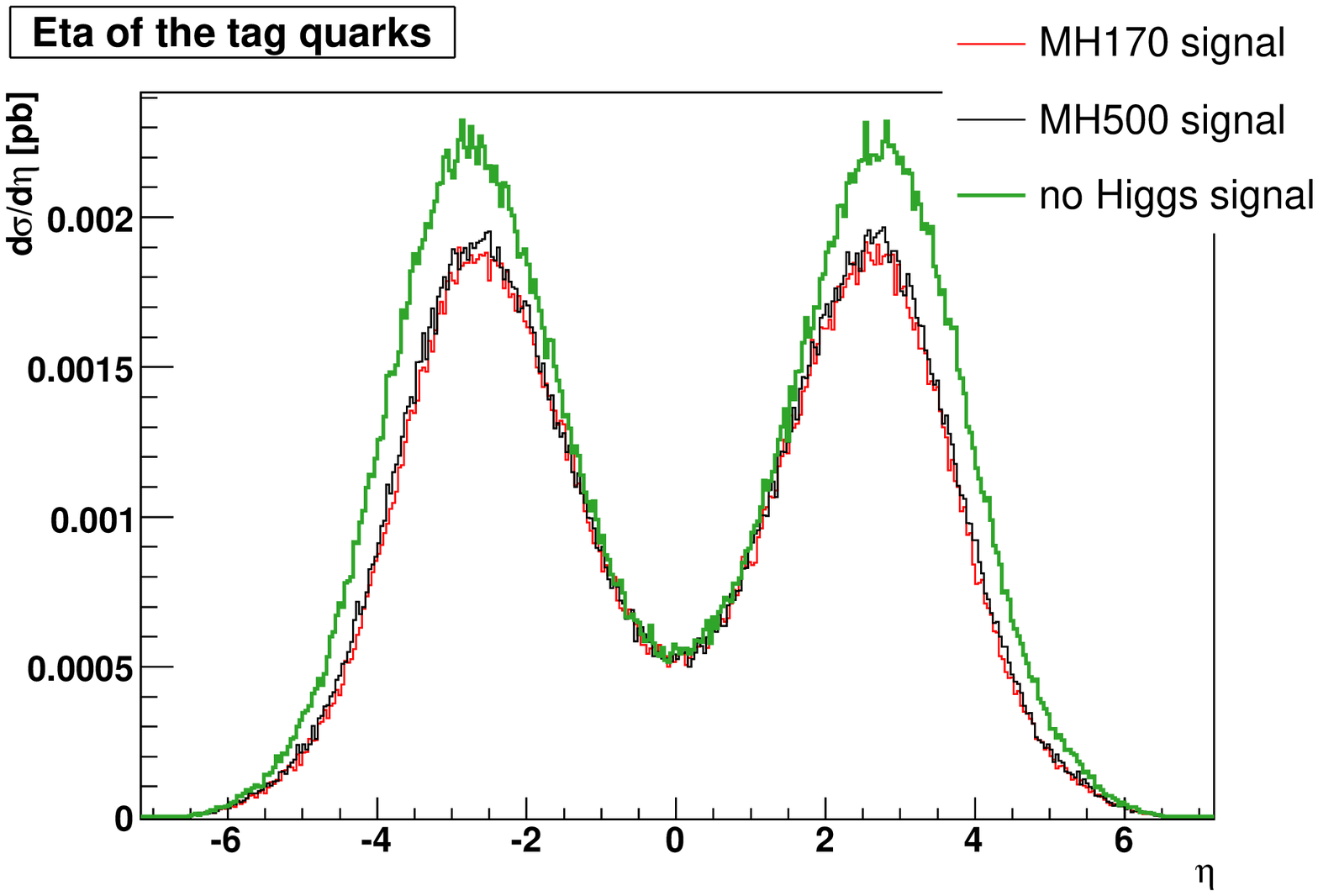,width=8cm}} 
}
\mbox{
{\epsfig{file= 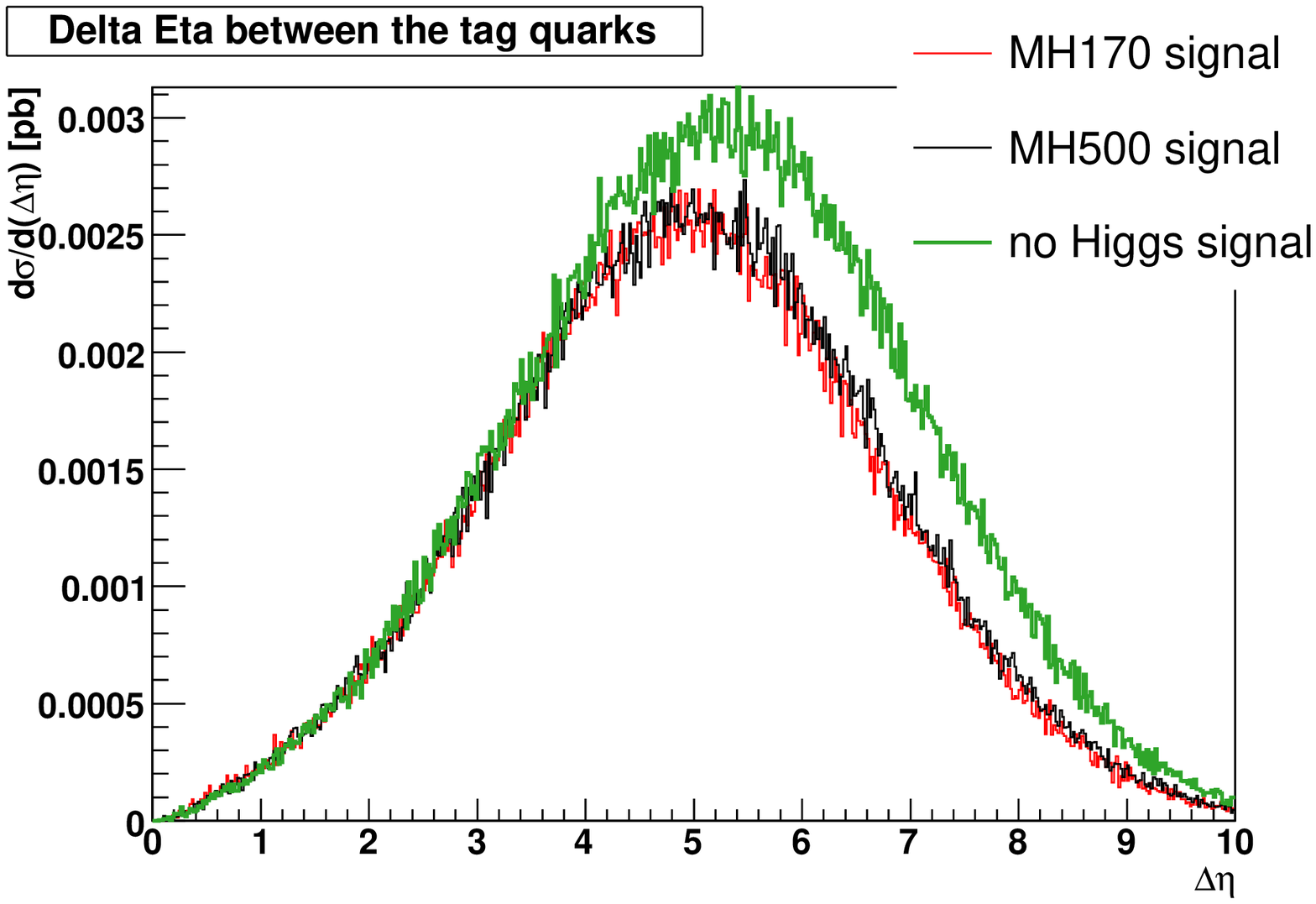,width=8cm}} 
{\epsfig{file=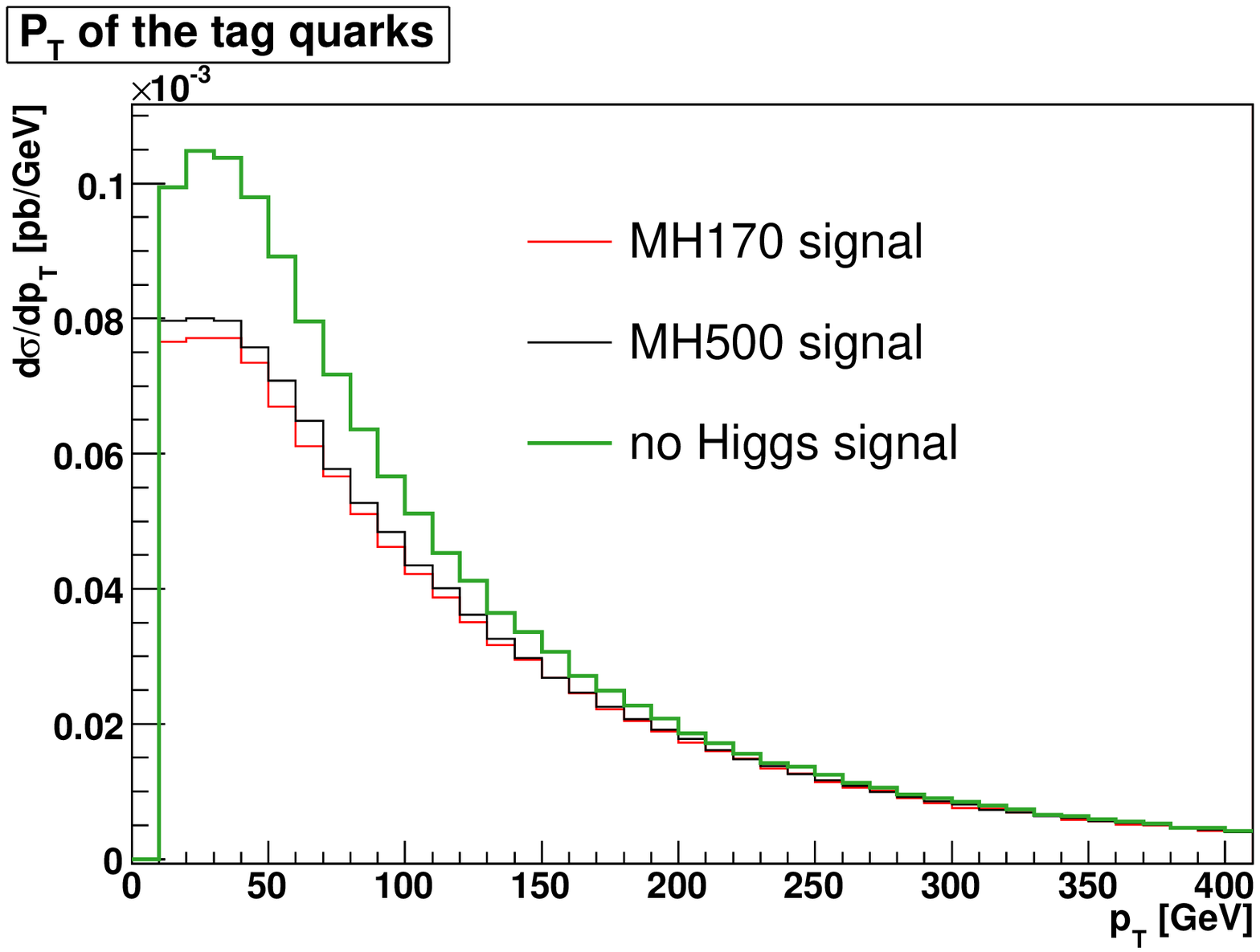,width=8cm}} 
}
\mbox{
{\epsfig{file= 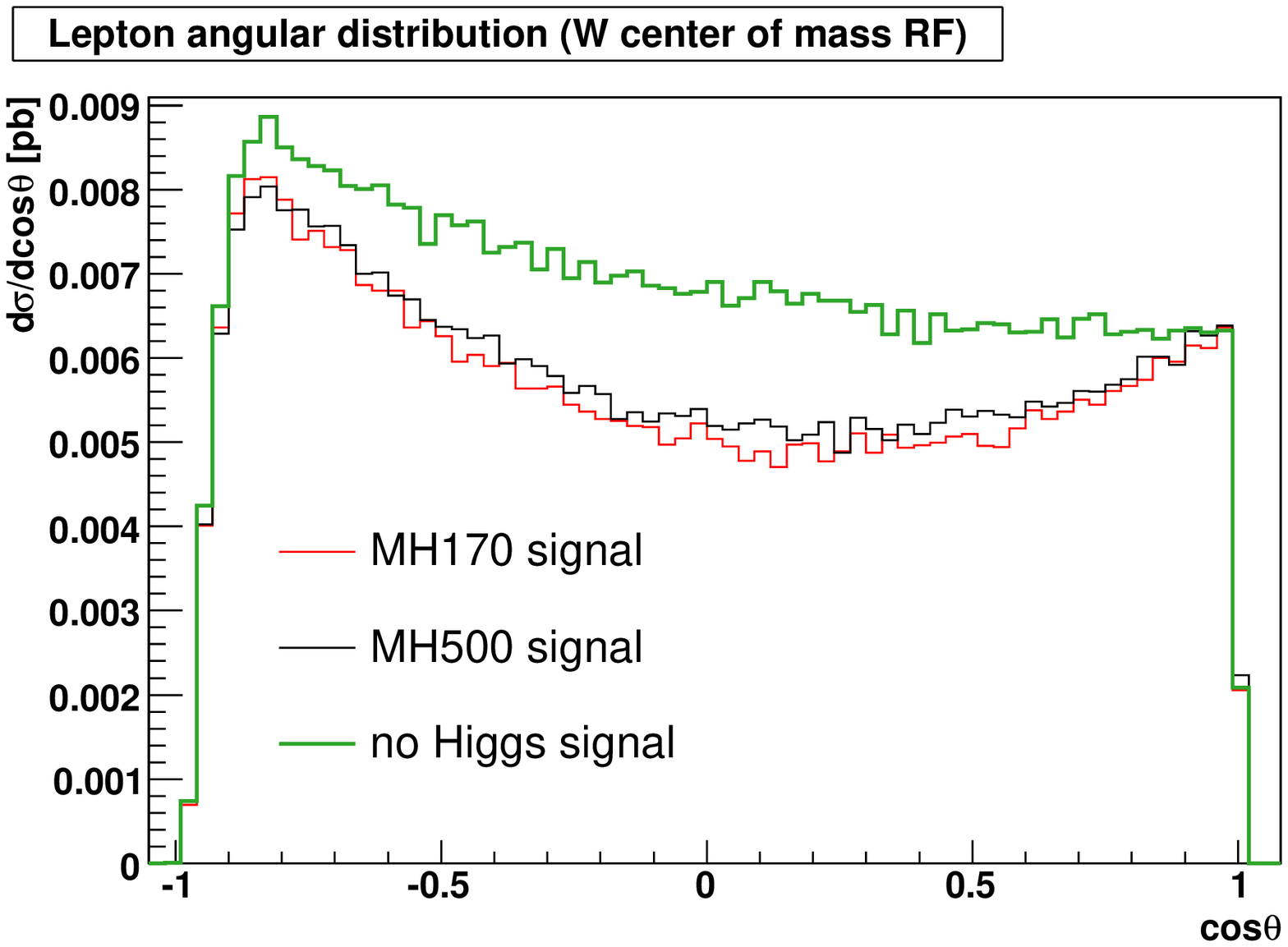,width=8cm}}
{\epsfig{file= 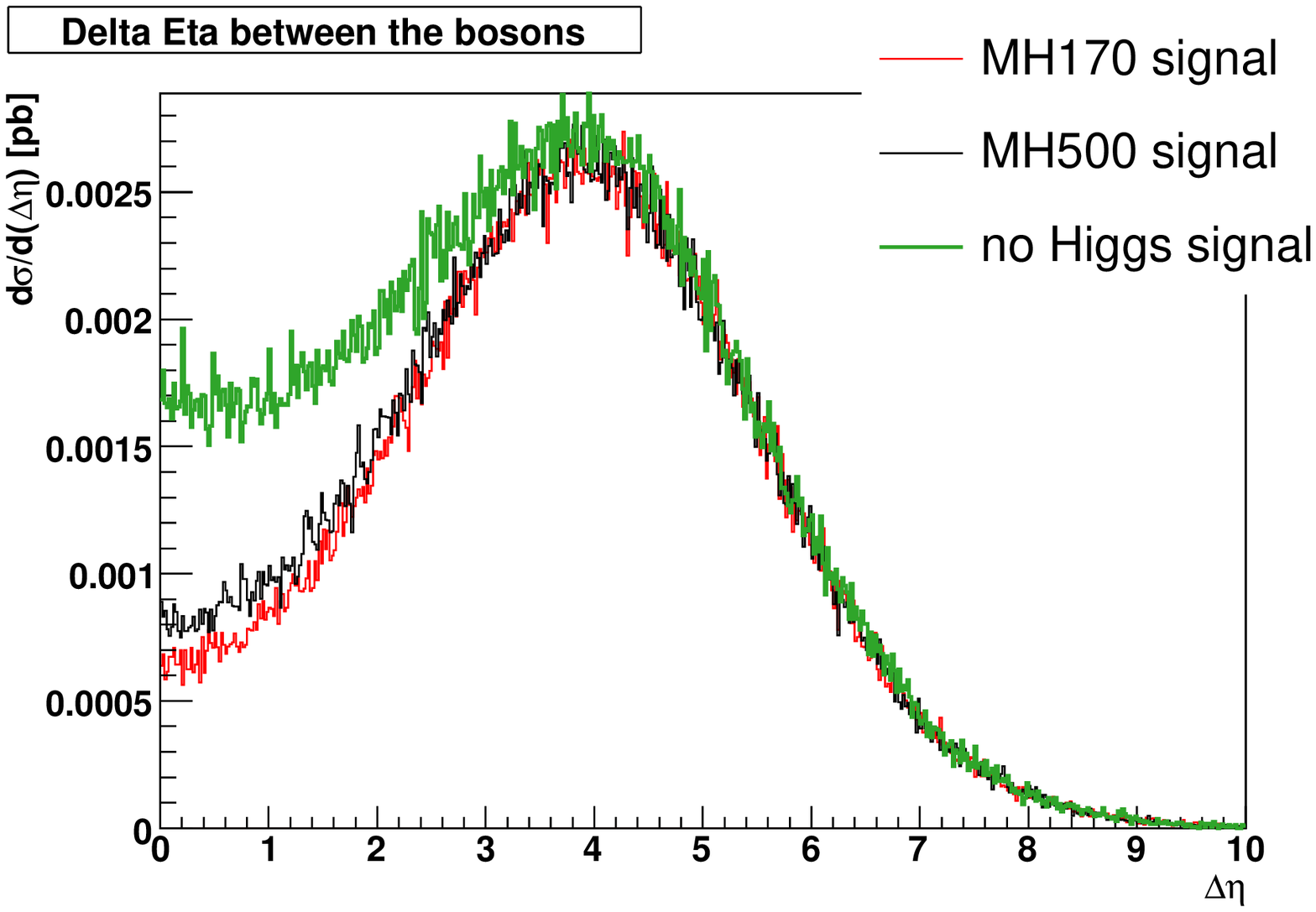,width=8cm}}}
\caption{ The invariant mass of the two vector bosons,
the pseudorapidity $\eta$ of the two central jets,
the $\Delta\eta$ of the two central jets,
the $\eta$ of the forward quarks
the $\Delta\eta$ of the two forward quarks,
the transverse momentum of the forward quarks,
the $\Delta\eta$ of the two vector bosons,
the cosine of the angle between the lepton and the W boson
in the W boson rest frame for $M(VW)>800$ \GeV. In green (full line) 
for the no-Higgs case, in black (dashed) for M(H)=500 \GeV and in red
for M(H)=170 \GeV.
The last four variables are used as input to the neural net.}
\label{dist:code}
\end{center}
\end{figure}

\section{The high mass region}
\label{sec:highmass}

An interesting physics possibility is to investigate whether
there exist or not an elementary Higgs boson by measuring the 
\VW cross section at large M(\VW). 
Preliminary studies performed at CMS with a fast
detector simulation \cite{CMS-0} showed that a resolution of about 10-15\% on
M(\VV) up to 2 \TeV is achievable with about 100 fb$^{-1}$.
The rise of the cross section related to unitarity violation
in the no-Higgs case is difficult to detect at the LHC, 
since the center-of-mass
energy is still rather low and the decrease of the proton distribution
functions at large $x$ has the dominating effect.
As we discussed in the previous section, if \WWL final states
could be isolated, the difference in the cross section at high \WW masses
would be sizeable, since the \WWL cross section 
decreases much more rapidly in the presence of a Higgs particle
than in the no-Higgs situation. 
In order to distinguish \WWL from \WWT we must exploit
the different behaviour of the final state in the two cases.

To this purpose kinematical distributions for M(H)=170 \GeV,
M(H)=500 \GeV and the
no-Higgs case have
been compared for M(\VV)$>$800 \GeV since the cross section 
at large M(\VV) for M(H)=170 \GeV and M(H)=500 \GeV is essentially due to
transversely polarized vector bosons, while
the cross section for the no-Higgs case is due to a
mixture of the two polarizations as shown in \fig{polariz1}.
\fig{dist:code} shows that the distributions are quite insensitive to the value
of the Higgs mass provided it is much smaller than
the invariant mass of the \VV
system. This raises the interesting possibility of defining Standard Model
predictions for high invariant mass production of \VV pairs. These predictions
will obviously suffer from the usual PDF and scale uncertainties, which could
however in principle be controlled by comparing with the cross section of some
appropriate ``standard candle'' process.
  
We have tried several sets of cuts and we believe that using Neural Network
is the most effective way of increasing the
separation between the no-Higgs case and the presence of a relatively light
Higgs. 
Two samples of events with a high invariant mass \VW pair,
for M(H)=500 \GeV and the 
no-Higgs case respectively, have been employed to train a Neural Network.
All events satisfy the cuts in \tbn{standard-cuts}.
A set of variables which discriminate between the two Higgs hypotheses
have been used  in the training. We have chosen four weakly correlated
variables: the
difference in pseudorapidity between the two bosons and between the two
tag quarks, the
transverse momentum of the tag quarks and the cosine of the angle
between the lepton and the \W boson in the \W center of mass
system. These (and other) kinematical 
variables are shown in \fig{dist:code} for the
no-Higgs case and for M(H)=500 \GeV and M(H)=170 \GeV.

\begin{table}[htb]
\begin{center}
\begin{tabular}{|l|c|c|c|c|c|} 
\cline{2-5}
 \multicolumn{1}{c}{} & \multicolumn{2}{|c|}{no-Higgs case}&\multicolumn{2}{|c|}{m(Higgs) = 500 GeV}&\multicolumn{1}{c}{}\\
\cline{2-6}
\multicolumn{1}{c|}{} & $\sigma$ & $\mathcal{L}$=$100fb^{-1}$
 & $\sigma$ & $\mathcal{L}$=$100fb^{-1}$ & ratio  \\
\hline
all events  & 13.6 fb & 1360 $\pm$ 37 & 11.6 fb & 1160 $\pm$ 34 & 1.2\\
\hline
NN $>$0.52 &  3.93 fb & 393 $\pm$ 20  & 2.70 fb & 270 $\pm$ 16 & 1.5\\
\hline
NN $>$0.54 & 3.17 fb & 317 $\pm$ 18  & 1.95 fb & 195 $\pm$ 14 & 1.6\\
\hline
NN $>$0.56 & 2.67 fb & 267 $\pm$ 16  & 1.47 fb & 147 $\pm$ 12 & 1.8\\
\hline
NN $>$0.58 & 2.28 fb & 228 $\pm$ 15  & 1.13 fb & 113 $\pm$ 11 & 2.0\\
\hline
\end{tabular}
\caption{Integrated cross section for M(\VW)$>$800 \GeV and number of
  expected events after one year at high luminosity.}
\label{nn-tab}
\end{center}
\end{table}

\begin{figure}
\begin{center}
\mbox{\epsfig{file=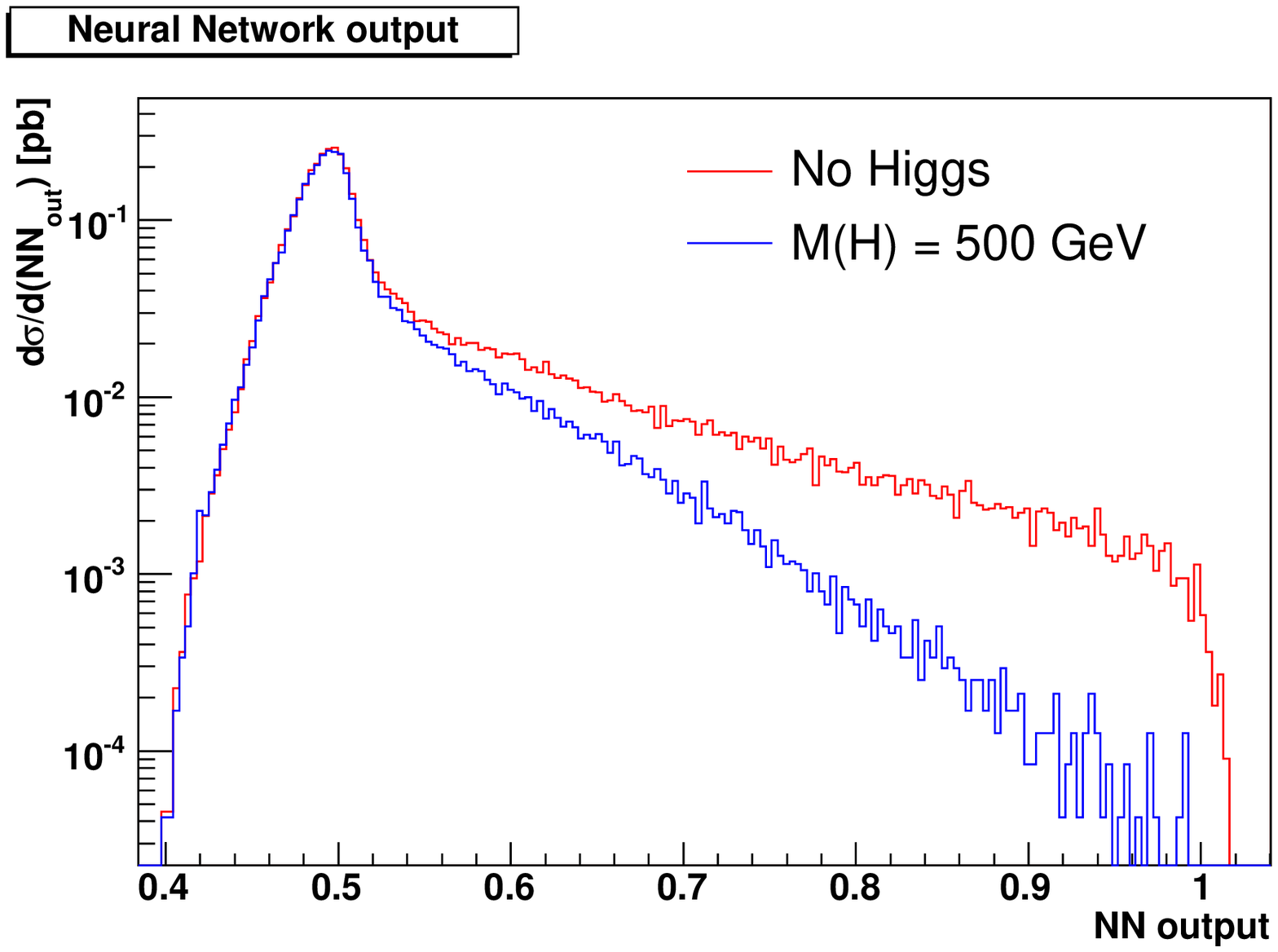,width=8cm}
\epsfig{file=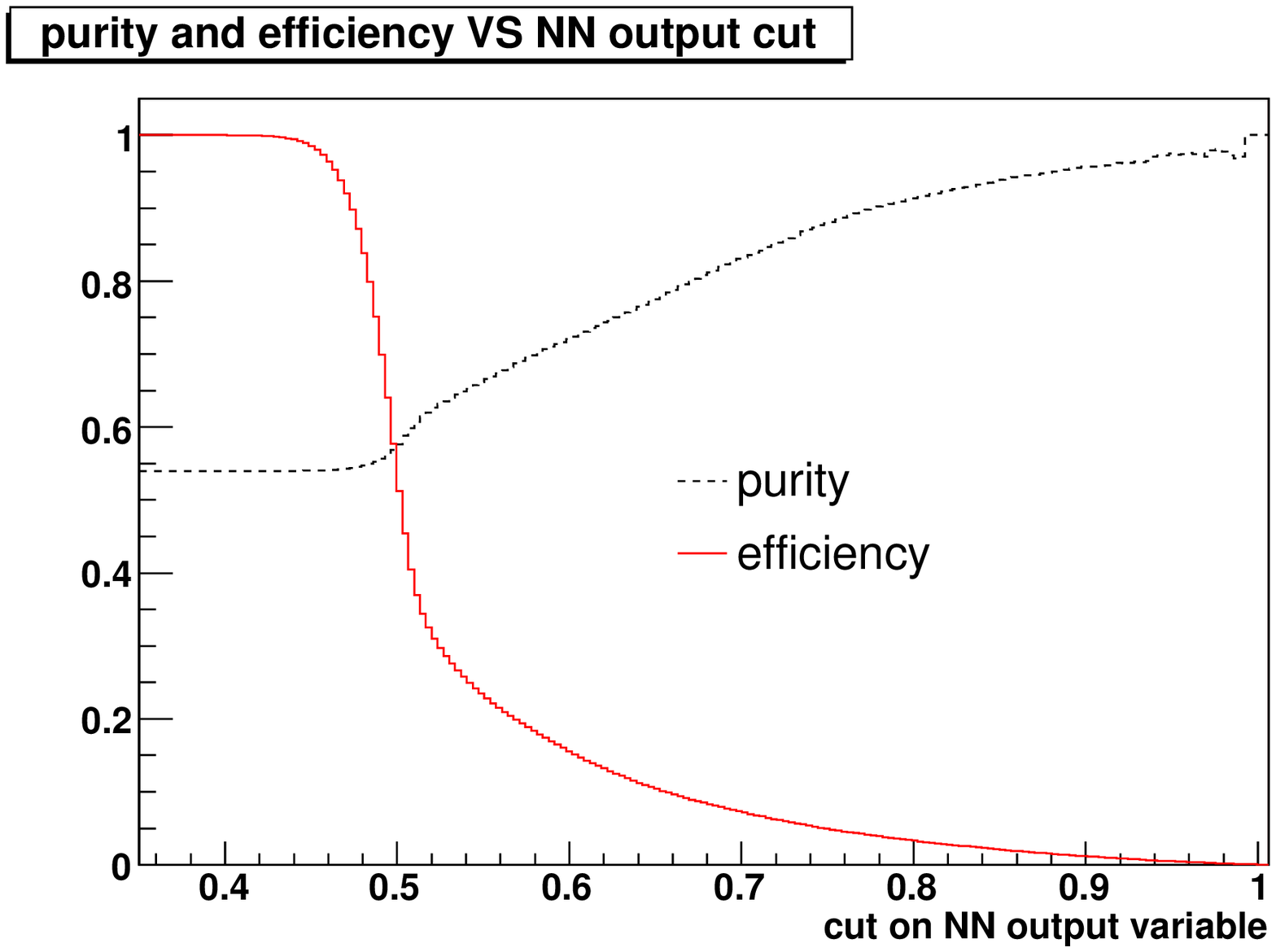,width=8cm}} 
\caption{ $d\sigma/dNN$ where NN is the neural network
output variable. in red (full line) 
for the no-Higgs case and in black (dashed) for M(H)=500 \GeV on the LHS.
The corresponding purity and efficiency on the RHS}
\label{nn-output}
\end{center}
\end{figure}

The neural network, a multilayer perceptron with  BFGS training, available
in the ROOT package \cite{root},
takes these four variables as input, has 
two intermediate layers,
with eight and four neurons respectively, and one output variable.
The differential cross section $d\sigma /dNN$ where NN is the neural network
output variable is shown on the left hand side of
 \fig{nn-output} for two event samples corresponding to the different Higgs
masses. Applying a cut on NN we can enhance
the separation between the heavy and light Higgs case.
The corresponding purity and efficiency is shown on the right hand side.

In \tbn{nn-tab} the integrated cross section and the number of events for
the two Higgs cases for M(\VW)$>$ 800 \GeV are shown for different values 
of the cut. The corresponding ratio is also presented. The predictions for 
M(H)=500 \GeV have a statistical uncertainty of about 10\% for NN $>$ 0.56,
while the very heavy Higgs case, with all the necessary caveats,
predicts a cross section which is larger by about a factor of two.
This is quite encouraging and suggests that further study should be worthwhile
in order to sharpen the SM predictions.
 
\begin{figure}
\begin{center}
\mbox{\epsfig{file=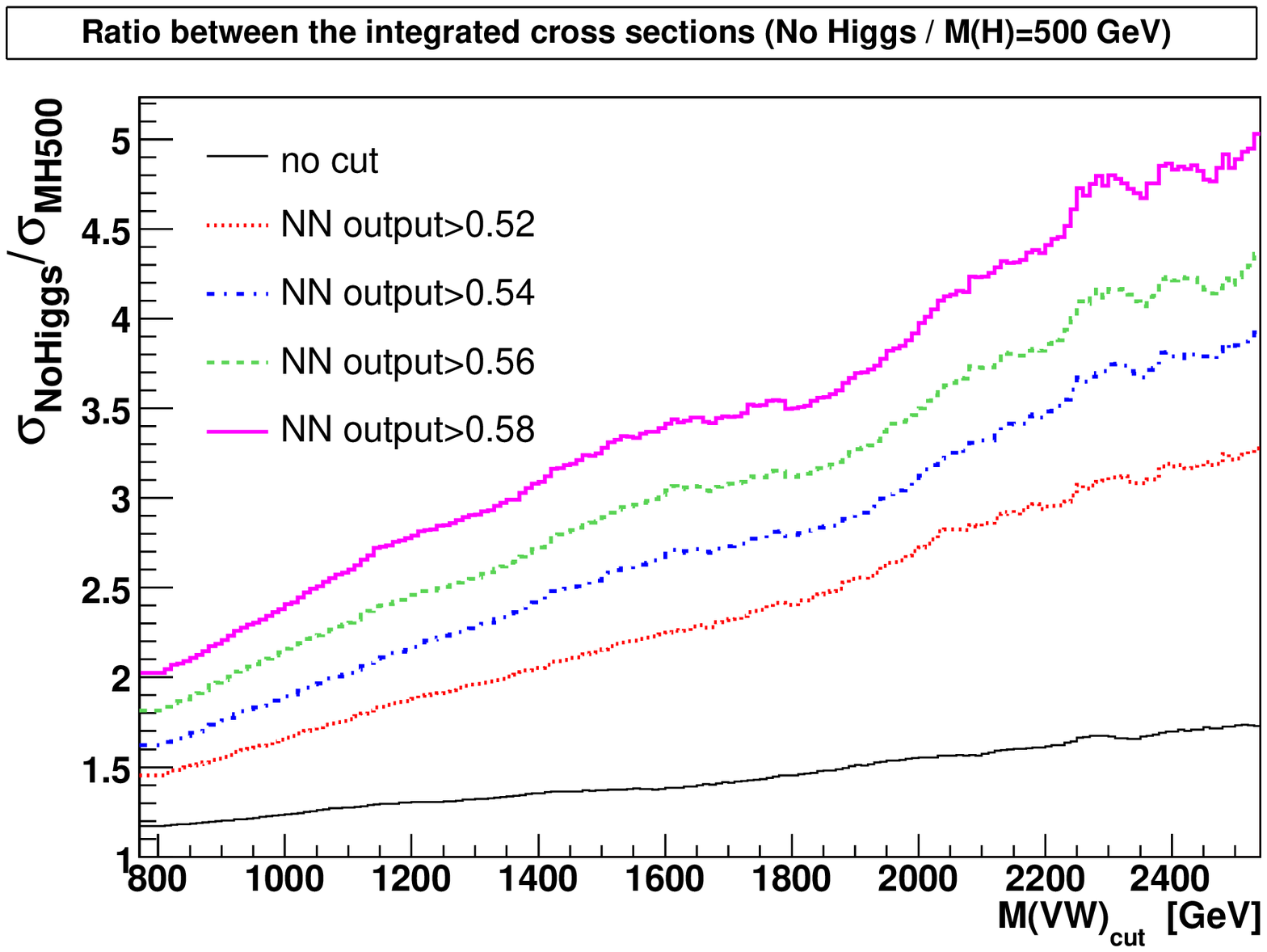,width=8cm}
\epsfig{file=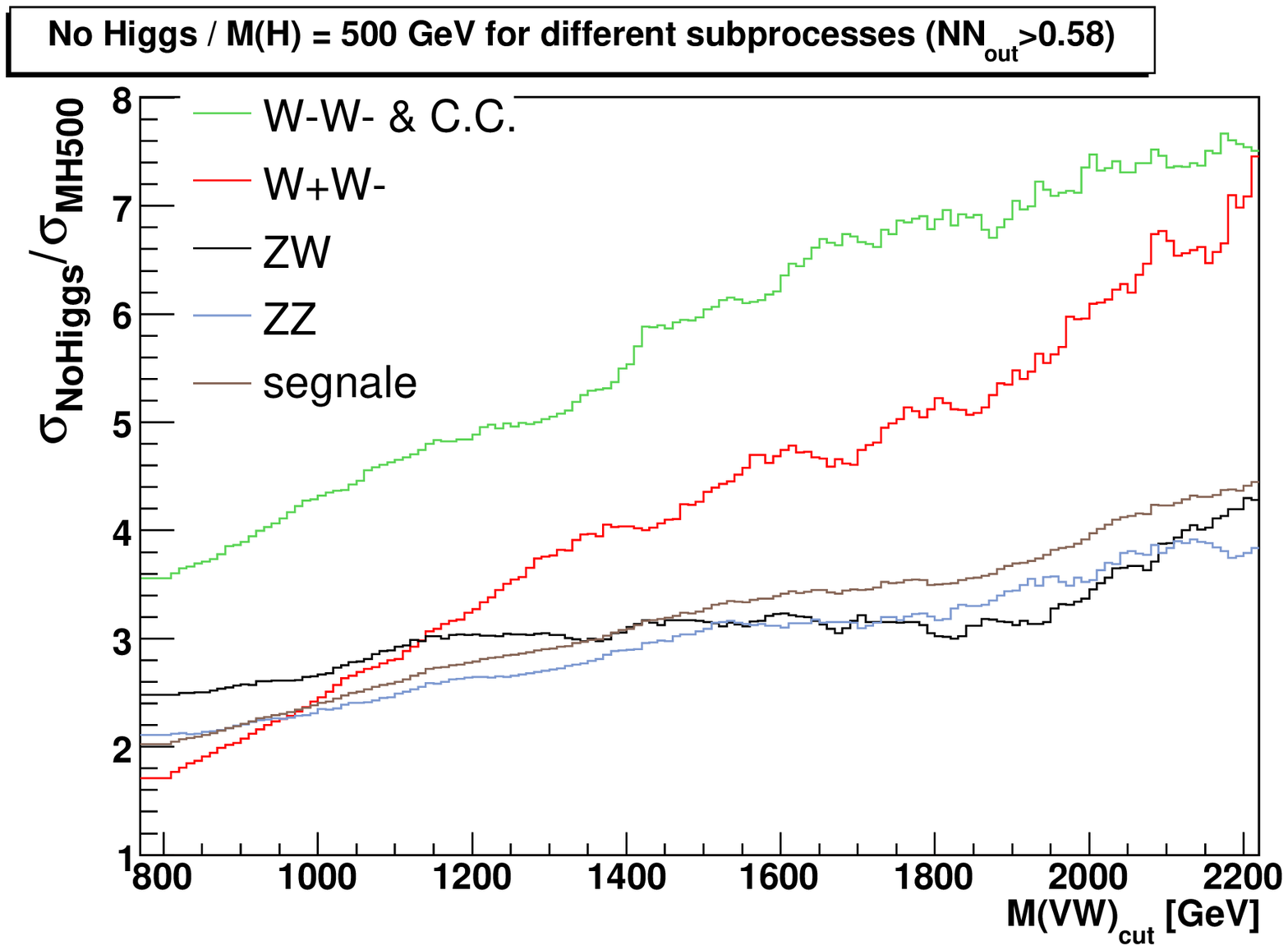,width=8cm}} 
\caption{ On the left, the ratio of cross sections, integrated between $M_{cut}$
and infinity, as a function of
$M_{cut}$, for different values of the neural network variable.
On the right,the ratio \eqn{ratiox} for
different groups of processes under the condition NN $>$ 0.58}
\label{xsec_ratio}
\end{center}
\end{figure}

In \fig{xsec_ratio} we show the ratio
\be
\frac
{\displaystyle{\int^\infty_{M_{cut}}dM_{VW}\frac{d\sigma_{noHiggs}}{dM_{VW}}}}
{\displaystyle{\int^\infty_{M_{cut}}dM_{VW}\frac{d\sigma_{M_H=500}}{dM_{VW}}}}
\label{ratiox}
\ee
for different cuts on the neural network output variable as a function of
$M_{cut}$, the lower limit of the integration range on $M_{VW}$.
The difference between the two cross sections is concentrated at large values
of NN and increases with increasing values of the output variable.
With a fixed lower cut on NN the ratio between integrated cross sections
increases with $M_{cut}$. With  $M_{cut} > 1$ \TeV and NN $>0.56$ the cross
section is about 2 fb for a very heavy Higgs, roughly twice the expected yield
for a light Higgs scalar.
On the right hand side of \fig{xsec_ratio} we show the ratio \ref{ratiox}
for different group of processes. The set which includes $W^\pm W^\pm$ is the one
with the largest separation, while the sets including $ZZ$ and $ZW$ scattering
are less sensitive to the cut on NN.

\section{Summary and future}

In this paper we have studied all $q_1 q_2 \rightarrow q_3 q_4 q_5 q_6 l \nu$
processes at order {$\O(\alpha_{em}^6)$} at the LHC
using for the first time a full fledged six fermion Monte Carlo event generator.
We have examined how simple kinematical cuts can be applied at generator level
to extract the \VV signal from the irreducible background.
In the high mass region we have compared the case of a relatively light Higgs
with the no-Higgs case as a guide to separate the LL component of
\VV scattering.
Employing a neural network approach it seems possible to obtain a good
separation of the two cases.

Work is in progress to extend \Phase to cover \mbox{${\mathrm qqZZ}~$}
final states.
The calculation of the full set of processes $2 \rightarrow 6$ at 
$\O(\alpha_{s}^2\alpha_{em}^4)$ is under way.
A phenomenological study of \VV processes with both vectors decaying
leptonically is in project.


\begin{thebibliography}{999}

\bibitem{HiggsLHC} Proceedings of the Large Hadron Collider Workshop, 
Aachen 1990, CERN Report 90--10, G. Jarlskog and D. Rein (eds.). 

\bibitem{djouadi-rev1} A.~Djouadi, {\it The Anatomy of Electro--Weak Symmetry Breaking.
Tome I: The Higgs in the Standard Model}, [hep-ph/0503172].

\bibitem{ATLAS-TDR} ATLAS Collaboration, {\it Detector and Physics Performance
Technical Design Report}, 
Vols. 1 and 2, CERN--LHCC--99--14 and CERN--LHCC--99--15.

\bibitem{Houches2003} K.A. Assamagan, M. Narain, A. Nikitenko, M. Spira, 
D. Zeppenfeld (conv.) {\it et al.}, Report of the Higgs Working Group,
Proceedings of the Les Houches Workshop on ``Physics at 
TeV Colliders'', 2003, [hep-ph/0406152]. 

\bibitem{lepewwg}
The LEP Collaborations (ALEPH, DELPHI, L3 and OPAL), 
the LEP Electroweak Working Group and the SLD Heavy Flavour Group, {\it A
combination of preliminary Electroweak measurements and constraints on the
Standard Model}, [hep-ex/0412015]; {\tt http://lepewwg.web.cern.ch/LEPEWWG}.

\bibitem{peskin-wells-01}
M.E.~Peskin, J.D.~Wells, \pr D64 2001 093003 , [hep-ph/0101342].

\bibitem{reviews}
M.S.~Chanowitz, {\it Strong WW scattering at the end of the 90's:
theory and experimental prospects}.    
In {\it Zuoz 1998, Hidden symmetries and Higgs phenomena} 81-109.
[hep-ph/9812215]

\bibitem{unitarization}
J.~Bagger {\it et al.}, \pr D52 1995 3878 ;
A.~Dobado, M.J.~Herrero, J.R.~Pel\'aez and E.~Ruiz~Morales,
\pr D62 2000 055011 ,[hep-ph/9912224];
J.M.~Butterworth,B.E.~Cox and J.R.~Forshaw, \pr D65 2002 96014 . 
[hep-ph/0201098]

\bibitem{history1}
M.J.~Duncan, G.L.~Kane and W.W.~Repko, \np B272 1986 517 ;
D.A.~Dicus and R.~Vega, \prl 57 1986 1110 ; 
J.F.~ Gunion, J.~ Kalinowski and A.~Tofighi--Niaki, \prl 57 1986 2351 .

\bibitem{history2}
 R.N. Cahn, S.D. Ellis, R. Kleiss and W.J. Stirling, Phys. Rev. D35 (1987) 1626;
 V. Barger, T. Han and R. Phillips, Phys. Rev. D37 (1988) 
2005 and D36 (1987) 295; R. Kleiss and J. Stirling, Phys. Lett. 200B (1988) 193;
V.\, Barger {\it et al.}, Phys.\ Rev.\ D42 (1990) 3052; {\it ibid.}  Phys.\, 
Rev.\, D44 (1991)  1426; {\it ibid.}  Phys.\ Rev.\ D46 (1992) 2028; 
D.~Froideveaux, in Ref.~\cite{HiggsLHC} Vol~II, p.~444;
M.~H.~Seymour, {\it ibid.}, p.~557;
U.~Baur and E.W.N.~Glover, Phys.\ Lett.\ B252 (1990) 683;  
D. Dicus, J. Gunion and R. Vega, Phys. Lett. B258 
(1991) 475; D. Dicus, J. Gunion, L. Orr and R. Vega, Nucl. Phys. B377 (1991) 
31; J.\,Bagger {\it et al.},\pr D49 1994 1246;
V.\, Barger, R.\, Phillips and D.\, Zeppenfeld, \pl B346 1995 106 ;
J.\,Bagger {\it et al.},\pr D52 1995 3878;
K. Iordanidis and D. Zeppenfeld, \pr D57 1998 3072 ; R. Rainwater and
D. Zeppenfeld, \pr D60 1999 113004 ; erratum ibid D61 (2000) 099901.  

\bibitem{EVBA}
M.S. Chanowitz and M.K. Gaillard,
\np B261 1985 379 .
M.S. Chanowitz and M.K. Gaillard, {\it Phys.  Lett.} 
142B, 85 (1984) and ref. 1;  G. Kane, W. Repko, B. Rolnick, {\it 
Phys.  Lett.} B148, 367 (1984); S. Dawson, \np B29 1985 42 .

\bibitem{ref:Phase}E.~Accomando, A.~Ballestrero, E.~Maina, 
\jhep 0507 2005 016 , [hep-ph/0504009].

\bibitem{method} A.~Ballestrero and E.~Maina, \pl B350 1995 225 ,
[hep-ph/9403244].

\bibitem{phact} 
A.~Ballestrero, {\tt PHACT 1.0 - \it Program for Helicity Amplitudes Calculations 
with Tau matrices'} [hep-ph/9911318] in {\it 
Proceedings of the 14th International Workshop on High Energy Physics 
and Quantum Field Theory (QFTHEP 99)}, 
B.B.~Levchenko and V.I.~Savrin  eds. (SINP MSU Moscow), pg. 303. 

\bibitem{ACAT-QFTHEP}
E.~Accomando, A.~Ballestrero, E.~Maina, Talk given at 9th International
Workshop on Advanced Computing and Analysis Techniques in Physics
Research (ACAT 03), Tsukuba, Japan, 1-5 Dec 2003.
{\it Nucl.Instrum.Meth.}A534:265-268,2004, [hep-ph/0404236];
E.~Accomando, A.~Ballestrero, E.~Maina, Proceedings of 18th International
Workshop on High-Energy Physics and Quantum Field Theory (QFTHEP 2004),
St. Petersburg, Russia, 17-23 Jun 2004. [hep-ph/0505225]

\bibitem{CTEQ5}
CTEQ Coll.(H.L.~Lai {\it et al.}) \epj C12 2000 375 .

\bibitem{pythia}
T.~Sj\"ostrand {\it et al.}, \cpc 135 2001 238 , [hep-ph/0010017];\\
T.~Sj\"ostrand, L.~L\"onnblab and S.~Mrenna, [hep-ph/0108264].

\bibitem{mad}
F.~Maltoni, T.~Stelzer, JHEP 0302 (2003) 027;
T.~Stelzer and W.~F.~Long, Comput. Phys. Commun. {\bf 81} (1994) 357;\\
H. Murayama, I. Watanabe and K. Hagiwara, KEK-91-11.

\bibitem{Atlas_HinWW} S.~Asai {\it et al.},
Eur.Phys.J.C32S2:19-54,2004, [hep-ph/0402254].

\bibitem{CMS-0}
R.~Bellan, PhD Thesis, http://www.to.infn.it/~bellan/works/tesi/tesi\_bellan.pdf

\bibitem{root}
http://root.cern.ch 

\end{thebibliography}
\end{document}